\documentclass[10pt]{article}
\usepackage{amsmath}
\usepackage{graphicx}
\usepackage{amsfonts}
\usepackage{amssymb}
\usepackage{epsf} 
\usepackage{latexsym}

\def\be{\begin{equation}}
\def\ee{\end{equation}}
\def\bea{\begin{eqnarray}}
\def\eea{\end{eqnarray}}
\def\Rho{\mbox{\Large $\rho$}}

\def\brho{\mbox{\boldmath$\rho$}}
\def\bpi{\mbox{\boldmath$\pi$}}

\def\tta{\mbox{\tt a}}

\def\ttB{\mbox{\tt B}}

\def\ttD{\mbox{\tt D}}

\def\mathfrakV{\mbox{\LARGE $\mathfrak{v}$}}
\def\MFF{\mbox{\LARGE $\mathfrak{f}$}}
\def\MFO{\mbox{\LARGE $\mathfrak{o}$}}
\def\MFM{\mbox{\LARGE $\mathfrak{m}$}}

\def\ttI{\mbox{\tt I}}
\def\ttJ{\mbox{\tt J}}

\def\ttN{\mbox{\tt N}}

\def\ttP{\mbox{\tt P}}
\def\ttQ{\mbox{\tt Q}}
\def\ttS{\mbox{\tt S}}

\def\ttg{\mbox{\tt g}}
\def\tta{\mbox{\tt a}}

\def\ttc{\mbox{\tt c}}

\def\ttg{\mbox{\tt g}}

\def\stta{\mbox{\scriptsize ${\tta}$}}

\def\sttc{\mbox{\scriptsize $\bar{\ttc}$}}

\def\sttg{\mbox{\scriptsize ${\ttg}$}}

\def\s{\mbox{s}}

\def\mn{\mbox{\scriptsize n}}

\def\NSI{Na\"{\i}ve Schr\"{o}dinger Interpretation }
\def\CPI{Conditional Probabilities Interpretation }
\def\NSII{Na\"{\i}ve Schr\"{o}dinger Interpretation}
\def\CPII{Conditional Probabilities Interpretation}

\def\uRR{{\underline{\mbox{R}}}}

\def\uPP{{\underline{\mbox{P}}}}

\def\foo{\footnote}
\def\hat{\widehat}

\def\beq{\begin{equation}}
\def\eeq{\end{equation}}
\def\bea{\begin{eqnarray}}
\def\eea{\end{eqnarray}}
\def\pa{\partial}
\def\d{\textrm{d}}

\def\ttD{\mbox{\tt D}}
\def\ttP{\mbox{\tt P}}

\def\ttQ{\mbox{\tt Q}}

\def\bttQ{\mbox{\boldmath {\tt Q}}}

\def\Circ{\mbox{\Large$\circ$}}

\def\tea{\mbox{\tt t}}
\def\stea{\mbox{\scriptsize\tt t}}
  
\def\Circ{\mbox{\Large$\circ$}}

\def\Star{\mbox{\Large$\ast$}}
\def\5Star{\mbox{\Large$\star$}}

\def\cr{\mbox{\scriptsize{\bf $\mbox{ } \times \mbox{ }$}}}

\def\suma{\sum\mbox{}_{\mbox{}_{\mbox{\scriptsize $i$}}}}
\def\sumi3{\sum\mbox{}_{\mbox{}_{\mbox{\scriptsize $i$=1}}}^3}
\def\sumin{\sum\mbox{}_{\mbox{}_{\mbox{\scriptsize $i$=1}}}^{n}}

\def\sumj3{\sum\mbox{}_{\mbox{}_{\mbox{\scriptsize $j$=1}}}^3}
\def\sumk3{\sum\mbox{}_{\mbox{}_{\mbox{\scriptsize $k$=1}}}^3}

\def\sss{\mbox{\scriptsize s}}

\def\mg{\mbox{g}}
\def\mh{\mbox{h}}
\def\mi{\mbox{i}}

\def\mn{\mbox{n}}

\def\mp{\mbox{p}} 
\def\mq{\mbox{q}}

\def\ms{\mbox{s}}
\def\mt{\mbox{t}}

\def\mC{\mbox{C}}
\def\mD{\mbox{D}}

\def\mF{\mbox{F}}

\def\mI{\mbox{I}}

\def\mL{\mbox{L}}
\def\mM{\mbox{M}}
\def\mN{\mbox{N}} 

\def\mP{\mbox{P}}
\def\mQ{\mbox{Q}}

\def\mT{\mbox{T}} 

\def\mV{\mbox{V}}
\def\mW{\mbox{W}}

\def\mZ{\mbox{Z}} 
\def\sa{\mbox{\scriptsize a}}
\def\sb{\mbox{\scriptsize b}}
\def\sc{\mbox{\scriptsize c}}
\def\sd{\mbox{\scriptsize d}}
\def\se{\mbox{\scriptsize e}}
\def\sf{\mbox{\scriptsize f}}
\def\sg{\mbox{\scriptsize g}} 
\def\sh{\mbox{\scriptsize h}} 
\def\si{\mbox{\scriptsize i}}
 
\def\sk{\mbox{\scriptsize k}}
\def\sll{\mbox{\scriptsize l}}  
\def\sm{\mbox{\scriptsize m}}
\def\sn{\mbox{\scriptsize n}} 
\def\so{\mbox{\scriptsize o}} 
\def\sp{\mbox{\scriptsize p}}
\def\sq{\mbox{\scriptsize q}}
\def\sr{\mbox{\scriptsize r}}

\def\st{\mbox{\scriptsize t}}
\def\su{\mbox{\scriptsize u}}

\def\sw{\mbox{\scriptsize w}}

\def\sy{\mbox{\scriptsize y}} 
 
\def\sA{\mbox{\scriptsize A}} 
\def\sB{\mbox{\scriptsize B}}

\def\sD{\mbox{\scriptsize D}}
\def\sE{\mbox{\scriptsize E}}
\def\sF{\mbox{\scriptsize F}}
\def\sG{\mbox{\scriptsize G}}
\def\sH{\mbox{\scriptsize H}}
\def\sI{\mbox{\scriptsize I}}
\def\sJ{\mbox{\scriptsize J}}

\def\sL{\mbox{\scriptsize L}} 
\def\sM{\mbox{\scriptsize M}} 
\def\sN{\mbox{\scriptsize N}} 

\def\sP{\mbox{\scriptsize P}} 
 
\def\sR{\mbox{\scriptsize R}}
\def\sS{\mbox{\scriptsize S}}
\def\sT{\mbox{\scriptsize T}}

\def\sW{\mbox{\scriptsize W}}
 
\def\sY{\mbox{\scriptsize Y}} 
\def\sZ{\mbox{\scriptsize Z}} 
\def\eph{\mbox{\scriptsize eph}}
 
\def\eph(B){\mbox{\scriptsize em(JBB)}}

\def\ttP{\mbox{\tt P}}

\def\eph(B){\mbox{\scriptsize emergent(JBB)}}

\def\ta{\mbox{\tiny a}}

\def\td{\mbox{\tiny d}}
\def\te{\mbox{\tiny e}}

\def\th{\mbox{\tiny h}}
\def\ti{\mbox{\tiny i}}

\def\tl{\mbox{\tiny l}}

\def\tm{\mbox{\tiny m}}
\def\tn{\mbox{\tiny n}}
\def\to{\mbox{\tiny o}}
\def\tp{\mbox{\tiny p}}

\def\tr{\mbox{\tiny r}}

\def\ttt{\mbox{\tiny t}}
\def\tA{\mbox{\tiny A}}

\def\tE{\mbox{\tiny E}}

\def\tG{\mbox{\tiny G}}

\def\tJ{\mbox{\tiny J}}

\def\tM{\mbox{\tiny M}}

\def\tP{\mbox{\tiny P}}

\def\tR{\mbox{\tiny R}}

\def\sbfM{\mbox{\bf \scriptsize\sffamily M}}

\def\bfM{\mbox{{\bf \sffamily M}}}

\def\bg{\mbox{{\bf g}}}

\def\bh{\mbox{{\bf h}}}

\def\fs{\mbox{\sffamily s}}

\def\FA{\mbox{\Large $\mathfrak{a}$}}

\def\fA{\mbox{\sffamily A}}
\def\fB{\mbox{\sffamily B}}
\def\fC{\mbox{\sffamily C}}

\def\fE{\mbox{\sffamily E}}

\def\fG{\mbox{\sffamily G}}
\def\fH{\mbox{\sffamily H}}

\def\fL{\mbox{\sffamily L}}
\def\fM{\mbox{\sffamily M}}
\def\fN{\mbox{\sffamily N}}

\def\fQ{\mbox{\Large $\mathfrak{q}$}}

\def\BigfQ{\mbox{\LARGE $\mathfrak{q}$}}

\def\sFG{\mbox{$\mathfrak{g}$}}

\def\fR{\mbox{\Large $\mathfrak{r}$}}
\def\fS{\mbox{\Large $\mathfrak{s}$}}
\def\fG{\mbox{\Large $\mathfrak{g}$}}

\def\FS{\mbox{\LARGE\tt s}}
\def\stS{\mbox{\large\tt s}}
\def\fP{\mbox{\sffamily P}}
\def\fT{\mbox{\sffamily T}}
\def\fU{\mbox{\sffamily U}}
\def\fV{\mbox{\sffamily V}}
\def\fW{\mbox{\sffamily W}}

\def\fZ{\mbox{\sffamily Z}}
\def\sfa{\mbox{\sffamily{\scriptsize a}}}
\def\sfb{\mbox{\sffamily{\scriptsize b}}}

\def\sfA{\mbox{\sffamily{\scriptsize A}}}
\def\sfB{\mbox{\sffamily{\scriptsize B}}}
\def\sfC{\mbox{\sffamily{\scriptsize C}}}

\def\sfT{\mbox{\sffamily{\scriptsize T}}}

\def\sfZ{\mbox{\sffamily{\scriptsize Z}}}
\def\tfA{\mbox{\sffamily{\tiny A}}}
\def\tfB{\mbox{\sffamily{\tiny B}}}

\def\tfZ{\mbox{\sffamily{\tiny Z}}}

\renewcommand{\H}{{\cal H}}                 % !!! Command redefined !!!

\def\K{Kucha\v{r} }

\begin{document}
\begin{titlepage}
\vspace{.7in}
\begin{center}
\Huge{\bf RELATIONALISM}\normalsize 

\vspace{.1in}

\normalsize

\vspace{.4in}

{\large \bf Edward Anderson$^*$}

\vspace{.2in}

\large {\em APC AstroParticule et Cosmologie, Universit\'{e} Paris Diderot CNRS/IN2P3, CEA/Irfu, 
Observatoire de Paris, Sorbonne Paris Cit\'{e}, 10 rue Alice Domon et L\'{e}onie Duquet, 75205 Paris Cedex 13, France. } \normalsize

\vspace{.2in}

\end{center}

\begin{abstract}

\mbox{ }

This article contributes to the debate of the meaning of relationalism and background independence, which has remained of interest in 
theoretical physics from Newton versus Leibniz through to foundational issues for today's leading candidate theories of quantum gravity.  
I contrast and compose the substantially different Leibniz--Mach--Barbour (LMB) and Rovelli--Crane (RC) uses of the word `relational'.  
Leibniz advocated primary timelessness and Mach that `time is to be abstracted from change'.  
I consider 3 distinct viewpoints on Machian time: Barbour's, Rovelli's and my own.  
I provide four expansions on Barbour's taking configuration space to be primary: to (perhaps a weakened notion of) phase space,  
categorizing, perspecting and propositioning.  
Categorizing means considering not only object spaces but also the corresponding morphisms and then functors between such pairs.
Perspecting means considering the set of subsystem perspectives; this is an arena in which the LMB and Rovelli approaches make contact.
By propositioning, I mean considering the set of propositions about a physical (sub)system.
I argue against categorization being more than a formal pre-requisite for quantization in general; however, perspecting is a categorical operation, and propositioning leads one to considering topoi, with Isham and Doering's work represents one possibility for a mathematically sharp implementation of propositioning. 
Further applications of this article are arguing for Ashtekar variables as being relational in LMB as well as just the usually-ascribed RC sense, relationalism versus supersymmetry, string theory and M-theory.  
The question of whether scale is relational is also considered, with quantum cosmology in mind.  

\end{abstract}

\vspace{2in} 

%
%Fundamental problems and general formalism    and    Quantum Gravity
 
\vspace{2in}   
 
\noindent $^*$ edward.anderson@apc.univ-paris7.fr

\end{titlepage}

%=========================================================================================================
%==========================================================================================================
\section{Introduction}
%==========================================================================================================
%==========================================================================================================

This article concerns relationalism.  
This word is used very differently by Barbour (and various collaborators) and I \cite{BB82, B94I, RWR, ABFKO, Phan, FORD, FileR} on the one hand, 
and by Crane \cite{Crane93, Crane} and Rovelli \cite{Rov96, Rovellibook}  (with subsequent use in Loop Quantum Gravity (`LQG') \cite{Montes, Rovellibook, Thiemann, GPPT, BojoBook, Rovellifqxi, Bojo1}) on the other hand.  
This article is mostly about the former (a successor to/considerable extension of \cite{BB82, B94I, RWR}), though it does contrast, and to some extent compose, this with the latter.  
Moreover, the position developed here most ly lies {\sl between} Barbour's \cite{B94I, fqxi} and Rovelli's \cite{Rovellibook, Rovellifqxi} extremes regarding the fine detail of the Machian notion of time as being abstracted from change.  
\noindent As regards other recent reviews, Pooley's \cite{Pooley12} is a 3-way comparison of which three Barbour's alone features in this review 
(I go for the approaches that enter Quantum Gravity programs, whereas Pooley goes for the past 2 decades' philosophizing).  
Barbour's article \cite{B11} and Gryb's Thesis \cite{GrybTh} largely only cover Barbour's approach.

\mbox{ }  

In Sec 2, I sketch the history of the absolute versus relational motion debate, which remains an active source of contention between a number of different approaches to the unsolved problem of Quantum Gravity (`QG') today.  
In fact, I consider QG to stand for {\it Quantum Gestalt}, a term that I coin in recognition of GR being not only a relativistic theory of gravitation but also, following Einstein, a freeing of physics from absolute/background structure; the expression `Quantum Gravity' fails to capture the latter.  
Thus I am taking `Quantum Gestalt' to be shorthand for Quantum Theory of the Relativistic Gravity--Background Independence Gestalt.
Certainly most LQG practitioners agree that what they are studying is also intended as a Quantum Theory of Background Independence; this is also to be understood of some approaches to M-theory (but not of perturbative String Theory).    
[As to other names, `Quantum GR' will not do in this role since it implies the specific Einstein field equations, whereas Quantum Gestalt remains open-minded as to which relativistic theory of gravitation is involved; this distinction parallels that between `Quantum GR' and `Quantum Gravity'.  
Quantum Gestalt is also in contradistinction with `Background Independent Quantum Gravity' since the latter carries connotations of the background 
independent and gravitational parts being {\sl split}, whereas in the former they are held to be {\sl joint}.  
Finally, much as one can model Quantum Gravity without background independence (at least for some purposes), one can likewise model Quantum Background Independence without relativistic gravitation; from the Quantum Gestalt perspective, these two ventures are complements, suggesting that one may learn considerably from the latter as well as from the former.  
Relational Particle Mechanics (RPM) \cite{BB82, B03, FORD}, once quantized \cite{FileR}, is such a theory of Quantum Background Independence without relativistic gravitation.]
Sec 2 also suggests some simple informal interpretations that might be taken to underlay some of the Leibniz- and Mach-type ideas that are themselves presented in subsequent sections.

In Sec 3, I consider configuration space $\fQ$ and Barbour's ascribing primality to it in his relational program \cite{BB82, Buckets, RWR, B03, ABFO, ABFKO, fqxi, B11}. 
Whilst Leibniz \cite{L} argued for time not to be primary, Mach \cite{M} argued for time to be an emergent concept, more specifically that 
`time is to be abstracted from change'.  
This program's subsequent build-up (see also \cite{B94I, FEPI, FileR}) of actions that are temporally relational (Leibnizian) and configurationally relational with respect to some group $\fG$ of transformations that are held to be unphysical, is reviewed and generalized in Secs 4 to 8.
This includes using Jacobi--Synge \cite{Lanczos, Synge} type actions, dispensing with velocity as a primary notion and recasting the canonical definition of momenta accordingly, as well as using Barbour and Bertotti's \cite{BB82} indirect implementation of configurational relationalism and the Jacobi--Barbour--Bertotti realization of a Machian emergent time.  
Sec 8 discusses the alternative direct implementation \cite{FORD, FileR} of configurational relationalism that is occasionally available.  
Sec 9 provides a number of examples, including 1- and 2-$d$ RPM's for which the direct implementation is available \cite{FORD, Cones, FileR}.  
It also covers how, whilst GR is usually formulated in a spacetime way that does not directly implement Mach's ideas as understood in the relational program, can nevertheless be cast in a manifestly relational form \cite{RWR, San, Phan, Lan2, FEPI} that is quite akin to geometrodynamics, and even more akin to the Baierlein--Sharp--Wheeler \cite{BSW} and Christodoulou Chronos Principle \cite{Christodoulou1, Christodoulou2} formalisms. 
Moreover, whilst this LMB (Leibniz--Mach--Barbour) program's literature usually uses this more traditional dynamical formulation of GR, I lay out 
how the Ashtekar variables approach is also LMB-relational.  
I also provide some comments on relationalism versus supersymmetry (see \cite{Arelsusy} for more), String Theory and M-Theory.  
As such, I make contact between LMB relationalism and much of modern Theoretical Physics.  
Given these examples, I am in a position to comment further on compatibility restrictions between the $\fQ$'s and the $\fG$'s in Sec 10.

I then consider four expansions of Barbour's `$\fQ$ is primary' postulate: to (perhaps weakened) phase space in Sec 11, 
to categorization in Sec 12, to propositioning in Sec 13 and to `perspectivalism' in Sec 14.  
Categorizing means considering not only object spaces but also the corresponding morphisms and then functors between such pairs.
Perspecting means considering the set of subsystem perspectives; this is an arena in which the LMB and RC approaches make contact.
By propositioning, I mean considering the set of propositions about a physical (sub)system; this makes contact with ideas of Isham, Doering and Butterfield \cite{bigcite2, ToposI, ToposTalk, ToposRev} concerning applying Topos Theory to Quantum Theory). 
Perspecting makes substantial contact with the other main line of work termed `relational' in the Quantum Gravity literature, i.e. Rovelli 
and Crane's works.    
In Sec 15 I consider a substantial generalization of Barbour's `best matching' indirect implementation of configurational relationalism, that 
also happens to generalize the mathematically well-known notion of group-averaging.

I next consider the corresponding situation at the quantum level in Secs 16 to 21. 
This is entwined with a number of facets of the notorious Problem of Time and of relevance as regards selection of strategies to attempt to 
resolve this.   
Sec 16 covers temporal and configurational relationalism within a quantum scheme along the lines of Isham 1984 \cite{I84}.  
I include analysis of the Problem of Time facets and strategies in accord to the extent to which their attitude to time is Leibniz--Mach.  
I illustrate this with examples in Sec 17.
I then consider  $\fQ$ primality at the quantum level in Sec 18, 
provide arguments against considering quantization in general as a functor in Sec 19, 
and discuss QM perspectivalism in Sec 20 with Rovelli and Crane inputs and QM propositioning in Sec 21 with contact with Isham and Doering's work.

My Conclusion (Sec 22) covers the presently-known principal applications of the current article. 
1) Pointers relating Machian time interpretations to the formulations of timeless approaches \cite{ARel2}, 
2) other Problem of Time applications, 
3)  the issue of whether scale should be among the concepts discarded by relationalism (which argument is tied to the semiclassical approach to 
Quantum Cosmology).  
4) This article comments on the foundations of, and relational notions applicable in, Ashtekar variables/LQG.  
5) It also suggests contention between Relationalism and Supersymmetry \cite{Arelsusy}. 
6) Finally, I outline various possibilities for unifying Isham--Doering propositioning with each of Rovelli--Crane perspecting and Records Theory.

%=========================================================================================================
%==========================================================================================================
\section{Some informalities}
%==========================================================================================================
%==========================================================================================================

%==========================================================================================================
\subsection{Historical background: absolute or relational motion debate}
%==========================================================================================================

This debate has been around since the inception of Newtonian Mechanics.\footnote{I use `relational' here rather than `relative' here to avoid 
%%%%%%%%%%%%%%%%%%%%%%%%%%%%%%%%%%%%%%%%%%%%%%%%%%%%%%%%%%%%%%%%%%%%%%%%%%%%%%%%%%%%%%%%%%%%%%%%%%%%%%%%%%%
confusion with Einstein's relativity, which also features prominently in the present article.    
Philosophers of physics have also taken to using `substantival' instead of `absolute'.}
%%%%%%%%%%%%%%%%%%%%%%%%%%%%%%%%%%%%%%%%%%%%%%%%%%%%%%%%%%%%%%%%%%%%%%%%%%%%%%%%%%%%%%%%%%%%%%%%%%%%%%%%%%%
%
In outline, Newton's \cite{Newton} traditional formulation of mechanics defined `true' motion to be 
relative to an absolute space that is all-pervasive, infinite, invisible and can not be acted upon.  
Newton also considered motion to occur {\sl in} time.  
His notion of time was additionally an absolute one.  
I.e., external, continuous, and uniformly flowing as `needed to transform kinematic geometry into physical dynamics' \cite{DOD}. 
(In fact, much of this conception of time was not new to Newton, c.f. Barrow \cite{Barrow}, Gassendi \cite{Gassendi} and even Ptolemy \cite{Ptolemy}.)
The virtue of Newton's scheme is working well in practise, nature immediately vindicating its practical consequences in detail.

However, Newton's formulation of mechanics has been argued to be bankrupt from a philosophical perspective. 
Absolute space does not comply with Leibniz's famous {\bf identity of indiscernibles} \cite{L}, by which e.g. our universe and a copy in which all material objects are collectively displaced by a fixed distance surely share all observable properties and thus are one and the same.  
Also, Leibniz already considered time not to be a separate external entity with respect to which things change.  
Bishop Berkeley \cite{Berkeley} and Mach \cite{M} (see also e.g. \cite{DOD, Buckets} for further discussion) added to these arguments, 
e.g. Mach holding absolute space to be a non-entity on account of its not being actable upon.  
The alternative is for mechanics to be {\bf relational}: spatial properties are to be entirely about the relations between 
material objects, and `time to be abstracted from change' (Mach's notion). 
Historically, there was a lack in alternative theories with such features, though the comparatively recent RPM's of Barbour--Bertotti (1982) 
and Barbour (2003) that are the subject of this article do now serve as such.

Furthermore, Leibniz, Bishop Berkeley and Mach's arguments are philosophically compelling enough that they 
ought to apply to not just mechanics but to physics as a whole.
As argued in Sec 9, GR is also relational, and indeed relationalism is one foundation from which GR 
{\sl can} be derived, for all that this was not the historical route to GR.  
For, while Einstein was interested in `Machian issues', the way in which he viewed these does not coincide with the Barbour-relational view \cite{WheelerGRT, DOD, RWR}, nor did Einstein's historical route to GR \cite{Einstein1, Einstein2} constitute a {\sl direct} implementation 
of Machian ideas (see Sec 9).

See e.g. \cite{B86, Buckets, Barbourphil3, B99, EOT, Hofer, Pooley012, Saunders12, ButterBar, Earman, Ryn, Sklar} 
as regards arguments for (and against) the value of Barbour and Bertotti's 1982 theory toward the absolute or relative motion debate.   
Relationalism would appear to have a good case for use in whole-universe situations, both from a 
philosophical perspective and because GR and at least some further GR-like theories do implement it.   
(These are good theories for whole-universe situations on a number of further grounds.)

%=============================================================================================================
\subsection{Underlaying of relational principles}
%=============================================================================================================

Some intuitive if vague underlying relational principles are as follows.  

\mbox{ }  

\noindent Relationalism 0) {\bf Relational physics is to be solely about the relations between tangible entities}

\mbox{ }

\noindent This is taken to be the universalization of Sec 2.1's statement of relationalism for mechanics to all physics.

\mbox{ } 

\noindent I use `{\bf tangible entities}' in place of `material objects' to make it clear that this is open to fields and `force mediators' 
as well as `matter building blocks', as befits modern physics.
The key properties of tangible entities are, rather, as follows. 

\mbox{ }

\noindent Relationalism 1) [Mach] These {\bf act testably and are actable upon}. 

\mbox{ }

\noindent Relationalism 2) [Leibniz] Any such which are {\bf indiscernible are held to be identical}. 

\mbox{ } 

%%%%%%%%%%%%%%%%%%%%%%%%%%%%%%%%%%%%%% NON-ENTITIES %%%%%%%%%%%%%%%%%%%%%%%%%%%%%%%%%%%%%%%%%%%%%%%%%%%%%%%%%%%%%%%%%%%%%%%%%%%%%%%%%%%%%%%%%%%%%%
%
\noindent Note 1) That which is not testably acting or actable upon is held to be, rather, a non-entity. 
[This is still held to be a type of entity as regards being able to philosophize about it or mathematically represent it, just not a 
{\sl physical} entity; absolute space is an obvious archetypal example.] 
This is less vague than saying `things', which could be {\sl merely} mathematical in nature.  

%%%%%%%%%%%%%%%%%%%%%%%%%%%%%%%%%%%%%% INDISCERNIBILITY VERSUS MATHEMATICAL REPRESENTATION %%%%%%%%%%%%%%%%%%%%%%%%%%%%%%%%%%%%%%%%%%%%%%%%%%%%%%%
%
\noindent Note 2) As regards Relationalism 2), in physics, relationalism posits that physical indiscernibility {\sl must} trump multiplicity of mathematical representation; this multiplicity still exists mathematically, but the true mathematics corresponding to the physics in question is the equivalence class spanning that multiplicity.
A range of attitudes to the identity of indiscernibles and alternatives to it are considered by Butterfield and Caulton \cite{BC11}
One would then only wish to attribute physical significance to computations of tangible entities which succeed in being independent 
of the choice of representative of the equivalence class; the archetype of this is Gauge Theory.  
Chapter 2 of \cite{FileR} also argues that dimension is partly undefined for a universe consisting of a sufficiently small number of particles.   
An important lesson from there is that a set of part-tangible and part-non entities is often altogether more straightforward to represent mathematically.

%%%%%%%%%%%%%%%%%%%%%%%%%%%%%%%%%%%%%%%%%% HETEROGENEITY %%%%%%%%%%%%%%%%%%%%%%%%%%%%%%%%%%%%%%%%%%%%%%%%%%%%%%%%%%%%%%%%%%%%%%%%%%%%%%%%%%%%%%%%%
%
\noindent Note 3) In this article's main sense of relational, one treats instantaneous configurations and time separately as befits the great conceptual heterogeneity between them.

%%%%%%%%%%%%%%%%%%%%%%%%%%%%%%%%%%%%%%%%%%%%%%%%% INDIRECTNESS %%%%%%%%%%%%%%%%%%%%%%%%%%%%%%%%%%%%%%%%%%%%%%%%%%%%%%%%%%%%%%%%%%%%%%%%%%%%%%%%%%%
%
\noindent Note 4) One largely accepts that the mathematics will be in general be most straightforward for mixtures of part-tangible- and part-non-entities.  

%%%%%%%%%%%%%%%%%%%%%%%%%%%%%%%%%%%%%%%%%%%%%%%%% THIS SEC IS BUT AN UNDERLAY %%%%%%%%%%%%%%%%%%%%%%%%%%%%%%%%%%%%%%%%%%%%%%%%%%%%%%%%%%%%%%%%%%%%
%
\noindent Note 5) I do not take this SSec's principles too seriously; it is the standard use of configuration space, and Relationalism 4--6) and 8), perhaps supplemented by Relationalism 3) and 7), that are the actually mathematically-implemented starting point of the LMB approach (and variants).  
As such, Relationalism 0--2) for now being vague or loose relative to Relationalism 3) to 8) is not a problem for the work actually done in this article, but only for this present SSec, which represents a tentative under{\sl laying} rather than a rigorous foundational under{\sl lying}.

%========================================================================================================
%========================================================================================================
\section{Configuration space} \label{SSSec: 1.1.1}
%========================================================================================================
%========================================================================================================

As regards the instantaneous configuration entities of a given theory, which I denote by ${\ttQ}^{\sfC}$, a key further concept is the 

\noindent{\sl configuration space} $\fQ$, i.e. the space of all the possible instantaneous configurations (each of the points in configuration space represents one state of the instantaneous configuration, i.e. a set of one value per ${\ttQ}^{\sfC}$).  

\mbox{ }

\noindent Example 1) Particle positions are configurations; see \cite{Lanczos} for a clear exposition of the 
configuration space notion for these.

\noindent Example 2) The more reduced inter-particle relative separations or relative cluster separations (see Sec 9.1)
are configurations.

\noindent Example 3) Sec 9.2 covers even more reduced relational mechanics configurations.  

\noindent Example 4) The values at each point of continuous extended objects, whether fields or geometrical objects that have values everywhere in space, which include both the 3-metrics (or the more reduced 3-geometries) of GR (see Sec 9.4) and objects that have their own separate {\it notion of space (NOS)} of extent such as strings or membranes. 

\mbox{ }

\noindent I denote the general NOS by $\Sigma_p$ for $p \leq d$ the dimension of space, and I coordinatize it by $x^{\mu_p}$. 
However, in this article all specific examples of this are for $p = d$ (field theories), in which case I 
drop the $p$ subscripts, or for $p = 0$, in which case $\Sigma_0$ is just a point corresponding to the particle position, which point ceases to possess any internal coordinatization. 
\noindent I denote integration over whatever NOS is appropriate by $\int_{\Sigma_p}\d \Sigma_p$. 
I take this to collapse to a multiplicative 1 for particle mechanics and other finite theories.  
\noindent I denote finite theory cases' instantaneous configurations by $Q^{\sfC}$.
For field theory cases, I denote the instantaneous configurations by $\mQ^{\sfC}(x^{\mu})$. 
Thus the typeface object $\ttQ^{\sfC}$ is a portmanteau of the upright space-dependent case for field theories and 
the slanty case for finite theories.   
That is the default notation in this article as regards typeface, straight and slanty symbols.
Such portmanteaux serve as a shorthand covering both cases at once to encompass both this article's RPM models and the 
geometrodynamical formulation of GR that they are closely analogous to. 
The $\fC$ in use above is a suitable multi-index, over particle and/or continuous extended object 
species.\footnote{I use $\fP\lfloor\ttQ^{\tfA}\rfloor$ as the portmanteau for $P(Q^{\tfA})$ for finite
%%%%%%%%%%%%%%%%%%%%%%%%%%%%%%%%%%%%%%%%%%%%%%%%%%%%%%%%%%%%%%%%%%%%%%%%%%%%%%%%%%%%%%%%%%%%%%%%%%%%%%%%%%%%% 
theories and $\mP(x^{\mu_p}; \mQ^{\tfA}]$ for infinite theories, where round brackets indicate function dependence, square brackets indicate 
functional dependence and ( ; ] is mixed function-functional dependence, with the former prior to the semi-colon and the latter after it. 
I use braces for actual brackets.  
I then use a special font (Large mathfrak) for such portmanteaux that come integrated over their 
corresponding notion of space (the action $\mbox{\tt S}$ is a such); these are therefore pure functionals in terms of what they depend on, 
but differ in the portmanteau way in terms of what features on the computational side of the equation.}
%%%%%%%%%%%%%%%%%%%%%%%%%%%%%%%%%%%%%%%%%%%%%%%%%%%%%%%%%%%%%%%%%%%%%%%%%%%%%%%%%%%%%%%%%%%%%%%%%%%%%%%%%%%%% 

\noindent N.B. configuration space is generally not just a set, being, rather, further equipped with topological space, metric space 
and geometrical structure.   
A broad example of the last of these is Finsler metric geometry $\fQ = \langle\fS, {\cal F} \rangle$ for 
$\fS$ a topological manifold and ${\cal F}$ the metric function.\footnote{I generally use
%%%%%%%%%%%%%%%%%%%%%%%%%%%%%%%%%%%%%%%%%%%%%%%%%%%%%%%%%%%%%%%%%%%%%%%%%%%%%%%%%%%%%%%%%%%%%%%%%%%%%%%%%%%%%%%
$\langle \mbox{ }  \rangle$ to denote `space of'.
I use lower-case Greek indices for spatial indices 1 to $d$.  
I use bold font as shorthand for configuration space indices and underline for spatial indices.}
%%%%%%%%%%%%%%%%%%%%%%%%%%%%%%%%%%%%%%%%%%%%%%%%%%%%%%%%%%%%%%%%%%%%%%%%%%%%%%%%%%%%%%%%%%%%%%%%%%%%%%%%%%%%%%%
%
(Semi-)Riemannian geometry is then a common subcase of this which is usually written in terms of the metric 
${\fM}_{\sfA\sfB}$ itself: $\fQ = \langle\fS, {\mbox{\boldmath${\fM}$}}\rangle$.
Some of the relational program's examples are in fact infinite-dimensional generalizations of such geometries. 
(But I still refer to such by the usual finite-dimensional geometries' names, i.e. I elevate those names to 
be finite and field-theoretic portmanteaux of the usual finite notions). 

\mbox{ } 

\noindent Relationalism 3) [Barbour] This concerns the possibility that the {\bf configurations are the primary entities}.  
I.e. the configuration space $\fQ$ has a primary role.

\mbox{ }

%%%%%%%%%%%%%%%%%%%%%%%%%%%%%%%%%%%%%%%%%%%%% PLAN AHEAD %%%%%%%%%%%%%%%%%%%%%%%%%%%%%%%%%%%%%%%%%%%%%%%%%%%%%%%%%%%%%%%%%%%%%%%%%%%%%%%%%%%%%%%%%
%
%\noindent Note 0) Secs 11-15 concern extending this structural postulate in four different ways.  

%%%%%%%%%%%%%%%%%%%%%%%%%%%%%%%%%%%%%%%%%%%%%%% MINIMALISM AND RELATIONALISM %%%%%%%%%%%%%%%%%%%%%%%%%%%%%%%%%%%%%%%%%%%%%%%%%%%%%%%%%%%%%%%%%%%%%
%
\noindent Note 1) Relationalism 3) is technically a not a relationalist postulate but a {\bf minimalist} one; such go quite often hand in hand with relational postulates.  
This does not necessarily mean that relationalists discard structures, but rather that they are prepared to consider the outcome of doing so.  
Some relational ideas are a subset of minimalism, e.g. doubting and discarding absolute space and absolute time.  
See also Secs 10 and 16 as regards whether scale is to be discarded from physics, 
Secs 2 and 9 as regards relationalists' questioning of Minkowski's view on spacetime and its subsequent primacy over how GR is conceived of.
[I.e. arguing that the dynamical tradition favours the primality of space, and that this, despite amounting to supposing less structure, 
nevertheless turns out to serve as an alternative foundation for GR.]

%===============================================================================================================================================
\section{Building relational actions and then using the Principles of Dynamics}
%===============================================================================================================================================

Relationalism 4) [Barbour] One aims to build up {\bf relational actions} from natural compound objects derived from the $\ttQ^{\sfC}$.  

\mbox{ } 

%%%%%%%%%%%%%%%%%%%%%%%%%%%%%%%%%%%%%%%% BARBOUR'S POD USE %%%%%%%%%%%%%%%%%%%%%%%%%%%%%%%%%%%%%%%%%%%%%%%%%%%%%%%%%%%%%%%%%%%%%%%%%%%%%%%%%%%%%%
%
\noindent Note 1) In making use of actions, we are working with, and drawing upon, the Principles of Dynamics.  
Many of the steps below follow from this as natural consequences of having an action, rather than of relationalism per se. 
But the particular actions used embody relationalism, and the some of the subsequent Principles of Dynamics workings assist in enforcing 
that. 
This valuing and making specific use of the Principles of Dynamics differs from Mach's own thinking; it is due to Barbour and concerns 
positing that relational actions are the most fundamental starting-point for (classical) whole-universe physics. 
  
%%%%%%%%%%%%%%%%%%%% STRUCTURAL COMPOSITION %%%%%%%%%%%%%%%%%%%%%%%%%%%%%%%%%%%%%%%%%%%%%%%%%%%%%%%%%%%%%%%%%%%%%%%%%%%%%%%%%%%%%%%%%%%%%%%%%%%%%%
% 
\noindent Note 2) This is a mathematical {\it structural composition} postulate; it is relational only in that the primary objects themselves are to be taken to be relational.   
 
%%%%%%%%%%%%%%%%%%%% LINKER: VELOCITIES NEEDED %%%%%%%%%%%%%%%%%%%%%%%%%%%%%%%%%%%%%%%%%%%%%%%%%%%%%%%%%%%%%%%%%%%%%%%%%%%%%%%%%%%%%%%%%%%%%%%%%%%
%
\noindent Note 3) The compound objects readily include NOS derivatives (whenever the NOS is nontrivial) and contracted objects.  
From what the reader knows about actions, one might expect velocities to be required too, and these involve an incipient notion of time, so we need to discuss that next before making progress with the construction of actions. 
Moreover, the manner of that progress may well be unexpected to the reader.  
In particular, actions conventionally involve velocities.

%========================================================================================================
%========================================================================================================
\section{Temporal relationalism and actions implementing it}\label{SSec: 1.1.2}
%========================================================================================================
%========================================================================================================

%========================================================================================================
\subsection{Temporal relationalism}\label{SSSec: 1.1.2}
%========================================================================================================

Relationalism 5) [Leibniz] A physical theory is {\bf temporally relational} if there is {\bf no meaningful primary notion of time for 
the system as a whole} (e.g. the universe) \cite{BB82, RWR, FORD}.  

\mbox{ }

\noindent This is implemented by 

\mbox{ }  

\noindent 1) {\bf manifest parametrization irrelevance}, alongside

\noindent 2) {\bf freedom from extraneous time-related variables}, such as external absolute Newtonian time or GR's `lapse' variable, 
at the level of the action principle.

%=======================================================================================
\subsection{A preliminary, more conventional view on constructing temporally relational actions} 
%=======================================================================================

Perhaps the reader is objecting that the preceding SSec does not help with obtaining a notion of velocity since it denies rather than 
provides a primary notion of time for the system as a whole.  
A simple way of getting round this is to point out that MPI actions are trivially equivalent to actions that are 

\mbox{ }  

\noindent $1^{\prime}$) {\bf manifestly reparametrization-invariant (MRI)}.  

\mbox{ }  

\noindent 
[For MRI, any choice of (monotonic) parameter will do, so that choice is empty, and any such choice is equivalent to making no choice by 
not parametrizing i.e. using the MPI action.] 
Next, MRI {\sl has} a parameter $\lambda$, so that, for all that it is a non-entity, one can interpret it as a {\it label time} non-entity, 
and write down 
\beq
\mbox{velocity} := \d \mbox{(configuration variable)}/\d (\mbox{label time}) \mbox{ i.e. } \mbox{ } \d\ttQ^{\sfA}/\d\lambda \mbox{ } .
\eeq 
One can then straightforwardly build the kinetic term 
$
\fT := ||\Circ\ttQ||_{\sbfM}\mbox{}^2/2 = \fM_{\sfA\sfB}\Circ\ttQ^{\sfA}\ttQ^{\sfB}/2
$.
I am assuming for now that this takes the most physically standard form that is homogeneous quadratic in the velocities; I lift this assumption 
later on in this SSec.   
${\fM}_{\sfA\sfB}$ is the configuration space metric portmanteau of ${M}_{\sfA\sfB}({Q}^{\sfC})$ for finite theories and 
${\mM}_{\sfA\sfB}({\mQ}^{\sfC}(x^{\mu}))$ for infinite theories (this is assuming {\it ultralocality}, i.e. no derivative dependence, collapsing the object from a functional to a mere function, which mathematical simplicity holds over the entirety of the standardly accepted fundamental theories of physics).
Its determinant is $\fM$ and its inverse is $\fN^{\sfA\sfB}$.  
$||\mbox{ }||_{\sbfM}\mbox{}$ denotes the corresponding ${\bfM}$-`norm'.
$\Circ := \ttD/\ttD\lambda$, the portmanteau of $\d/\d\lambda$ in the finite case and $\pa/\pa\lambda$ in the field-theoretic 
case.\footnote{This is intended as 
%%%%%%%%%%%%%%%%%%%%%%%%%%%%%%%%%%%%%%%%%%%%%%%%%%%%%%%%%%%%%%%%%%%%%%%%%%%%%%%%%%%%%%%%%%%%%%%%%%%%%%%%%%%%%%%%%%%%%%%%%%%%%%%%%
type of `dot symbol'; I often use large preceding derivative symbols rather than overhead derivative symbols for graphical ease 
of accommodation of the subtleties of configurational relationalism 
by hanging suffixes on the large-symbol versions.
I use $\delta/\delta$ for functional derivatives, 
     $\ttD/\ttD$ as ordinary, partial derivative portmanteau, 
     $\nabla/\nabla$ as partial, functional derivative portmanteau, 
     $D/D$ as ordinary, functional derivative portmanteau. 
I hang `cov', `abs' suffices on these for covariant and absolute derivatives, using $\mD$ for $\pa$ and ${\cal D}$ for $\delta$ in these contexts.
Finally, I use 
$\triangle$ for Laplacians (suitably suffixed to say of which type), 
$\mathbb{D}$ for measures and oversized
$\mbox{\large $\delta$}$ for variations.}
%%%%%%%%%%%%%%%%%%%%%%%%%%%%%%%%%%%%%%%%%%%%%%%%%%%%%%%%%%%%%%%%%%%%%%%%%%%%%%%%%%%%%%%%%%%%%%%%%%%%%%%%%%%%%%%%%%%%%%%%%%%%%%%%%
%
One can then form the Lagrangian portmanteau, $\fL\lfloor \dot{\ttQ}^{\sfA}, \ttQ^{\sfA} \rfloor$ of the Lagrangian $L(Q^{\sfA}, \dot{Q}^{\sfA})$ and the {\sl ultralocal} Lagrangian density $\mL(\dot{\mQ}^{\sfA}; \mQ^{\sfA})$ for field theories, and thus, integrating over $\lambda$ and the NOS, the relational action.
Compliance with MRI does make the Lagrangian portmanteau in question {\it look} somewhat unusual, i.e. not be of difference-type form 
$\fL = \fT - \fV$, but rather of {\it product form} $\fL = 2\sqrt{\fT\fW}$.  
For now we take on trust that this ${\fW} = {\fE} + {\fU}$ for ${\fU} = -{\fV}$, where $\fV$ is the potential energy portmanteau: potential $V(Q^{\sfA})$ for finite theories and potential density ${\mV}(x^{\mu}; \mQ^{\sfA}(x^{\mu})]$ for field theories.  
Thus it is assumed to be A) independent of the velocities [this is another mathematical simplicity that happens to be in accord with the standardly-accepted fundamental physics]. 
B) time-independent [which makes good sense for fundamental whole-universe setting as opposed to the setting for dissipative subsystems or approximate modelling e.g. involving friction].
This time-independence of the potential, alongside time-independence of the kinetic metric is part of what is covered by the 
`no extraneous time-like objects' part of the given mathematical implementation of temporal relationalism.
$\fE$ is the energy-like portmanteau: total energy $E$ for finite theories and some kind of total energy density 
for field theories (`density' here includes a Jacobian factor).  
[$\fU$ and $\fW$ are, for now, just to be considered as the notational tidyings defined above;
we will justify this trust in Sec \ref{SSSec: B}.]
The MRI relational action is thus 
\be
\FS^{\sM\sR\sI}_{\sJ} = \int\d\lambda\int_{\Sigma_p}\d\Sigma_p\fL^{\sM\sR\sI}_{\sJ} = 
2\int\d\lambda\int_{\Sigma_p}\d\Sigma_p\sqrt{{\fT}{\fW}} \mbox{ } ,    
\label{action}
\ee
where the J stands for `Jacobi' since the finite case of this is Jacobi's principle.

%================================================================================================================================================
\subsection{Relational dethroning of velocity as a primary notion}
%================================================================================================================================================

A more manifestly relational manoeuvre, however, is to acknowledge that if there is no meaningful notion of time, 
then conventional physics' definition of velocity, 
\be
\mbox{velocity := d(configuration variable)/d(some notion of time)} \mbox{ } , \label{Flutt}
\ee
{\sl reads that} there is consequently no meaningful primary notion of velocity! 
Subsequently, nothing defined in terms of velocities such as kinetic terms, Lagrangians or canonical momenta, is guaranteed 
to meaningfully exist either.  
What {\sl does} have tangible physical content, however, are the
\beq
\mbox{d(configuration variable)} \mbox{ } \mbox{ i.e. the} \mbox{ } \d\ttQ^{\sfA}
\eeq
themselves, so that velocities are replaced by differentials of the configuration variables.  
I.e., changes {\sl in time} are meaningless in relational physics, the tangible content resides, rather, among the 
changes of one configuration variable with respect to another, 
\beq
\mbox{d(configuration variable 1)/d(configuration variable 2) } 
\mbox{ } \mbox{ i.e. } \mbox{ } \d\ttQ_1/\d\ttQ_2 \mbox{ } . \label{erina}
\eeq

%============================================================================================
\subsection{The subsequent relational view on constructing temporally relational actions}
%============================================================================================

As a follow-up of this dethroning of velocity from the realm of relational physics, the usual homogeneous quadratic kinetic term is supplanted 
by d$\fs^{2}$/2 for $\mbox{(kinetic arc element)$^2$} := \d \fs^2 := ||\d \bttQ ||_{\sbfM}\mbox{}^2 = \fM_{\sfA\sfB}\d\ttQ^{\sfA}\d\ttQ^{\sfB}$.    
I note that this is a purely geometrical quantity.  
The action is, in complete generality, to be homogeneous linear in the d(configuration variable) so as to comply with MPI/MRI;   
in the homogeneous quadratic case this amounts to taking the square root.
Then expression under the action's integral in the MPI action is of the form $\d\widetilde{\fs}$, where the tilde indicates 
a rescaling by some weight that is homogeneous of degree zero in the d$\ttQ$. 
[This is usually taken to be independent of the weights so as to match the conventional approach to fundamental physics having velocity-independent potential.]
I use 
$
\mbox{(physical arc element)} := \d\widetilde{\fs} := 2\sqrt{\fW}\d \fs = 2\sqrt{\fW}||\d \bttQ ||_{\sbfM}  
$
and then the action is 
\beq
\FS^{\sM\sP\sI}_{\sJ} := \int\int_{\Sigma_p}\d {\Sigma_p}\d \widetilde{\fs} \mbox{ } . 
\label{GeneralAction}
\eeq
\mbox{ } \mbox{ } Such assembly of an action may be subject to some limitations, whether from implementing physical or philosophical 
principles (see e.g. Sec \ref{SSSec: 1.1.4}), or from purely mathematical simplicity postulates such as `no derivatives higher than first (or, in a few special cases, that include GR, second)'.
\noindent The above particular action coming from the homogeneous quadratic physics only assumption is of {\it Jacobi type}.  
In the case of mechanics itself, the NOS integration collapses to 1 and the above is precisely the Jacobi action principle, 
which is well-known and established to be cleaner in a number of ways than Euler--Lagrange's (alongside having a number 
of relational whole-universe features and applications).  
In parallel to the Lagrangian portmanteau, $\d\fs$ and $\d\widetilde{\fs}$ are in fact portmanteaux of arclength elements in the finite case 
and nontrivially Riemannian elements (no signature restriction implied) in the field-theoretic case.
\noindent The MPI formulation does not have a primary concept of Lagrangian - it has the arc element instead.  
In other words, it is a `{\bf geodesic principle}' i.e. a variational principle that takes the mathematical form of finding the geodesic 
corresponding to a Riemannian (or sometimes more general) geometry.    
\noindent By the above, and how the integrand being a product and not a difference like the conventional Lagrangian 
does {\sl not} unacceptedly alter the physics resulting from the action, are explained below.

The MRI or MPI action of the above form for homogeneous quadratic $\fT$ or $\d\fs^2$ physics is the Jacobi-type 
action principle; in the finite case of mechanics, it is the Jacobi action principle itself. 
This is more commonly encountered in the MRI form, though I note that Jacobi himself did use the MPI geodesic principle form.

%====================================================================
\subsection{More general temporally relational actions}\label{SSSec: A}
%====================================================================

\beq
\FS^{\sM\sP\sI}_{\sJ\sS} = \int\int_{\Sigma_p}\d \Sigma_p \d\widetilde{\fs} = \int\int_{\Sigma_p}\d \Sigma_p \d\lambda \fL  = \FS^{\sM\sR\sI}_{\sJ\sS}
\mbox{ } 
\eeq
complies with temporal relationalism, and has mutual compatibility, by 
$
\fL\left\lfloor \frac{\ttD\ttQ}{\ttD \lambda}, \ttQ \right\rfloor\d\lambda  = 
\frac{\d \widetilde{\fs}}{\d \lambda}\d\lambda = {\d \widetilde{\fs}} = \frac{\d s}{\d \mu}\d\mu = \fL\left\lfloor \frac{\ttD\ttQ}{\ttD \mu}, \ttQ,\right\rfloor\d\mu 
$.  
[This uses Euler's theorem for homogeneous functions, and takes $\lambda$ and $\mu$ to be any two 
monotonically-related parameters; also, by this equivalence, I henceforth drop MPI and MRI labels.]   
The `square root of homogeneous quadratic' Jacobi-type case is then an obvious subcase of this.
Moreover, the above generalization includes passage from Riemannian geometry (no signature connotation intended) to 
Finsler geometry (no nondegeneracy connotation intended; $\fL$ is cast in the role of metric function ${\cal F}$).
It is still a geodesic principle, just for a more complicated notion of geometry.  
In fact, the Jacobi-type action more commonly arises in the literature (see e.g. \cite{Hsiang1, Lanczos}) not from the desire to be 
relational but rather from the desire to provide the natural mechanics for a given geometry. 
Synge's work starting with \cite{Synge} generalized this to the more general geometry above \cite{Lanczos}.  
The geometry in question enters by playing the role of configuration space geometry.
In their honour, I denote the geometrically-natural construction of a mechanics given a metric geometry by $\ttJ\ttS: \langle{\fS }, 
{\cal F}\rangle \longrightarrow \FS^{\sM\sP\sI}_{\sJ\sS}$.

%==============================================================
\subsection{Temporally relational formulation of the notion of conjugate momentum}  
%==============================================================  

The standard definition of conjugate momentum is
$
\ttP_{\sfA} := \nabla\fL/\nabla\dot{\ttQ} \mbox{ } .
$
Then computing this from the MRI form of the Jacobi action gives
$
\ttP_{\sfA} = \sqrt{\fW/\fT}\fM_{\sfA\sfB}\Circ\ttQ^{\sfA} \mbox{ } .  
$
However, manifest relationalism has been argued to have no place for velocities or Lagrangians, so either the above definition of momentum, 
or the conceptual role played by momentum, need to be reformulated. 
In this case, it turns out that the notion of conjugate momentum itself survives, by there being an equivalent manifestly-relational form 
for the defining formula, 
$
\ttP_{\sfA} := \nabla\d\widetilde{\fs}/\nabla\d\ttQ^{\sfA} \mbox{ } .  
$
Then computing this gives
$
\ttP_{\sfA} = \sqrt{2\fW}\fM_{\sfA\sfB}\d\ttQ^{\sfB}/|| \d\bttQ||_{\sbfM} \mbox{ } , 
$
which is manifestly within the span of notion (\ref{erina}).

%==============================================================================================================
\subsection{Outline of the Dirac method} 
%==============================================================================================================

Next, one inspects to see if there are any inter-relations among these momenta due to the form of the action: {\it primary}   
{\it constraints} \cite{Dirac}.  
Then variation of the action with respect to the base objects provides some evolution equations and perhaps some {\it secondary constraints} 
\cite{Dirac}. 
One then finds the evolution equations.
Finally, one checks whether even more constraints may arise by the requirement that the evolution equations propagate the constraints; this is 
best done at the level of Poisson brackets (or Dirac brackets if required by the presence of so-called {\it second-class} constraints).

%========================================================================================================
\subsection{Quadratic constraint: momenta as `direction cosines'}\label{SSSsec: 1.1.7}
%========================================================================================================

The MRI implementation of temporal relationalism leads to constraints via the following general argument of Dirac \cite{Dirac} 
(as straightforwardly modified by me to concern differentials rather than velocities, and MPI instead of MRI actions).    
Since MPI actions are homogeneous of degree 1 in the differentials,    
the $k$ momenta arising from these are homogeneous of degree 0 in the differentials.  
Hence the momenta are functions of at most $k - 1$ independent ratios of differentials. 
Thus the momenta must have at least 1 relation between them (which is by definition a primary constraint).  

\noindent Moreover, for the above Jacobi-type action, there is precisely one such constraint (per relevant NOS point). 
This is due to 

\noindent A) the square-root form of the action, by which the momenta are much like direction cosines and `their squaring to 1' property 
by which they are not all independent.  
Here, instead, these square (using the $\bfM$ matrix) to $2\fW$.
Thus, the kinetic term in terms of momenta (a legitimately relational object since the momenta are), which is half of the above square, 
is equal to $\fW$.

\noindent B) in cases with nontrivial space of extent, due to the particular {\it local} ordering of having the square root (and sum over 
$\fA$, $\fB$ implicit in the kinetic arc element or kinetic term) {\sl inside} the integral-over-space-of-extent sign as opposed to outside it.  
All-in-all, one has an energy-type constraint which is quadratic and not linear in the momenta, 
\beq
{\cal Q}\mbox{uad} := {\fN}^{\sfA\sfB}{\ttP}_{\sfA}{\ttP}_{\sfB}/2 + \fW = 0 \mbox{ } .   
\label{GEnCo}
\eeq

%======================================================================================================================
\subsection{Evolution equations}
%======================================================================================================================

Again, the usual form of the evolution equations makes reference to Lagrangians, times and velocities:
\beq
\frac{\ttD}{\ttD\lambda}\left\{\frac{\nabla\d\fL}{\nabla \dot{\ttQ}^{\sfA}}\right\}  = \frac{\nabla\d\fL}{\nabla\ttQ^{\sfA}} \mbox{ } .  
\eeq
There is however a manifestly relational version of this available, 
$
\d \{\nabla\d\widetilde{\fs}/\nabla \d\ttQ^{\sfA}\}  = \nabla\d\widetilde{\fs}/\nabla\ttQ^{\sfA}
$.  
\beq
\mbox{I.e. expanding,} \hspace{1.2in} 
\frac{\sqrt{2\fW}}{||\d\bttQ||_{\sbfM}}  \d 
\left\{
\frac{\sqrt{2\fW}}{||\d\bttQ||_{\sbfM}}  \d \ttQ^{\sfA}
\right\} 
+ \Gamma^{\sfA}\mbox{}_{\sfB\sfC} 
\frac{\sqrt{2\fW}}{||\d\bttQ||_{\sbfM}} \d \ttQ^{\sfB}
\frac{\sqrt{2\fW}}{||\d\bttQ||_{\sbfM}} \d \ttQ^{\sfC} = \fN^{\sfA\sfB}
\frac{\nabla\fW}{\nabla \ttQ^{\sfB}} \mbox{ } ,  \hspace{4in}
\label{Evol}
\eeq
where $\Gamma^{\sfA}\mbox{}_{\sfB\sfC}$ are the configuration space Christoffel symbols.

%======================================================================================================================
\subsection{Emergent Jacobi time}\label{Mach-1}
%======================================================================================================================

Both the conjugate momenta and the evolution equations are simplified if one uses
\beq
\Star := 
\frac{D}{D\tea^{\se\sm(\sJ)}} :=  
\sqrt{\frac{\fW}{\fT}} \Circ := 
\frac{1}{\ttN}\Circ := 
\frac{1}{\dot{\ttI}}\Circ =  
\frac{D}{D\ttI} 
\mbox{ } .
\label{emtime}
\eeq

\noindent Here, $\tea^{\se\sm(\sJ)}$ is to be interpreted as an emergent time, the portmanteau of position-independent 
time $t$ for finite systems and position-dependent time $\mt(x^{\mu})$ for infinite systems; I term it the {\bf Jacobi emergent time}. 
It, and modifications of it, have long appeared in Barbour's work (e.g. \cite{BB82, B94I}). 
Though I have also seen it in earlier Russian literature \cite{Ak} for mechanics, and as Christodoulou's Chronos Principle 
\cite{Christodoulou1} for GR, so I leave open the question of who its first proponent was.  

\noindent Inverting so as to obtain an expression for the emergent Jacobi time itself, 
\beq
\tea^{\se\sm(\sJ)} - \tea^{\se\sm(\sJ)}(0) = \int||\d\bttQ||_{\sbfM}/\sqrt{2\fW} \mbox{ } .
\eeq
This implements 

\mbox{ } 

\noindent Relationalism 6) [Mach] {\bf Time is to be abstracted from change}: 
\beq
\tea_{\sM\sa\sc\sh} = F[\d\ttQ] \mbox{ } .
\eeq
\noindent Relationalism 7) [Leibniz--Barbour]  {\bf THE MOST PERFECT time is to be abstracted from THE TOTALITY OF change} (my phrasing, 
and which is implemented here via {\sl all} of the $\bttQ$ are involved via the $\fW$ and the $\d\bttQ$): 
\beq
\tea_{\sL\sM\sB} = F[\mbox{all } \d\ttQ] \mbox{ } .
\eeq
I refer to Relationalism 6, 7) together as the LMB view of time.  

\mbox{ }  

\noindent In (\ref{emtime}), the first step defines a shorthand, the second is the MRI computational formula:  the third relates the lapse $\ttN$ 
to it, the fourth recasts lapse as a truer {\it velocity of the instant} notion $\dot{\ttI}$ \cite{FEPI}. 
The fifth recasts this as the even truer {\it differential of the instant} notion \cite{FileR} $\d \ttI$, by which the instant $\ttI$ itself 
is identified with the emergent time $\tea^{\se\sm(\sJ)}$ that labels that instant.

The above generalized lapse notion $\ttN$ is a portmanteau of $N$ for finite theories and $\mN(x^{\mu})$ for field theories. 
Similarly, the above generalized instant notion $\ttI$ is a portmanteau of $I$ for finite theories and $\mI(x^{\mu})$ for field theories. 
%
%\mbox{ }
%
The momenta in terms of $\Star$ are then just $\ttP_{\sfA} = \fM_{\sfA\sfB}\Star\ttQ^{\sfB}$, while the evolution equations are 
${D}_{\sa\sb\sss}\mbox{}^2 {\ttQ}^{\sfA} = \Star\Star\ttQ^{\sfA} + 
\Gamma^{\sfA}\mbox{}_{\sfB\sfC}\Star\ttQ^{\sfB}\Star\ttQ^{\sfC} = \nabla\fW/\nabla\ttQ_{\sfA}$. 
Note that this is but a {\bf parageodesic equation} with respect to the kinetic metric (meaning it has a forcing term arising from the $\fW$), 
but is clearly going to be a true geodesic equation with respect to the physical (tilded) metric.

Of use in practical computations, one can supplant one of the evolution equations (finite case) or one per space point (infinite case) with the Lagrangian form of the quadratic `energy-type' constraint, 
\beq
{\fM}_{\sfA\sfB}\Star{\ttQ}^{\sfA}\Star{\ttQ}^{\sfB}/2 + \fW = 0 \mbox{ }  .  
\label{ENERGY}
\eeq
As first remarked upon in the subcase of mechanics \cite{B94I}, the above reveals $\tea^{\se\sm(\sJ)}$ to be a  recovery on relational premises of the same quantity that is more usually assumed to be the absolute external Newtonian time, $t^{\sN\se\sw\st\so\sn}$;  there is a conceptually-similar recovery in the general case which covers a further number of well-known notions of time (see Sec 2).   

\noindent Another form for the parageodesic equation is 
$
\Star \ttP_{\sfA} = \nabla\fW/\nabla\ttQ^{\sfA}
$
\noindent As $\tea^{\se\sm(\sJ)}$  has been cast as an expression solely in terms of the $\ttQ^{\sfA}$ and $\d \ttQ^{\sfA}$, 
\beq
\mbox{d(configuration variable/d(emergent Jacobi time)} = \d\ttQ^{\sfA}/\d\tea^{\se\sm(\sJ)} 
\eeq
{\sl do} have tangible physical content, through being a type of (\ref{Flutt}) and not of (\ref{erina}).  
Thus in the relational approach the above equations entirely make sense from a temporally relational perspective, unlike in the absolute 
Newtonian counterparts of them that have the same mathematical form but pin an absolute time interpretation on the times present.    
The emergent time is provided {\sl by} the system.  
For now, it furthermore gives the appearance of being provided by the whole of the subsystem.
These last two sentences fit Mach's own conception of time as per Sec 2.1. 
Also, the usually-assumed notion of time as an independent variable is un-Leibnizian and un-Machian.  
However, though it is to be overall abstracted from motion, once this is done it {\sl is} a convenient choice for (emergent) independent 
variable.  

\mbox{ } 

\noindent  We shall from now on work with MPI and MRI forms held to be interchangeable, with preference for writing MPI ones. 
We will likewise present specific equations in terms of $\Star$ in the geodesic/Lagrangian picture (the geodesic picture is in terms of configuration variables and their differentials in parallel to how the Lagrangian picture is in terms of configuration variables and their velocities).

%==========================================================================================================
\subsection{Inter-relating the Jacobi product and Euler--Lagrange difference actions}\label{SSSec: B}
%==========================================================================================================

The more well-known difference alias Euler--Lagrange-type actions are, rather,  
\be
\FS_{\sE\sL} = \int\d \tea\int_{\Sigma_p}\d \Sigma_p \fL = \int\d \tea\int_{\Sigma_p}\d \Sigma_p\{{\fT}_{\stea} - {\fV}\} \mbox{ } . 
\label{Lagaction}
\ee
Here ${\fT}_{\stea}$ is the kinetic energy formulated in terms of $\ttD/\ttD\tea$ derivatives for $\tea$ conventionally an 
external notion of time (absolute time for mechanics, coordinate time for GR, which is a label time because GR is already-parametrized). 
See e.g. \cite{Lanczos} or  Sec 2 of \cite{FileR} for how (\ref{action}) $\Rightarrow$ (\ref{Lagaction}) in the case of mechanics by Routhian 
reduction, and  \cite{SemiclI} or Sec 2 of \cite{FileR} for (\ref{Lagaction}) $\Rightarrow$ (\ref{action}) by the emergence of some lapse-like 
quantity.\foo{Such 
%%%%%%%%%%%%%%%%%%%%%%%%%%%%%%%%%%%%%%%%%%%%%%%%%%%%%%%%%%%%%%%%%%%%%%%%%%%%%%%%%%%%%%%%%%%%%%%%%%%%%%%%%%%%%%%%%%%%%%
mathematics conventionally appears in the mechanics literature under the name of the {\it parametrization procedure}.  
Namely, one may adjoin the original notion of time's time variable to the configuration space $\mbox{\large $\mathfrak{q}$}$ $\longrightarrow$ 
$\mbox{\large $\mathfrak{q}$} \times \sfT$ by 
rewriting one's action in terms of a label-time parameter $\lambda \in \sfT$ (see e.g. \cite{Lanczos}).  
However, in the relational context in which Barbour and I work, one rather at this stage adopts 1) above.}
%%%%%%%%%%%%%%%%%%%%%%%%%%%%%%%%%%%%%%%%%%%%%%%%%%%%%%%%%%%%%%%%%%%%%%%%%%%%%%%%%%%%%%%%%%%%%%%%%%%%%%%%%%%%%%%%
%
Various advantages for product-type actions over difference-type actions stem from this as regards 
consideration of whole-universe fundamental physics, which is the setting for Quantum Cosmology.  
The above justifies the identification of $\fW$ as the combination of well-known physical entities $\fE - \fV$.

%========================================================================================================
%========================================================================================================
\section{Configurational relationalism and its indirect implementation}\label{SSSec: 1.1.3}
%========================================================================================================
%========================================================================================================

\noindent Relationalism 8) [my extension and part-reformulation of Barbour]. 
The action is to be {\bf configurationally relational} as regards a group of transformations if a $\fG$-transformed world-configuration is indiscernible from one that is not.  
I.e., one is to build into one's theory that a certain group of transformations $\fG$ acting upon the theory's configuration space $\fQ$ are 
to be irrelevant, i.e. physically meaningless \cite{BB82, RWR, Lan, Phan, FORD, FEPI, Cones}, transformations.  
More specific detail of the $\fQ$--$\fG$ pairing is postponed until after we have seen details of specific examples, to Sec \ref{Patti}.

\mbox{ }

\noindent One way to implement this is to use not `bare' configurations and their composites as above, but rather 
their arbitrary-$\fG$-frame-corrected counterparts.
This is the only known way that is sufficiently widespread for a relational program to underlie the 
whole of the classical fundamental physics status quo of GR coupled to the Standard Model (see Sec 9).
The corrections are with respect to auxiliary variables $\ttg^{\sfZ}$ [a portmanteau of the finite case $g^{\sfZ}$ and the position-dependent 
case $\mg^{\sfZ}(x^{\mu})$ in field theory] that are paired with the infinitesimal generators of $\fG$.

\noindent Originally the $\ttg^{\sfZ}$ employed  were in the form of multiplier coordinate corrections to the velocities 
\beq
\dot{\ttQ}^{\sfA} - \sum\mbox{}_{\mbox{}_{\mbox{\scriptsize $\fZ$}}}
                               \stackrel{\rightarrow}{\fG_{\sttg^{\tfZ}}} {\ttQ}^{\sfA} \mbox{ } .  
\eeq
Here $\stackrel{\longrightarrow}{\fG}$ denotes the infinitesimal group action.
However, this formulation is not consistent with manifest temporal relationalism since the 
multiplier corrections spoil this. 
Nor is it relational insofar as it is formulated in terms of meaningless label-time velocities.
Nor is it an arbitrary-$\fG$ frame approach.  
The next SSec amends these things, and then the rest of the present SSec applies just as well within this amendment.

The above also seems to be taking a step in the wrong direction as regards $\fG$ being irrelevant in passing 
from the already-frame-redundant $\fQ$ to the principal fibre bundle\foo{This 
%%%%%%%%%%%%%%%%%%%%%%%%%%%%%%%%%%%%%%%%%%%%%%%%%%%%%%%%%%%%%%%%%%%%%%%%%%%%%%%%%%%%%%%%%%%%%%%%%%%%%%%%%
being a bona fide bundle may require excision of certain of the degenerate configurations. 
Recall also that a fibre bundle being principal means that the fibres and the structure group coincide \cite{Nakahara}.}
%%%%%%%%%%%%%%%%%%%%%%%%%%%%%%%%%%%%%%%%%%%%%%%%%%%%%%%%%%%%%%%%%%%%%%%%%%%%%%%%%%%%%%%%%%%%%%%%%%%%%%%%%
%
over $\fQ$, $P(\fQ, \fG)$.     
For, from the perspective of just counting degrees of freedom, this can be locally regarded as the product space $\fQ \times \fG$, which has more degrees of freedom, as opposed to passing to the quotient space $\fQ/\fG$, which has less.  
However, we shall also resolve this contention below.

Moreover, the essential line of thought of this SSec is the {\sl only} known approach to configurational relationalism that is general enough to 
cover the Einstein--Standard Model presentation of Physics.\foo{For 
%%%%%%%%%%%%%%%%%%%%%%%%%%%%%%%%%%%%%%%%%%%%%%%%%%%%%%%%%%%%%%%%%%%%%%%%%%%%%%%%%%%%%%%%%%%%%%%%%%%%%%%%%
mechanics in 1- and 2-$d$ \cite{FORD}, I have found that working directly on the least redundant/most relational configuration space 
provides workable relational theories. 
Moreover, these coincide with the restriction to those dimensions of Barbour's theories as formulated in arbitrary $\sFG$-frame form and then reduced (Sec 3 of \cite{FileR}).} 
%%%%%%%%%%%%%%%%%%%%%%%%%%%%%%%%%%%%%%%%%%%%%%%%%%%%%%%%%%%%%%%%%%%%%%%%%%%%%%%%%%%%%%%%%%%%%%%%%%%%%%%%%

\mbox{ } 

\noindent This SSec's notion of configurational relationalism is my \cite{Lan} portmanteau that spans both of the following.

\mbox{ }

\noindent 1) {\bf Spatial relationalism}, as in the traditional mechanics setting and the (geometro)dynamical formulation of GR. 
%
%and which was Barbour's notion prior to this extension.  

\mbox{ }  

\noindent 2) {\bf Internal relationalism}, as in electromagnetism, Yang--Mills Theory and the associated scalar and fermion gauge theories.  

\mbox{ }

\noindent This wide range of cases is afforded by correspondingly wide ranges of $\fQ$ and $\fG$. 
(Although adopting a $\fQ$ may carry connotations of there being some underlying NOS with which that is compatible.  
Also, given a $\fQ$, there are consistency limitations on what variety of $\fG$ can then have -- see Sec \ref{SSec: QG-Comp}).
For the scaled version of RPM, $\fG$ is the Euclidean group of translations and rotations, Eucl($d$), while for pure-shape version it is 
the similarity group of translations, rotations and dilations, Sim($d$).

This extension and partial reformulation of Barbour's spatial relationalism shows that it and the conventional 
notion of Gauge Theory (usually internal but also applicable to spatial concepts) bear a very close relation.  
Following on from Note 3 of Sec 2, Gauge Theory is indeed Leibnizian and indeed an example of how such can usually only be implemented indirectly by mathematics of a mixed set of tangible entities and non-entities (respectively true dynamical degrees of freedom and gauge degrees of freedom).
%
%susy causes a major difference here?  
%
\noindent Configurational relationalism is generalized in Sec 15.  

\mbox{ }

\noindent Finally, I here propose to extend the meaning of configurational relationalism to have a second clause: 
{\bf `no extraneous space structures'}.
This matches temporal relationalism's second clause of no extraneous time variables, and also serves to exclude the Nambu--Goto string 
action (which, without this extra criterion, can indeed be included under the previous literature's formulation of relationalism).
For such reasons \cite{bigcite1, bigcite2, bigcite3, bigcite4, bigcite5, bigcite6, bigcite7, bigcite8, bigcite9, I93, Kieferbook, Dirac}, theories in fixed-

\noindent background  contexts are `less relational' (or only relational in a weaker sense).  
It is, rather, with background-independent M-theory (or at least some limiting classical action for this) that the most purely relational 
program would be expected to make contact (see Sec \ref{SSec: Stri-RPM}). 
\noindent N.B. How generally this clause should be imposed requires further discussion in Sec 9.11.

%========================================================================================================
%========================================================================================================
\section{Combining temporal and configurational relationalism}\label{SSSec: 1.1.4}
%========================================================================================================
%========================================================================================================

Typically,
the potential term ${\fV}$ is manifestly a good $\fG$-scalar but the kinetic term ${\fT}$ has correction terms due to `$\fG$-transformations 
and differentials not commuting'.   
Consider for now theories whose configuration space carry kinetic arc elements that are homogeneous quadratic and whose associated metric 
has at most dependence on the ${\ttQ}^{\sfA}$ [i.e. a (semi)Riemannian as  opposed to Finslerian or even more general geometry].  
Then one can render the action MRI by supplanting the previous $\fT$ by 
\be
{\fT} := ||\Circ_{\sg}{\bttQ}  ||_{\mbox{\scriptsize\boldmath${\sbfM}$}}\mbox{}^2/2
\mbox{ } \mbox{ for } \mbox{ } \Circ_{\sttg}{\ttQ}^{\sfA} := \Circ{\ttQ}^{\sfA} - 
\sum\mbox{}_{\mbox{}_{\mbox{}_{\sfZ}}}\stackrel{\rightarrow}{\fG}_{\dot{\sttg}^{\tfZ}}{\ttQ}^{\sfA} \mbox{ } . 
\label{Taction}
\ee  
The auxiliary variable in use here is now interpreted as the velocity corresponding to a cyclic coordinate; 
I term the ensuing variational principle 
{\ttJ\ttB\ttB}[$P(\langle \fS$, ${\cal F} \rangle$, $\fG$)], the BB standing for Barbour--Bertotti.  
[As a map, {\ttJ\ttB\ttB} = {\ttJ\ttS} $\circ \mbox{ } \fG$-bundle, for 

\noindent $\fG$-bundle: $\langle \fS, {\cal F} \rangle \times \fG \longrightarrow 
P(\langle \fS$, ${\cal F} \rangle$, $\fG$).]
On relational grounds, however, I prefer the equivalent MPI form, the argumentation behind which never involves the label-time non-entity 
or the subsequently-dethroned notion of velocity.

Here, instead of the above, I supplant the previous configurational-relationally unsatisfactory arclength (density) $\d \fs$ by 

\noindent
\beq
\mbox{the $\fG$-corrected arclength } \hspace{1.3in}
\d \fs_{\sJ\sB\sB} = ||\d_{\sttg}\bttQ||_{\sbfM}  \mbox{ } \mbox{ for } \mbox{ } \d_{\sttg}{\ttQ}^{\sfA} := \d{\ttQ}^{\sfA} - 
\sum\mbox{}_{\mbox{}_{\mbox{}_{\sfZ}}}\stackrel{\rightarrow}{\fG}_{\d{\sttg}^{\tfZ}}{\ttQ}^{\sfA} \mbox{ } .  \hspace{3in}  
\eeq 
The auxiliary variable in use here is now interpreted as the differential corresponding to a cyclic coordinate. 
One can then repeat the preceding SSec's treatment of conjugate momenta, quadratic constraint, evolution equations and the 
beginning of the treatment of emergent time `by placing $\ttg$ suffices' on the $\d$, $\ttI$, $\tea^{\se\sm}$ and $\Star$ (and, if one ever makes  
use of them, the $\Circ = \ttD/\ttD\lambda$ and $\ttN$; I also find it helpful in considering the formulae inter-relating these to place a $\ttg$ 
suffix on the $\fT$ as a reminder of the $\ttg$-dependence residing within it).
\noindent $\tea^{\se\sm(\sJ\sB\sB)}_{\sttg}$ is no longer a relationally-satisfactory notion of time due to manifest configurational 
non-relationalism. 
As such, I term this a $\fG$-dependent proto-time.
I also name it, and its eventual $\fG$-independent successor, after JBB rather than just after Jacobi, to reflect the 
upgrade to a configurational-relationally nontrivial context.

%========================================================================================================
\subsection{Linear constraints: enforcers of configurational relationalism}\label{SSSec: 1.1.8}
%========================================================================================================

The novel feature in the variational procedure is, rather, that variation with respect to each $\fG$-auxiliary produces one independent 
secondary constraint.  
In this article's principal examples, each of these uses up {\sl two} degrees of freedom.
(That this is the case is testable for in a standard manner \cite{Dirac}, and amounts to these constraints being first- and not second-class.)
Thus one ends up on the quotient space $\fQ/\fG$ of equivalence classes of $\fQ$ under $\fG$ motions.  
This resolves the abovementioned controversy, confirming the arbitrary $\fG$-frame method both to indeed 
implement configurational relationalism and to be an indirect implementation thereof.  
I denote these linear constraints by ${\cal L}\mi\mn_{\sfZ}$: 
\beq
0 = \nabla \fL/ \nabla \dot{\ttc}^{\sfZ} \mbox{ (or } \mbox{ } \nabla\d\widetilde{\fs}/\nabla\d\ttc^{\sfZ}) := {\cal L}\mi\mn_{\sfZ} = 
\ttP_{\sfA}\frac{\mbox{\Large $\delta$}}{\mbox{\Large $\delta$}\d\ttc^{\sfZ}}\{\stackrel{\rightarrow}{\fG_{\d\sttc}}\ttQ^{\sfA}\} \mbox{ } ,   
\label{LinZ}
\eeq
where the last form manifestly demonstrates these being purely linear in the momenta.
The variational procedure behind obtaining these is justified in Chapter 2 of \cite{FileR}.

One then applies the Dirac procedure to be sure that the constraints breed no further constraints (see Sec \ref{Patti} for 
what the consequences are if they do).

%========================================================================================================
\subsection{Outline of `best matching' as a procedure} \label{SSSec: 1.1.10}
%========================================================================================================

Barbour's `best matching' \cite{BB82, B03} amounts to 

\noindent Best Matching 1) construct an arbitrary $\fG$-frame corrected action.  

\noindent Best Matching 2) Vary with respect to the $\fG$-auxiliary $\ttg^{\sfZ}$ to obtain the linear constraint ${\cal L}\mbox{in}_{\sfZ} = 0$.

\noindent Best Matching 3) Solve the Lagrangian form of ${\cal L}\mbox{in}_{\sfZ} = 0$ for $\ttg^{\sfZ}$. (This is often an impasse.) 

\noindent Best Matching 4) Substitute this back in the action to obtain a new action.  

\noindent Best Matching 5) Elevate this new action to be one's primary starting point.  

\mbox{ }  
  
\noindent Note 1) Best Matching 2) to 4) can be viewed as a minimization (or, at least, as an extremization).
One is searching for a minimizer to establish the least incongruence between adjacent physical configurations.
See Sec 9.5 for more.  

\noindent Note 2) Best Matching 2) to 5) can be viewed as a configuration space reduction procedure (Fig \ref{Red}).  
%
%FFFFFFFFFFFFFFFFFFFFFFFFFFFFFFFFFFFFFFFFFFFFFFFFFFFFFFFFFFFFFFFFFFFFFFFFFFFFFFFFFFFFFFFFFFFFFFFFFFFFFFFFFFFFFFFFFFFFFFFFFFFFFFFFFFFFFFFFFFFFFFFFFF
{            \begin{figure}[ht]
\centering
\includegraphics[width=0.4\textwidth]{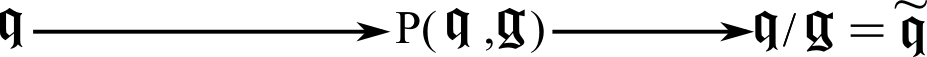}
\caption[Text der im Bilderverzeichnis auftaucht]{        \footnotesize{One passes from the configuration space $\fQ$ to the principal 
bundle by adjoining $\fG$ generators and then to the corresponding quotient space if one succeeds in performing reduction.} }
\label{Red} \end{figure}          }
%FFFFFFFFFFFFFFFFFFFFFFFFFFFFFFFFFFFFFFFFFFFFFFFFFFFFFFFFFFFFFFFFFFFFFFFFFFFFFFFFFFFFFFFFFFFFFFFFFFFFFFFFFFFFFFFFFFFFFFFFFFFFFFFFFFFFFFFFFFFFFFFFFF

%========================================================================================================
\subsection{Emergent Jacobi--Barbour--Bertotti time} \label{SSSec: 1.1.6}
%========================================================================================================

As a distinct application of Best Matching 3), substitute its solution into the  $\fG$-dependent proto-time $\tea^{\se\sm(\sJ\sB\sB)}_{\sttg}$
to render it configurational-relationally acceptable. 
[Albeit this is at the cost of being technically implicit and subject to the threats of nonuniqueness and of the extremization not being 
solvable in practise.]
I.e. the JBB emergent time is
\beq
\tea^{\se\sm(\sJ\sB\sB)} - \tea^{\se\sm(\sJ\sB\sB)}(0) = \stackrel{\mbox{\scriptsize extremum $\ttg \mbox{ } \in \mbox{ }$} \sFG}
                                                               {\mbox{\scriptsize of $\stS^{\tr\te\tl\ta\ttt\ti\to\tn\ta\tl}$}}                                                              
\left(                                                              
                                                               \int||\d_{\sg}\bttQ||_{\sbfM}/\sqrt{\fW} 
\right)  \mbox{ } .
\label{Kronos}
\eeq
\noindent Note 1) JBB time does look rather Machian [Relationalism 6)], though a `boson' caveat will appear in Sec 9.7.

\noindent Note 2)  Contrarily, the Internal Time approach to the Problem of Time (c.f. Appendix A) {\sl lacks} in Machianity.  
This is by its finding specific degrees of freedom that make up such a time, rather than by everything contributing to the time.

%==================================================================================================================================================
%==================================================================================================================================================
\section{Direct implementation of configurational relationalism}\label{SSSec: 1.1.11}
%==================================================================================================================================================
%==================================================================================================================================================

Configurational relationalism can also be implemented, at least in lower-$d$ RPM's, by directly constructing one's formulation to be $\fG$-invariant \cite{FORD, Cones}. 
One can view this as working directly on the relational configuration space.
%)
There is now no need for any arbitrary $\fG$-frame variables, nor then do any linear constraints arise nor 
are these to be used as a basis for reduction to pass to a new action. 
Here, instead, one's action is already directly $\fG$-invariant i.e. on $\fQ/\fG =: \widetilde{\fQ}$ by construction; this is 
reflected by the more complicated form taken by the kinetic arc element/kinetic term.  
This approach \cite{FORD, Cones} builds on work of Kendall \cite{Kendall} on the spaces of shapes and the `cones' \cite{Cones} 
over these so as to include scale, by then using the Jacobi--Synge approach to build a natural mechanics from a metric geometry.  
I.e. $\ttJ\ttS$(shape space) and $\ttJ\ttS(\mC$(shape space)) = $\ttJ\ttS$(relationalspace) where C stands for `the cone over' (a notion explained in Chapter 3 of \cite{FileR}). 
More generally, see Sec 9.2 for this terminology and Secs 9.6 and 9.8 for comparison between it and configuration space notions in GR.    

\mbox{ } 

\noindent This {\bf direct relationalspace construction} can be viewed as a gestalt of both the geometrization of mechanics that motivated Jacobi and Synge themselves and of the establishment of a temporally relational theory as above.\footnote{I use {\it relationalspace} 
%%%%%%%%%%%%%%%%%%%%%%%%%%%%%%%%%%%%%%%%%%%%%%%%%%%%%%%%%%%%%%%%%%%%%%%%%%%%%%%%%%%%%%%%%%%%%%%%%%%%%%%%%%%%%%%%%%%%%%%%%%%%%%%%%%%%%%%%%%%%%%%%%%%%
as the portmanteau of relational space for scaled RPM and shape space for pure-shape RPM, i.e. each case's non-redundant configuration space.
See Chapter 3 of \cite{FileR} for detailed discussion of the direct relationalspace approach.}
%%%%%%%%%%%%%%%%%%%%%%%%%%%%%%%%%%%%%%%%%%%%%%%%%%%%%%%%%%%%%%%%%%%%%%%%%%%%%%%%%%%%%%%%%%%%%%%%%%%%%%%%%%%%%%%%%%%%%%%%%%%%%%%%%%%%%%%%%%%%%%%%%%%%

\mbox{ } 

\noindent Note 1) Chapter 3 of \cite{FileR} establishes that the direct relationalspace implementation coincides with the configuration space reduction procedure `Best Matching 2-5)' in the case of 1- and 2-$d$ RPM's for which both are explicitly calculable.  

\noindent Note 2) {\it This provides a second foundation for RPM's that is independent of, but the output from which is coincident with, Barbour's work}  \cite{FORD, Cones}. 
Indeed, it coincides with what arises from Barbour's formulation upon performing reduction (see also Chapter 3 of \cite{FileR}).

\noindent Note 3) This formulation possesses emergent time as per (\ref{emtime}), quadratic constraint (\ref{GEnCo}), no linear constraints, 
evolution equations in form (\ref{Evol}) and energy constraint (\ref{ENERGY}) [the last two now have no $\ttg$ subscripts].

%===================================================================================================================================================
%===================================================================================================================================================
\section{Examples of Relational Theories}
%===================================================================================================================================================
%===================================================================================================================================================

%========================================================================================================
\subsection{Scaled RPM action in relative Jacobi coordinates} \label{SSSec: ERPM-Rel-Jacs}
%========================================================================================================

As relative Jacobi coordinates are inter-particle (cluster) separations, I take $\fQ = \fR(N, d)$ = $\mathbb{R}^{\sn d}$, and $\fG$ = Rot($d$) 
-- the $d$-dimensional rotation group.  
Then \cite{06I} using mass-weighted Jacobi coordinates $\rho^i$ and rotational auxiliary variable $\underline{B}$, 
\be
\FS^{\sE\sR\sP\sM}_{\sJ-\sJ\sA} = 
\sqrt{2}\int\sqrt{ E - V(\underline{\rho}^j\cdot\underline{\rho}^k \mbox{ alone)}} \, \d s^{\sE\sR\sP\sM}_{\sJ-\sJ\sA}      
\mbox{ } 
\label{wasT}
\mbox{ for } \mbox{ }  
\d s^{\sE\sR\sP\sM}_{\sJ-\sJ\sA} = ||\d_{{\underline{B}}}\brho|| \mbox{ } , \mbox{ } \mbox{ } 
\d_{\underline{B}}\underline{\rho}^{i} := \d\underline{\rho}^i - \d{\underline{B}} \cr \underline{\rho}^{i} \mbox{ } .  
\ee
\be
\mbox{Then the conjugate momenta are}  \hspace{0.7in}
\pi_{\alpha i} = \delta_{\alpha\beta}\delta_{ij}\Star_{\underline{B}}\rho^{\beta i} \mbox{ } ,
\mbox{ } 
\mbox{ where }
\mbox{ } 
\Star_{\underline{B}} := {\d }/{\d t^{\se\sm(\sJ\sB\sB)}_{\underline{B}}} := \sqrt{2W}\d/||\d_{\underline{B}}\brho|| \mbox{  }  . \hspace{4in}
\eeq 
The surviving constraints are, as a primary constraint, ${\cal E} := ||\bpi||^2/2 + V = E$,  and, as a secondary constraint from variation with  respect to $\underline{B}$, $\underline{\cal L} =  \sumin \underline{\rho}^{i} \cr \underline{\pi}_{i} =  0$.

\beq
\mbox{The emergent JBB time is now given by } \hspace{0.5in}
t^{\se\sm(\sJ\sB\sB)} - t^{\se\sm(\sJ\sB\sB)}(0) = 
\stackrel{\mbox{\scriptsize extremum $\underline{B} \mbox{ } \in \mbox{ } \mbox{Rot($d$)}$}}
         {\mbox{\scriptsize of $\stS^{\tE\tR\tP\tM}_{\tJ-\tJ\tA}$}}                                                              
\left(                                                              
                                                               \int||\d_{\underline{B}}\mbox{\boldmath$\rho$}||/\sqrt{2W} 
\right)
\mbox{ } . \hspace{4in}
\eeq

This RPM model gives back the zero angular momentum portion of Newtonian Mechanics for the whole universe, or an arbitrary angular momentum 
Newtonian Mechanics for an island universe subsystem.

%========================================================================================================
\subsection{Reduced/Relationalspace examples}
%========================================================================================================

The mathematically simplest \cite{FileR} RPM's in this respect are the $N$-stop metrolands (`$N$ particles in 1-$d$ models'), followed by the $N$-a-gonlands (`$N$ particles in 2-$d$'), the first two of which have are called, rather, triangleland \cite{08I, 08II, +Tri, 08III}  and quadrilateralland \cite{QuadI, QuadII}.  
The scaled $N$-stop metroland's configuration space is $\mathbb{R}^n$ \cite{Cones}, for $n = N - 1$, with standard flat-space metric most lucidly expressed in (ultra)spherical polar coordinates that pick out the shape--scale split and constitute the cone over the shape space $\{n - 1\}$-sphere.  
More generally, the RPM with $\fQ = \mathbb{R}^{Nd}$ has translation-corrected configuration space $\mathbb{R}^{nd}$ and then scaling-corrected configuration space (`{\it preshape space}') $\mathbb{S}^{nd - 1}$.  
$$
\mbox{The scaled $N$-a-gon's configuration space is the cone \cite{Cones} over $\mathbb{CP}^{N-2}$ \cite{Kendall84, FORD}} \hspace{0.3in}
S^{\sE\sR\sP\sM} = \sqrt{2}\int\sqrt{E - V(\rho, {\mZ}^{\sfA})}\d s^{\sE\sR\sP\sM}_{\sJ-\sF\sS\sA}   \mbox{ } , \hspace{4in}
$$
$$
\mbox{for}  \hspace{1.5in}
\d s^{\sE\sR\sP\sM \, 2}_{\sJ-\sF\sS\sA}  = \d\rho^2 + \rho^2\d s^2_{\sF\sS} \mbox{ } , \mbox{ } \mbox{ } 
\d s^2_{\sF\sS} = \big\{\{1 + |{\mZ}|_{\sc}^2\}|\d {\mZ}|_{\sc}^2 - |({\mZ} ,\d\overline{{\mZ}}\}_{\sc}|^2\big\}/
\{1 + |{\mZ}|_{\sc}^2\}^2  \mbox{ } .  \hspace{4in}
$$ 
I use $|{\mZ}|_{\sc}\mbox{}^2 := \sum_{\stta}|{\mZ}^{{\stta}}|^2$, $( \mbox{ } , \mbox{ } )_{\sc}$ 
for the corresponding inner product, overline to denote complex conjugate and $|\mbox{ }|$ to denote complex modulus.
The conjugate momenta are then $\pi_{\rho} = \rho^{*}$ and $\pi^{\sZ}_{\sfA} = \rho^2M_{\sfa\sfb}Z^{\sfb\,*}$.  
${\cal E}:= \{\pi_{\rho}^2 + N^{\sfa\sfb}\pi^{\sZ}_{\sfa} \pi^{\sZ}_{\sfb}/\rho^2\}/2 + V = E$ for $N^{\sfa\sfb}$ the inverse of the 
Fubini--Study metric.  
$$
\mbox{The emergent time is then given by }  \hspace{1in}
t^{\se\sm(\sJ\sB\sB)} - t^{\se\sm(\sJ\sB\sB)}_0 = \int \d s\left/\sqrt{2\{E - V\}} \right. \mbox{ } .   \hspace{4in}
$$
Next, in the specific triangleland case
\beq 
S^{\sE\sR\sP\sM}_{\triangle} = \sqrt{2}\int\sqrt{E - V(I, \Theta, \Phi)}\d s^{\sE\sR\sP\sM}_{\sJ-\sA}  
\mbox{ }
\mbox{ for } \mbox{ } 
\d s^{\sE\sR\sP\sM \, 2}_{\triangle(\sf\sll\sa\st)}  = \d\rho^2 + \rho^2\{\d\Theta^2 + \mbox{sin}^2\Theta\d^2\Phi\}/4
\eeq 
which becomes, up to an overall factor of $1/4I$ that can be passed on to the $E - V$ term,     
\beq  
\d s^{\sE\sR\sP\sM\, 2}_{\triangle(\sf\sll\sa\st)} = \d I^2 + I^2\{\d\Theta^2 + \mbox{sin}^2\Theta\d^2\Phi\}  
\eeq  
Here, $\rho$ is the configuration space radius, equal to $\sqrt{I}$ for $I$ the total moment of inertia of the system.
$\Phi$ is the relative angle between the base and the median that are privileged by one's coordinate choice, 
and $\Theta$ is a function of the ratio of the lengths of these (see e.g. \cite{QuadI} for more details).

The momenta are then $\pi_{I} = I^{*}$, $\pi^{\Theta} = I^2\Theta^*$ and $\pi^{\Phi} = I^2\mbox{sin}^2\Theta\,\Phi^*$.  
${\cal E}:= \{\pi_{I}^2 + \{\pi_{\Theta}^2 + \pi_{\Phi}^2/\mbox{sin}^2\Theta\}/I^2\}/2 + V/4I = E/4I$ for $N^{\sfa\sfb}$ the inverse of the FS metric. 
Finally, the emergent JBB time is now given by 
\beq
t^{\se\sm(\sJ\sB\sB)} - t^{\se\sm(\sJ\sB\sB)}_0 = \int \d \fs^{\sE\sR\sP\sM}_{\triangle(\sf\sll\sa\st)}/\sqrt{\{E - V\}/2I} \mbox{ } .  
\eeq

%==============================================================================================================================
\subsection{Standard spacetime presentation of GR: a relational critique}
%==============================================================================================================================

The usual covariant spacetime tensor presentation of the Einstein field equations follows from the Einstein--Hilbert action
\beq
S^{\sG\sR}_{\sE\sH} = \int \d^4x\sqrt{|g|} \, \mbox{Ric}(\bg) \mbox{ }    
\eeq
for $\bg$ the spacetime metric with determinant $g$ and Ricci scalar Ric($\bg$).  
This came about due to Einstein reconceptualizing the nature of space and time, in good part due to the compellingness of Mach's arguments. 
However, there is considerable confusion as to how GR is and is not Machian.   
SR transmutes rather than eliminates absoluteness issues -- one passes from Newton's set of 
unexplainedly privileged notions to another set, the `SR inertial frames'.  
Doing so amounts to a destruction of simultaneity, which is replaced by the universal significance of null cones.  
In SR it is often said that space and time can also be regarded as fused into the spacetime of Minkowski.   
(However, SR breaks only isolation of space and time, not their distinction \cite{Broad}.)   
Whitrow \cite{Whitrow} and Barbour \cite{B11} are also dismissive of Minkowski's considering the individual notions 
of time and space to be  ``{\it doomed to fade away.}")
Moreover, this is itself a fixed background structure, possessing a timelike Killing vector and privileged spatial 
frames that precede the introduction of actual material physics into one's worldview.  
However, next, in GR matter now exerts an influence on the form of space and time, 
by which it is in general curved rather than flat as Minkowski spacetime is.   
While in some senses that has a Machian character, Einstein's inception of the theory did not concretely 
build up on Mach's ideas \cite{WheelerGRT, DOD}, so whether the theory actually implements these has been a source of quite some argument. 
%
%(This is not helped by Mach's ideas being all of multiple, poorly understood and detached from concrete mathematical implementations).   
%
We shall see below, however, that directly building up on some of Mach's ideas happens to lead one to (a portion of) GR, arriving to it in its 
(geometro)dynamical form \cite{Buckets, DOD, RWR, Phan}, so that Einstein's GR does, in any case, happen to contain this philosophically 
desirable kernel.

%================================================================================================================================
\subsection{GR is a relational theory. 0. Geometrodynamics as a precursor}
%================================================================================================================================

While the role of spacetime in GR has often been touted, one should not forget that GR also admits a dynamical interpretation in terms of evolving spatial 3-geometries, i.e. geometrodynamics.  
In this approach to GR, one has to preliminarily choose a residual NOS in the sense of a spatial hypersurface of fixed topology $\Sigma$. 
This, I take to be a compact without boundary one for simplicity, and 3-$d$ as this suffices to match current observations.  
Under the ADM split of the spacetime metric\foo{I use capital Greek letters as spacetime indices.  
%%%%%%%%%%%%%%%%%%%%%%%%%%%%%%%%%%%%%%%%%%%%%%%%%%%%%%%%%%%%%%%%%%%%%%%%%%%%%%%%%%%%%%%%%%%%%%%%%%%%%%%%%%%%%%%%%%%%%%%%%%%%%%%%%%%%%%%%%%%%%%%%%%% 
%
The spatial topology $\Sigma$ is taken to be compact without boundary. 
$h_{\mu\nu}$ is a spatial 3-metric thereupon [within the perspective of (36), the {\it induced metric} upon this by the spacetime metric 
$g_{\Gamma\Delta}$, with determinant $h$, covariant derivative $D_{\mu}$, Ricci scalar Ric($h$) and conjugate momentum $\pi^{\mu\nu}$.  
$\alpha$ is the lapse and $\beta^{\mu}$ is the shift.
$\pounds_{\beta}$ is the Lie derivative with respect to $\beta^{\mu}$.  
$t$ is GR coordinate time.  
$\Lambda$ is the cosmological constant.
The GR configuration space metric is ${\fM}^{\mu\nu\rho\sigma} := \{h^{\mu\rho}h^{\nu\sigma} - h^{\mu\nu}h^{\rho\sigma}\}$, i.e. the undensitized inverse DeWitt supermetric with determinant ${\fM}$ and inverse ${\fN}_{\mu\nu\rho\sigma}$ that is itself the undensitized DeWitt supermetric, $h_{\mu\rho}h_{\nu\sigma} - h_{\mu\nu}h_{\rho\sigma}/2$. 
I note that this metric, and thus the GR kinetic term, is {\sl indefinite}.  
The densitized versions are $\sqrt{h}$ times the former and $1/\sqrt{h}$ times the latter. 
To represent this as a configuration space metric (i.e. with just two indices, and downstairs), use DeWitt's 2 index to 1 index map \cite{DeWitt67}.}
%%%%%%%%%%%%%%%%%%%%%%%%%%%%%%%%%%%%%%%%%%%%%%%%%%%%%%%%%%%%%%%%%%%%%%%%%%%%%%%%%%%%%%%%%%%%%%%%%%%%%%%%%%%%%%%%%%%%%%%%%%%%%%%%%%%%%%%%%%%%%%%%%%
\beq
g_{\Gamma\Delta} =
\left(
\stackrel{    \mbox{$ \beta_{\mu}\beta^{\mu} - \alpha^2$}    }{ \mbox{ }  \mbox{ }  \beta_{\delta}    }
\stackrel{    \mbox{$\beta_{\gamma}$}    }{  \mbox{ } \mbox{ }  h_{\gamma\delta}    }
\right)
\mbox{ }    
\eeq
with respect to a foliation by such spatial hypersurfaces, the Einstein--Hilbert action splits into the ADM--Lagrangian action 
\beq
\FS^{\sG\sR}_{\sA\sD\sM-\sL} = 
\int_{\Sigma}\d\mt\int\d^3x\sqrt{h}\,\alpha 
\left\{
\mT^{\sG\sR}_{\sA\sD\sM-\sL}/\alpha^2 +   \mbox{Ric}(h)   - 2\Lambda
\right\} \mbox{ } , \mbox{ } \mbox{ for }
\label{ADM}
\mT^{\sG\sR}_{\sA\sD\sM-\sL} = ||\dot{h}_{\mu\nu}     - \pounds_{\beta}h_{\mu\nu}||^2_{\mbox{\scriptsize ${\bfM}$}}/4 \mbox{ } .  
\eeq
This action encodes the GR Hamiltonian constraint 
\beq
{\cal H} := {\cal N}_{\mu\nu\rho\sigma}\pi^{\mu\nu}\pi^{\rho\sigma} - \sqrt{h}\{\mbox{Ric}(h) - 2\Lambda\} = 0
\label{Hamm}
\eeq
from variation with respect to the lapse $\alpha$ and the momentum constraint 
\beq
{\cal M}_{\mu} := - 2D_{\nu}{\pi^{\nu}}_{\mu} = 0  
\label{Momm}
\eeq
from variation with respect to the shift $\beta^{\mu}$.
The latter is interpretable geometrically as GR not being just a dynamics of 3-metrics (`metrodynamics') but furthermore that moving the spatial points around with 3-diffeomorphisms does not affect the physical content of the theory.    
The remaining information in the metric concerns the `underlying geometrical shape'.  
This is how GR is, more closely, a dynamics of 3-geometries in this sense: geometrodynamics. 
This description is less (rather than completely non-) redundant (the partial redundancy being due to the Hamiltonian constraint not yet being addressed).
Some useful notes about (in some cases precursors to) Problem of Time facets are as follows.  

\mbox{ } 

\noindent Note 1) The GR constraints ${\cal M}_{\mu}$  and ${\cal H}$ {\sl close} in the form of the Dirac Algebra \cite{Dirac, HKT}.

\noindent Note 2) In the dynamical approach to GR, one often considers not one spatial hypersurface $\Sigma$ but a {\it foliation} of the spacetime by such spatial hypersurfaces.  
Moreover, classical GR has the further property of {\it refoliation invariance}: the evolution between two given spatial hypersurfaces
is independent of which foliation interpolating the two is used.  
Both the closure of Note 1) and foliation-independence are held to be valuable properties of classical GR.  
Moreover, both of these have relational character -- the first via Sec 10 and the second due to foliation dependence in GR amounting to 
coordinate-dependent physics, at least from the point of view in which conventional GR's spacetime notion is taken as given.  

\noindent Note 3) The {\bf Global Problem of Time} is due to a single time function not always being definable over the whole of spacetime/space. 
Classically, this is already present but is a relatively straightforward issue of the finiteness of coordinate charts and the subsequent standard meshing condition being required in changes involving passing from one GR coordinate time to another.  

\noindent Note 4) Quantities commuting with both ${\cal M}_{\mu}$  and ${\cal H}$ are {\it Dirac observables}; quantities commuting only 
with the former are \K observables.  
The {\bf Problem of Observables} facet of the Problem of Time is that these are hard to find (especially the Dirac ones), though Rovelli's own opinion on this matter is that this be sidestepped by considering only a weaker notion of partial observables (see Sec \ref{cl-pers}). 
LMB relationalism favours \K's stance \cite{BF08}, or, at least a partly-Dirac stance \cite{AHall} (i.e. commutativity with ${\cal H}$ still being required alongside permission to view ${\cal H}$ as distinct from a gauge constraint, that in itself constituting a deviation from Dirac's writings \cite{Dirac}). 
In fact, following Bell \cite{Speakable} I prefer talking in terms of {\it beables} rather than observables, especially in the context of Quantum Cosmology.

%========================================================================================================
\subsection{GR is a relational theory. 1. Sandwich and Chronos antecedents}\label{SSec: 1.3}
%========================================================================================================

I continue here with discussion of the {\it thin sandwich approach}, in which Baierlein, Sharp and Wheeler \cite{BSW, WheelerGRT} regarded $h_{\mu\nu}$ and $\dot{h}_{\mu\nu}$ as knowns.
[I.e. the `thin' limit of taking the bounding `slices of bread' $h_{\mu\nu}(\lambda = 1)$ and $h_{\mu\nu}(\lambda = 2)$ as knowns.] 
This is solved for the spacetime `filling' in between.
In more detail, 

\mbox{ }  

\noindent BSW 0) They take as primary a conventional difference-type action for geometrodynamics.             
  
\noindent BSW 1) They vary this with respect to the ADM lapse  $\alpha$.
  
\noindent BSW 2) They solve resulting equation for $\alpha$.   
  
\noindent BSW 3) They substitute this back in the ADM action to obtain a new action: the BSW action,
\beq
\FS^{\sG\sR}_{\sB\sS\sW} = 
\sqrt{2}\int\d \mt \int_{\Sigma}\d^3x\sqrt{h}\,\d\s^{\sG\sR}_{\sA\sD\sM-\sL}\sqrt{\mbox{Ric}(h) - 2\Lambda} \mbox{ } .   
\eeq

\noindent BSW 4) They then take this action as one's primary starting point.

\noindent BSW 5) They vary with respect to the shift $\beta^{\mu}$ to obtain the GR linear momentum constraint, ${\cal M}_{\mu} = 0$.

\noindent BSW 6) They posit to solve the Lagrangian form of ${\cal M}_{\mu} = 0$ for the shift $\beta^{\mu}$.

\noindent BSW 7) They posit to then substitute this into the computational formula for the lapse $\alpha$.   

\noindent BSW 8) Finally, they substitute everything into the formula for the extrinsic curvature to construct the region of spacetime 
contiguous to the original slice datum. 

\mbox{ } 

\noindent Note 1) Unfortunately the p.d.e. involved in this attempted elimination is the {\it thin sandwich equation}; this is difficult and not much is known about it (obstructions based on this are termed the {\bf Thin Sandwich Problem}, which is one of the facets of the Problem of Time). 
Counterexamples to its solubility have been found \cite{TSC1, counterTS2}, while good behaviour {\sl in a restricted sense} took many years 
to establish \cite{TSC2}.  

%\noindent Note 2) I also gave myself the liberty of adding a cosmological constant to the potential factor of this action. 
%
%\noindent This does not significantly change the above procedure or the thin sandwich problem.  

\noindent Note 2) BSW 0--3) are a multiplier elimination; they can be reformulated as a cyclic velocity elimination  
i.e. a passage to the Routhian \cite{Lanczos, FileR} directly equivalent to the Euler--Lagrange action to Jacobi action move mentioned in the Introduction, or a cyclic differential elimination.

\noindent Note 3) BSW 4--6) directly parallel Best Matching 1--3) modulo the upgrade from multiplier shift to 
cyclic velocity/differential of the instant. 
Thus in the geometrodynamics case, Best Matching is hampered in practise by the Thin Sandwich Problem.  
A general phrasing for this is the obstruction, whether practical-computational or provable-analytical, 
to 
\beq
\mbox{solving ${\cal L}\mi\mn_{\sfZ} = 0$ for whichever form of $\ttg^{\sfZ}$ auxiliaries it contains} \mbox{ } .   
\label{BMMMM}
\eeq
Moreover, `sandwich' is clearly a geometrodynamical, or more generally Diff(3)- or foliability-specific name, whereas `best matching' is a more universal one, so in fact I choose the name {\bf Best Matching Problem} for the general theoretical problem (\ref{BMMMM}) of which the Thin Sandwich Problem is but the particular case corresponding to theories with some kind of NOS Diff-invariance.  
This reflects that the sandwich problem is both generalized, and placed on relational terms, in the present article.  

\noindent Note 4) BSW did not consider the culmination move Best Matching 4) of this which (merely formally) produces an action on superspace.  

\noindent Note 5) BSW 7) is, however, a primitive version of the $\tea^{\se\sm(\sJ\sB\sB)}$ computation (via the $\ttN$ to $\dot{\ttI}$ to 
$\d\ttI = \d  \tea^{\se\sm(\sJ\sB\sB)}$ and hence $\tea^{\se\sm(\sJ\sB\sB)}$ progression).  
There is a presupposed (BSW) versus emergent (relational approach) difference in the status of the computed object too.

\noindent Note 6)  As far as I am aware, Christodoulou \cite{Christodoulou1, Christodoulou2} was the first person to cast BSW 7) as an emergent time 
[amounting to the differential form of (\ref{Kronos}), which he termed the `{\it chronos principle}'].  
Moreover, like Wheeler, he did not base this on relational first principles. 
[Though Christodoulou was certainly aware in doing so of some of what Barbour and I have held to be the Leibniz--Mach position on relational time: ``{\it They contain the statement that time is not a separate physical entity in which the changing of the physical system takes place.  
It is the measure of the changing of the physical system itself that is time}"].  
He deduced the generalization of (\ref{BMMMM}) to include gauge fields minimally-coupled to GR involving shift and the Yang--Mills generalizations of the electric potential \cite{Christodoulou3}. 
I credit Alevizos for pointing out this connection between Barbour's and Christodoulou's work.  

\noindent Note 7) BSW 8) is  subsequent step concerning dynamical evolution which for geometrodynamics has nontrivial geometrical content.  

\noindent Note 8) I have thus clearly laid out how the technical content of Barbour's Best Matching notion resides within a 
very immediate generalization of the original BSW paper.  
Wheeler did not, however, have the relational significance that underpins  this procedure, or the cyclic auxiliary coordinate formulation 
of the next SSec.

%=========================================================================================================================================
\subsection{GR as a relational theory. 3. Actual relational reformulation}
%=========================================================================================================================================

The Barbour-type indirect formulation of RPM's (\ref{GeneralAction},\ref{Taction}) makes these parallels particularly clear. 
In the geometrodynamical 

\noindent counterpart of this, $\fQ$ = Riem($\Sigma$) -- the space of Riemannian 3-metrics on some 
spatial manifold of fixed topology $\Sigma$ (taken to be compact without boundary for simpleness).  
This choice of 3-metric objects is a simple one (one geometrical object, rather than more than one), moreover one that is 

\noindent A) useful for physics given its significance in terms of lengths.

\noindent B) Mathematically rich -- one then gets a connection for free -- the metric connection -- 
as well as the notions of curvature and of Hodge star.

\noindent Nevertheless, one is entitled to view this choice as a possibly inessential simplicity postulate.  

\noindent The group of irrelevant motions $\fG$ is now Diff($\Sigma$), i.e., the diffeomorphisms on $\Sigma$.  
This configurational relationalism is implemented indirectly by introducing auxiliary variables that represent arbitrary-Diff($\Sigma$) 
corrections. 
Superspace($\Sigma$) = Riem($\Sigma$)/Diff($\Sigma$).  
Temporal relationalism is implemented by building a MRI/MPI action \cite{RWR, Lan, Phan} [this is similar to the above BSW action but now properly combining the temporal and configurational relationalisms].    
The relational formulation of geometrodynamics is valuable in providing guidance in yet further investigations of alternative conceptual foundations 
for GR \cite{BB82, B94I, RWR, Phan, FEPI}, and as regards addressing the Problem of Time in Quantum Gravity (see Sec \label{SSec: 1.7} and Part III of \cite{FileR}).  
Now,\foo{$\dot{\mF}^{\mu}$ is the velocity of the 
%%%%%%%%%%%%%%%%%%%%%%%%%%%%%%%%%%%%%%%%%%%%%%%%%%%%%%%%%%%%%%%%%%%%%%%%%%%%%%%%%%%%%%%%%%%%%%%%%%%%%%%%%%%%%%%%%%%%%%%%%%%%%%%%%%%%%%%%%%%%%%%%%%
frame; in the manifestly relational formulation of GR, this cyclic velocity plays the role more usually played by the shift Lagrange multiplier coordinate.}
%%%%%%%%%%%%%%%%%%%%%%%%%%%%%%%%%%%%%%%%%%%%%%%%%%%%%%%%%%%%%%%%%%%%%%%%%%%%%%%%%%%%%%%%%%%%%%%%%%%%%%%%%%%%%%%%%%%%%%%%%%%%%%%%%%%%%%%%%%%%%%%%%%
\beq
\FS_{\sG\sR}^{\sr\se\sll\sa\st\si\so\sn\sa\sll} =  2\int\d\lambda\int\d^{3}x\sqrt{h}\sqrt{ \mT_{\sG\sR}^{\sr\se\sll\sa\st\si\so\sn\sa\sll}  \{\mbox{Ric}(\bh) - 2\Lambda\}  } = \sqrt{2}\int\d\lambda\int\d^{3}x\sqrt{h}\sqrt{\mbox{Ric}(\bh) - 2\Lambda}
\d\ms_{\sG\sR}^{\sr\se\sll\sa\st\si\so\sn\sa\sll}
\eeq
\beq
\mbox{for } \mbox{ } \mbox{ }  
\mT_{\sG\sR}^{\sr\se\sll\sa\st\si\so\sn\sa\sll}   := ||\Circ_{\sF}\bh||_{\mbox{\scriptsize \boldmath${\bfM}$}}\mbox{}^2/4 
\mbox{ }  \mbox{ or } \mbox{ }
\d\ms_{\sG\sR}^{\sr\se\sll\sa\st\si\so\sn\sa\sll} := ||\d_{\sF}\bh||_{\mbox{\scriptsize \boldmath${\bfM}$}}/2 
\mbox{ and } 
\Circ_{\sF}h_{\mu\nu} := \dot{h}_{\mu\nu} - \pounds_{\dot{\sF}}h_{\mu\nu}   
\mbox{ }  \mbox{ or } \mbox{ }
\d_{\sF}h_{\mu\nu} := \d{h}_{\mu\nu} - \pounds_{\d{\sF}}h_{\mu\nu} 
\label{GRaction}.  
\eeq
In this case, the action is a JBB$[P(\langle \mbox{Riem}(\Sigma), \mbox{ } \mbox{\boldmath${\bfM}$}\rangle, \mbox{ } \mbox{Diff}(\Sigma))]$ 
for a particular potential proportional to Ric($\bh$).   

\mbox{ } 

\noindent Note 1) The relational action bears many similarities to the better-known BSW one, but supersedes it via attaining manifest temporal relationalism.  
The two become indistinguishable for minisuperspace:  
%
%likewise the pair of actions (\ref{ADM}, \ref{AAAAAc}).  
%
${\fM}^{\mu\nu\rho\sigma}(h_{\gamma\delta}(x^{\omega}))$ collapses to an ordinary $6 \times 6$ matrix ${\fM}_{\Gamma\Delta}$ or 
further in the diagonal case (a $3 \times 3$ matrix ${\fM}_{\sfA\sfB}$) -- the `minisupermetric'.  

\noindent 
$$
\mbox{Then the conjugate momenta are} \hspace{0.15in} 
\pi^{\mu\nu} =  \sqrt{h}{\fM}^{\mu\nu\rho\sigma}\Star_{\sF}h_{\rho\sigma} 
\label{tumvel}
\mbox{ } \mbox{ where } \mbox{ } 
\Star_{\underline{\sF}} := {\d }/{\d \mt^{\se\sm(\sJ\sB\sB)}_{\underline{\sF}}} := \sqrt{2\{\mbox{Ric}(\bh) - 2\Lambda\}}\d/
||\d_{\sF}\bh||_{\mbox{\scriptsize\boldmath${\bfM}$}} \mbox{ } .  \hspace{4in}  
$$  
MPI/MRI gives GR's Hamiltonian constraint (\ref{Hamm}) as a primary constraint \cite{Dirac}; this has the momenta contribute quadratically but not linearly.

Also, variation with respect to the auxiliary Diff($\Sigma$)-variables $\mF^{\mu}$ gives the GR momentum constraint (\ref{Momm}).
\noindent Just as for the standard ADM approach, the momentum constraint is interpretable geometrically as GR being a geometrodynamics. 
In the relational approach, GR has an analogue of emergent JBB time (in this context crediting also Christodoulou \cite{Christodoulou2}), 
\beq
\mt^{\se\sm(\sJ\sB\sB)}(x^{\mu}) - \mt^{\se\sm(\sJ\sB\sB)}_0(x^{\mu}) = 
\stackrel{    \mbox{\scriptsize extremum} \mbox{ } \sF^{\mu} \mbox{ }  \in  \mbox{ }  \mbox{\scriptsize Diff}(\Sigma)        }   
         {    \mbox{\scriptsize of} \mbox{ } \stS^{\mbox{\tiny relational}}_{\tG\tR}    }  
\left. \int ||\d_{\sF}\bh||_{\mbox{\scriptsize \boldmath${\bfM}$}} \right/   \sqrt{     \mbox{Ric}(h) - 2\Lambda    }  \mbox{ } .
\label{GRemt}
\eeq
This represents the same quantity as the usual spacetime-assumed formulation of GR's {\it proper time}, 
and, in the (predominantly) homogeneous cosmology setting, the cosmic time. 
It also coincides with GR's own version of emergent semiclassical (alias WKB) time (see Appendix A and Sec 16.4 for a brief explanation of what this is).  

\mbox{ } 

\noindent Interpretation 4) Next, Wheeler asked \cite{Battelle} why ${\cal H}$ takes the form it does and whether this could follow from first principles (`7th route' to GR) rather than from mere rearrangement of the Einstein equations.  
To date, there are two different answers which tighten wide classes of ans\"{a}tze down to the GR form (see \cite{Phan} for a comparison of these).  

\mbox{ } 

\noindent Answer A) [Hojman--Kucha\v{r}--Teitelboim] \cite{HKT} is from {\it deformation algebra} first principles  (this assumes embeddability into spacetime).

\noindent Answer B) [Barbour--Foster--O'Murchadha and I] \cite{RWR, AB, San, OM02, OM03, Van, Than, Lan, Phan, Lan2} the `{\it relativity without relativity}' approach, which lies within the relational program, arrives at actions similar to the BSW one, and is the subject of the next two SSecs.

%========================================================================================================
\subsection{Relativity without relativity (and addition of matter)}\label{SSSec: RWR}
%========================================================================================================

Furthermore, adopting the relational first principles, {\sl without} assumption of additional features derived in ADM's approach, leads to the 
{\sl recovery} of the BFO-A action of GR as one of very few consistent choices within a large class of such actions. 
E.g. one does not need to assume the GR form of the kinetic metric or of the potential. 
One then has a family of trial actions $\ttJ\ttB\ttB[P(\langle \mbox{Riem}(\Sigma), \mbox{ } \mbox{\boldmath${\fM}$}^W\rangle, \mbox{ } \mbox{Diff}(\Sigma))]$ for $\mbox{\boldmath${\fM}$}^{W}$ the general ultralocal supermetric ${\fM}^{W}_{\mu\nu\rho\sigma} := h_{\mu\rho}h_{\nu\sigma} - 
Wh_{\mu\nu}h_{\rho\sigma}$ and a potential ansatz such as $\mW^{\st\sr\si\sa\sll} = A\,\mbox{Ric(\bh)} + B$ for $A$, $B$ constant,   
\beq
\FS^{\sR\sW\sR}_{\st\sr\si\sa\sll} = \sqrt{2}\int\sqrt{\mW^{\st\sr\si\sa\sll}}\d \ms^{\st\sr\si\sa\sll}_{\sq\su\sa\sd}  \mbox{ } .  
\eeq
Then relational postulates alongside a few simplicities already give this since the Dirac procedure \cite{Dirac} prevents most other choices of potential term $\d \ms^{\st\sr\si\sa\sll}$ from working \cite{RWR, San, Than, Lan, Phan, OM02, OM03} alongside fixing $W$ to take the DeWitt value 1 of GR as one of very few possibilities.
This amounts to a classical precursor of the Spacetime Reconstruction Problem facet of the Problem of Time, which, once again, the machinery 
of the classical GR constraints resolves.

The above relational formulation of GR is furthermore robust to the inclusion of a sufficiently full set of fundamental matter sources so as to describe nature.
Thus electromagnetism is included in \cite{RWR}, Yang--Mills Theory in \cite{AB}, 
spin-1/2 fermions in \cite{Van} and scalar and spin-1/2 fermion gauge theories \cite{Van, Than, Arelsusy}.  
Moreover, it does {\sl not} do so in a way by which relational postulates alone {\sl pick out these matter theories}. 
(\cite{Phan, Lan, Lan2} demonstrate how earlier claims to the contrary \cite{RWR} in fact contain additional tacit simplicity assumptions.  
A further gap, not yet addressed, is that the possibility of second-class constraints is not properly 
catered for in any uniqueness by exhaustion proof so far given.)

One way some of these things are accommodated is through lifting the homogeneous quadraticity restriction \cite{Lanczos, Van} (which is not per se relational).
In particular this allows for mechanics with linear `gyroscopic' terms \cite{Lanczos}, as well as moving charges, spin-1/2 fermions 
and the usual theories coupling these to gauge fields, scalars and GR. 
The form of action here is of Randers type \cite{Randers, ARel2}  
\beq
\FS^{\sR\sW\sR}_{\st\sr\si\sa\sll-\sA} = \int\int_{\Sigma}\d\Sigma \big\{\sqrt{2}\sqrt{\mW^{\st\sr\si\sa\sll}}\d \ms^{\st\sr\si\sa\sll}_{\sq\su\sa\sd} + \d\ms^{\st\sr\si\sa\sll}_{\sll\si\sn} \big\} \mbox{ } ,
\label{62}
\eeq
with $\d s^{\st\sr\si\sa\sll}_{\sll\si\sn}$ then not contributing to the quadratic constraint or to the emergent JBB time (thus this is not entirely universal, though the potential associated with such fields does sit inside the square root and thus itself does contribute to these things).  
$\d \ms^{\st\sr\si\sa\sll}_{\sll\si\sn}$ does however contribute to the linear constraints.

\mbox{ } 

\noindent A further implementation of Mach [E.A.: Relationalism 7B)] is that {\bf time is abstracted from BOSONIC change.}  

\mbox{ }

\noindent I further discuss these variants on relationalism involving boson--fermion distinctions (and the supersymmetric case) in \cite{Arelsusy}.  

\mbox{ }

\noindent Note: (\ref{62}) also renders clear how to extend Christodoulou's Chronos Principle working to include fermions.  

\mbox{ }  

\noindent  
Other simplicities involve neglecting the possibility of metric--matter cross-terms and other terms that one knows appear in the canonical 
formulation of more general fields, as well as the restriction on allowed orders of derivatives that is much more ubiquitous in theoretical physics.  
By such means, some more general/complicated possibilities can be included among the relationally-formulable theories:

\noindent 1) Brans--Dicke theory \cite{RWR}, 

\noindent 2) Proca theory \cite{Lan}, and 

\noindent 3) local-SR-cone-violating and equivalence principle violating vector--tensor theories \cite{Lan2}.

\noindent Thus, contrary to the claims/speculations in \cite{RWR}, relationalism does {\sl not} derive or imply the equivalence principle 
or a universal null cone. 
(The former effectively requires some form of geometrodynamical equivalence principle postulate \cite{Lan2, HKT}, while the latter requires a number of nonrelational simplicity assumptions.)

%=======================================================================================================================================
\subsection{Conformogeometrodynamical formulations on relational foundations}
%=======================================================================================================================================

\beq 
\mbox{York time} \hspace{0.7in}
\mt^{\sY\so\sr\sk} := \mbox{$\frac{2}{3}$}h_{\mu\nu}\pi^{\mu\nu}/\sqrt{h} = c(\lambda \mbox{ coordinate label time alone, i.e. a spatial hypersurface constant}) 
\mbox{ } \hspace{4in}
\label{CMCsl}
\eeq 
is \cite{Paris, 06II, SemiclI} more or less \cite{FileR} paralleled by an `Euler time' variable       
$
t^{\sE\su\sll\se\sr} := \suma \uRR^i \cdot \uPP_i.
$
This is all underlied for both GR and RPM's by scale--shape splits, the role of scale being played by 
$\sqrt{I} = \rho$ or $I$ for RPM's and by such as the scalefactor $a$ or $\sqrt{h}$ in GR.  
In both cases it is then tempting to use the singled-out scale as a time variable but this runs into monotonicity problems. 
These are avoided by using as times the quantities conjugate to (a function of) the scale, such as  
$\mt^{\sY\so\sr\sk}$ or $t^{\sE\su\sll\se\sr}$.

Moreover, GR in the conformogeometrical formulation on CMC slices \cite{York72, York73} can be reformulated \cite{ABFKO} as a relational theory in which $\fG$ consists both of Diff($\Sigma$) and a certain group of conformal transformations. 
Conf($\Sigma$) are conformal transformations; 
CRiem($\Sigma$) is pointwise superspace =  Riem($\Sigma$)/Conf($\Sigma$).  
CS($\Sigma$) is conformal superspace = Riem($\Sigma$)/Diff($\Sigma$) \mbox{\textcircled{S}} Conf($\Sigma$), for \textcircled{S} 
denoting semi-direct 
product \cite{Mackey, I84}.  
\{CS + V\}($\Sigma$) is conformal superspace to which has been adjoined a single global degree of freedom: the spatial volume of the universe. 
This can be viewed as Riem($\Sigma$)/Diff($\Sigma$) \textcircled{S} VPConf($\Sigma$), for VPConf($\Sigma$) the global volume-preserving conformal transformations.    
CS($\Sigma$) and (CS + V)($\Sigma$) have on a number of occasions been claimed to be the space of true dynamical degrees of freedom of GR 
\cite{York73, York74, FM96, ABFKO, BONew}.  
Following from the alternative theory of \cite{BO99} and \cite{ABFO}, \cite{ABFKO} used volume-preserving conformal transformations. 
and \cite{Kos2, Kos1, Merc, Kos3, Gomes, Kos4, GrybT, Gom2} additionally then considered a linking theory approach. 
Of the latter, I comment that using a Hamiltonian formulation is cleaner, but also that these papers use in excess of the usual amount of canonical theory; I do not know whether this is necessary; the current article presents the case for questioning excess structures in relational approaches.

%=====================================================================================================
\subsection{Relationalism and the Ashtekar variables formulation of GR}
%=====================================================================================================

Proponents of Ashtekar variables often use the word `relational' of their approach. 
However, this word has not hitherto been systematically used in their program in the full LMB or LMB-A sense; 
instead they use this word with less detail and/or to carry (a subset of) the perspectival meanings of Sec 14.  
Nevertheless, I state here that the Ashtekar variables scheme can be formulated in the full LMB or LMB-A relational scheme.
This is modulo a small glitch at the start of such an investigation, which nevertheless turns out to be entirely surmountable.  
Namely, that the pure gravity case of the Ashtekar variables approach is not castable in temporally relational form at the level of 
the classical Lagrangian.  
From the point of view in which spacetime is presupposed \cite{Van}, in which one sets about elimination of the lapse to form the temporally relational action, this glitch is due to the following.  
The lapse-uneliminated action is purely linear in the lapse.
This is because the well-known `pure-$\mT$' character of Ashtekar's canonical action for pure GR, in contrast to the more usual `$\mT - \mV$' 
form of the geometrodynamical action.
Thus the variation with respect to the lapse produces an equation independent of the lapse. 
Thus the lapse cannot now be eliminated from its own variational equation. 
However, it is clear that addition of matter fields breaks this pure linearity in the lapse, so this unusual accident that blocks the elimination of the lapse from its own variational equation ceases to occur and a temporally relational action can be obtained.

Thus that the Barbour/LMB-relational literature's looks to be fixated with GR in geometrodynamical form as regards which examples it uses 
is in fact down to a matter of taste/of each paper's applications, rather than being in any way being indicatory of LMB-relationalism only applying to the older geometrodynamics and not the newer Ashtekar variables work that underlies LQG.
It is then surely of foundational importance for Ashtekar variables programs for them to be revealed to be relational in this additional 
well-motivated and historically and philosophically well-founded way in addition to the various other meanings that they have hitherto been pinning on the word `relational' in the Ashtekar variables/LQG program.
Pointing this out is one of the main applications of this article. 
Configurational relationalism for Ashtekar variables formulations has  $\fG = SU(2) \times$ Diff(3).
Then variation with respect to the additional $SU(2)$-auxiliaries produces the $SU(2)$ Yang--Mills--Gauss constraint that is ubiquitous in the Ashtekar variables literature, whilst variation with respect to the Diff(3)-auxiliaries produces the Ashtekar variables form of the GR momentum 
constraint.   
The Hamiltonian constraint arises as a primary constraint, and the usual form for a JBB time emerges.    
An Ashtekar variables formulation of a type of GR shape dynamics has recently appeared \cite{BST}.

%========================================================================================================
\subsection{Is there a supersymmetric extension of RPM and does it toy-model supergravity?} \label{SSec: SusyQ}
%========================================================================================================

Compared to theories preceding it, supersymmetry is conceptually strange in how the product of two supersymmetry operations produces a {\sl spatial} transformation, and possibly also via how the supersymmetry constraint in supergravity is the square root of the Hamiltonian constraint
\cite{SqrtTeitelboim}. 
As such, there is some chance that the supersymmetry transformation is one of not conceptually sound, or conceptually deeper than, other transformations, and the relational perspective could be one that had more to say about this matter.  
Thus, {\sl the relational program might be able to provide its own prediction} as to whether to expect supersymmetry in nature. 
Conversely, insisting on supersymmetry might expose the current conceptualization of Relationalism to be insufficient.  
See Secs \ref{SSec: QG-Comp}, \ref{SSE} and \cite{Arelsusy} for more.

%========================================================================================================
\subsection{RPM's versus String Theory}\label{SSec: Stri-RPM}
%========================================================================================================

Aside from Sec 6's use of the preclusion of perturbative strings to sharpen up criteria for relationalism, there are some conceptual parallels between passing from studying point particles to studying strings and passing from point particles to studying relational quantities only. 
These are comparable from a conceptual perspective; whether each of these is rich, tractable and has anything to say about Quantum Gravity is another matter.  
The latter is moreover more conservative -- the relational quantities come from careful thought about the original problem rather than replacing it  with a distinct problem as in String Theory.  
This gives less new structure leading perhaps to less mathematical richness, but it it would appear to be a safer bet as regards the 
``{\it hypotheses non fingo}" tradition.  

\noindent Exporting the configurational relationalism/best matching ideas to other parts of physics could conceivably be of interest even if  not accompanied by the further demands of no background spatial structures or of temporal relationalism.    
E.g. one could consider what form these ideas take for target space theories of which strings on a given background.    
Finally, one would expect nonperturbative M-theory to have {\sl both} of these features (relationalism {\sl and} extended objects).

\mbox{ }

\noindent Provided that supersymmetry can be accommodated into (a suitable generalization of) Relationalism, I conjecture that relational notions 
along the lines of those in this article are necessary for any truly GR-embracing notion of background independence for M-theory\foo{To avoid  
%%%%%%%%%%%%%%%%%%%%%%%%%%%%%%%%%%%%%%%%%%%%%%%%%%%%%%%%%%%%%%%%%%%%%%%%%%%%%%%%%%%%%%%%%%%%%%%%%%%%%%%%%
confusion, I note here that some other uses of the term `background (in)dependence' in the String Theory literature have a different meaning, 
namely concerning the effect of choice of vacuum on string perturbations.} 
%%%%%%%%%%%%%%%%%%%%%%%%%%%%%%%%%%%%%%%%%%%%%%%%%%%%%%%%%%%%%%%%%%%%%%%%%%%%%%%%%%%%%%%%%%%%%%%%%%%%%%%%%  
Since background-independence amounts to a Problem of Time, it would then be exceedingly likely for M-theory to have one of those too.  
Thus study of the Problem of Time is likely to be a valuable investment from the perspective of developing and understanding M-theory too.  
Geometrodynamics (very easily generalizable to arbitrary-$d$) could be seen as a first toy model of spatially 10-$d$ M-theory.  
There may be better RPM's and geometrodynamics models for this particular setting, e.g. spatially 10-$d$ supergravity (which is a 
low-energy/`semiclassical' limit of M-theory). 

\mbox{ }  

\noindent The situation with extended objects that occupy/furnish a multiplicity of notions of space is, in addition to supersymmetry, an  interesting frontier in which to further explore the relational postulates.

%========================================================================================================
%========================================================================================================
\section{Further compatibility restrictions}\label{Patti}\label{SSec: QG-Comp}
%========================================================================================================
%========================================================================================================

Now we have covered some examples, we can return to this...

\noindent Relationalism 9) [E.A.] 
\noindent A) {\bf Nontriviality} $\fG$ cannot be too big [if dim($\fG$) $\geq$ dim($\fQ$) -- 1 then the relational procedure will yield a theory with not enough degrees of freedom to be relational, or even an inconsistent theory]. 

\noindent B) Further {\bf structural compatibility} is required.   

\mbox{ } 

\noindent Example 1) If one uses such as Sim($d_1$) or Eucl($d_1$) with $\fQ(N, d_2)$, it is usually for $d_1 = d_2$ (or at least $d_1 \leq d_2$).  

\noindent Example 2) If one uses Diff($\Sigma_1$) to match to Riem($\Sigma_2$), it is usually for $\Sigma_1 = \Sigma_2$, or at least for the 
two to be obviously related (e.g. taking out Diff($\mathbb{S}^1$) from Riem($\mathbb{S}^1 \times \mathbb{S}^1$)

\mbox{ }  

\noindent C) More concretely, $\fG$ {\bf has to have a (preferably natural, faithful) group action on} $\fQ$.

\mbox{ }

\noindent Relationalism 10) [Barbour and E.A. ] In looking to do fundamental modelling, one may well have a strong taste for {\bf lack of extraneity}, e.g. always taking out all rotations rather than leaving a preferred axis.  

\noindent I) Aut($\FA$) is then a very obvious designate for this, though some subgroup of Aut($\FA$) might also be desirable, 
and there is also the issue of automorphisms of up to which level of structure.  
Then Aut($\FA$) and its subgroups definitely comply with A) and stand a good chance of suitably satisfying criteria B) and C).

\noindent II) One may also have a strong taste for eliminating as many extraneous properties as possible [Barbour's view], 
though under some circumstances there may be strong practical reasons to not eliminate some [E.A.'s counter-view]. 

\mbox{ }

\noindent Example 1) One may wish to retain the scale, thus using Eucl($d$) rather than Sim($d$) for RPM's, or and superspace($\Sigma$) or \{CS + V\}($\Sigma$) instead of CS($\Sigma$) for geometrodynamics, so as to have a viable cosmological theory and possibly for reasons of time provision.  
Scale is possibly undesirable from the relational perspective via being a single heterogeneous addendum to the shapes \cite{Piombino, BONew}.
The cone structure makes scale especially extraneous in RPM's -- it has the same sort of dynamics irrespective of what the shapes are.
It may then seem quite conceptually undesirable for scale to play so prominent a roles in Cosmology. 
However, 1) there are currently no credible alternatives to this as regards explaining Cosmology. 
2) This heterogeneity can be useful by alignment with choices that would be harder to make without it, renders scale more presentable from a conceptual perspective.
This is the position I will take in the upcoming `Conformal Nature of the Universe' Conference at the Perimeter Institute.  
See Sec 16 for how various approaches fare in this regard.

\noindent Example 2) Quotienting out Diff($\Sigma$) and Conf($\Sigma$) does not interfere with the existence of spinors, but for further/other choices one has a lack of such guarantees \cite{PenRind}, likely compromising one's ability to model spin-1/2 fermion matter. 
Moreover, it is well-known that some choices of $\Sigma$ by themselves preclude the existence of fermions (see e.g. \cite{Nakahara}), and that orientability of $\Sigma$ is also often desired.
(But supersymmetric extensions are quite clearly also compatible in this sense \cite{Arelsusy}.)

\mbox{ }

\noindent In more detail, one has a background NOS ${\FA}$, which has properties deemed redundant, all or some of which can be modelled as such. 
There is then some level of structure $\langle {\FA}, {\cal P}_{\sm\so\sd\se\sll\sll\se\sd} \rangle$ within the $\langle {\FA}, 
{\cal P}_{\sT\so\st\sa\sll}\rangle = \langle {\FA}, \langle{\cal P}_{\sm\so\sd\se\sll\sll\se\sd}, 
{\cal P}_{\su\sn\sm\so\sd\se\sll\sll\se\sd} \rangle\rangle$.  
One then builds the configuration space of actually mathematically manipulable part-tangible part-intangible entities to lean on the structure of ${\FA}$, so I denote it in this Sec by $\fQ({\FA})$. 
Then finally one passes to 
\beq
\fQ({\FA})/\mbox{Aut}(\langle {\FA}, {\cal P}_{\sm\so\sd\se\sll\sll\se\sd} \rangle) \mbox{ } .
\eeq 
BB82 has a means of modelling continuous transformations at the level of Riemannian geometry.
As per Chapter 2 of \cite{FileR}, one can also conceive of ${\FA}$ in such a way that discrete isometries can be considered as well as continuous ones.  
If one fails to be able to model ${\cal P}_{\sT\so\st\sa\sll}$, then the automorphisms in question will not be freeing one of all the levels 
of structure/aspects of $\FA$. 
This is clearly the case for superspace($\Sigma$): quotienting out Diff($\Sigma$) from Riem($\Sigma$) 
clearly does not free one from what choice of $\Sigma$ one made in setting up Riem($\Sigma$).  
This is because $\Sigma$ pertains to the level of topological structure, and rendering the diffeomorphisms irrelevant is only a freeing 
from background {\sl metric} structure.  

The relationalist who takes minimalism to an extreme would then aim to remove all sources of indiscernibility and actors that cannot be acted upon.
%
%However, if a feature is held to play a role, it is kept.
%
See Secs 10 and 16 as regards the status of scale, and Chapter 14 of \cite{FileR} as regards whether spatial topology in GR is discernible and can be acted upon; another example is how dimension is not primary for RPM's (Chapter 2 of \cite{FileR} explains how 3-particle RPM's cannot 
distinguish between dimension 2 or higher).  

\mbox{ } 

\noindent The objective of relationalism is thus 
\beq
\fQ(\FA)/\mbox{Aut}(\langle \FA, {\cal P}_{\mbox{\scriptsize held to be } \su\sn\sp\sh\sy\si\sc\sa\sll}\rangle), 
\mbox{ or possibly even } \mbox{ }  
\fQ(\FA)/\mbox{Aut}(\langle\FA, {\cal P}_{\st\so\st\sa\sll} \rangle)   \mbox{ } .  
\eeq
\noindent Relationalism 11) [BFO--A].
Note that $\fQ$ (the entity taken to have some tangible physical content) has the {\bf a posteriori right to reject} \cite{RWR, Lan, Phan, Lan2} 
a proposed $\fG$ by triviality or inconsistency that go beyond 1), arising instead via, for the moment, the Dirac procedure yielding further constraints.

\mbox{ } 

\noindent Example 1) If one attempts to use $\fG =$ id with Riem($\Sigma$) and a BSW-like trial action (like \'{O} Murchadha \cite{OM02} or I \cite{San}), 
one finds that enlarging $\fG$ to Diff($\Sigma$) is {\sl enforced} by the ${\cal M}_{\mu}$ arising as an integrability of ${\cal H}$.  
This also exemplifies how not all subgroups of Aut are guaranteed to serve.  

\mbox{ }

\noindent Note 2) Additionally, a given $\fQ$, $\fG$ pair still constitutes a substantial ambiguity as to the form of the action principle.
This can sometimes be truncated with simplicity postulates like that the action is not to contain 
higher than first (or occasionally second) derivatives, but ambiguities beyond that still often remain.   

\mbox{ }  

\noindent Example 2) The potential is completely free in scaled RPM and free up to homogeneity in pure-shape RPM. 

\noindent Example 3) One can have GR or Euclidean GR or strong gravity for the Riem($\Sigma$), Diff($\Sigma$) pair 
with action restricted to contain at most first label-time derivatives and quadratically and at most second spatial derivatives.  

\mbox{ } 

\noindent Note 3) This scheme retains considerable freedom.  
Relationalism does {\sl not} exert a highly unique control over the form that theoretical physics is to take [though Relationalism 10) and 11) can sometimes help with this].  

\mbox{ }  

\noindent Relationalism 11) Quotienting out the $\fG$ is to {\bf remove all the structure in the} $\fQ$ {\bf that is actually extraneous}. 
However, precisely what structures/features are extraneous in the case of the universe we live in is not so clear.  
[E.g., again, whether to discard scale.]  
One passes to 
\beq
\fQ({\FA})/\mbox{Aut}(\langle {\FA}, {\cal P} \rangle) \mbox{ } 
\eeq 
for ${\cal P}$ the structures up to a certain mathematical level (e.g. metric geometry or topology). 
One might ideally wish for relationalism to remove {\sl all} levels of structure of $\FA$.  

\mbox{ }

\noindent In particular, $\fG$ = id is too trivial for most of theoretical physics, whilst a number of further GR features 
come from $\fG$ = Diff or some generalization (Diff $\times$ Conf, the SuperDiff of Supergravity) or possibly some slight weakening, such as by 
attitude to the large diffeomorphisms \cite{Giu} or restriction to the transverse diffeomorphisms \cite{Alvarez}.

\mbox{ }

I next consider four structural expansions on Barbour's `$\fQ$ is primary' postulate [Relationalism 3)].    

\mbox{ }

\noindent 1) One usually considers {\bf phase space} Phase, or at least {\bf adjoining momenta} $\ttP$ to the configurations $\ttQ$.
This generalizes  $\fQ$ to a more general concept of {\bf state space}, which I denote by $\fS$.
\noindent I will consider these expansions at the classical level first, and then at the quantum level, where such as Hilbert spaces of quantum states play the role of $\fS$.  
 
\mbox{ } 

\noindent 2) {\bf Categorizing}.  One may well wish to consider whichever $\fS$ as objects alongside corresponding morphisms $\MFM$ to form the 
category ($\fS$, $\MFM$). 
Examples of morphisms include the point transformations Point for $\fQ$ and the canonical transformations Can for $\fS$. 

\mbox{ }  

\noindent 3) {\bf Perspecting}.  
Taking, say, $\fS$ to be primary in fact amounts to (especially in the case of whole-universe studies) considering some set of subsystems of $\fS$.

\mbox{ } 

\noindent 4) {\bf Propositioning}.  
If, say, we take $\fS$ to be primary, Physics is in fact about the propositions about $\fS$, Prop($\fS$), rather than about $\fS$ itself.  

\mbox{ }  

\noindent As we shall see, some additional issues concerning how to compose some of these expansions is crucial.

%====================================================================================================================================================
%====================================================================================================================================================
\section{State space extension of $\fQ$-primality?}\label{SSE}
%====================================================================================================================================================
%====================================================================================================================================================

I consider passing from conceiving $\fQ$ to be primary to suggesting that $\ttQ$ and $\ttP$ are operationally distinct; I refer to the ensuing 
state space as {\bf RigPhase}.  
From the more usual perspective, this is a privileged {\it polarization} \cite{Woodhouse, Ashtekar} of Phase (phase space) that maintains the distinction between the physical $\ttP$'s and $\ttQ$'s.  
On the other hand, a distinct and far more common viewpoint is that phase space
\beq
\mbox{Phase} = (T^*(\fQ), \{ \mbox{ }  , \mbox{ } \}) 
\eeq
is central. 
Here, $T^*(\fQ)$ is the cotangent bundle (for simple bosonic theories) of momenta over the configurations and \{  ,  \} 
is the Poisson bracket, so that this is a Poisson algebra. 
N.B. the status of RigPhase in this program is as an {\sl option} that is more structurally minimalist.  

\mbox{ } 

\noindent 
RigPhase Motivation 1) As argued in Sec 3, considering more minimalist possibilities is very much part and parcel of relational thinking.    
One such avenue then would be to investigate what happens if one weakens the amount of structure assumed in the study of the 
`associated spaces' of the principles of dynamics. 
There are very strong reasons why $\fQ$ (or something with an equivalent amount of physical information in it) is indispensable, 
but some parts of the conventional structure of phase space are more questionable.  
Much as one can envisage preferred-foliation counterparts of GR spacetime, one can also construe of preferred-polarization versions of phase space.
Whereas relational perspectives usually view preferred foliations of spacetime as less physical, I contend that, at the level of the associated 
abstract spaces of the Principles of Dynamics, the relationalist's ontological hierarchy as regards preferred versions of these spaces is the reverse of that for spacetime itself.
My argument then is that the operational distinction between configuration and impact measurements can be encoded by using a preferred  polarization centred about the {\sl physical} $\fQ$ (as opposed to any other nontrivially canonically-related {\sl mathematical} $\fQ$).  
\noindent As a general point, one should not necessarily postulate that physical entities that are operationally distinct {\sl have} to be 
mixable just because transformations that mix them can be mathematically defined.  
One should not confuse `is often useful to transform' with giving unwarrantedly unconditional physical significance to the transformations in 
question.   

\mbox{ }

\noindent I argue the practical implementation of RigPhase to be as follows. 
I hold that the physicality of $\fQ$ is {\sl not} the centre of this argument; that is, rather,  operational distinguishability between the 
$\ttP$'s on the one hand and the $\ttQ$'s on the other.   
If one includes fermions, one can operationally tell apart fermionic quantities that manage to be positions and momenta at once from quantities 
that solely manifest themselves as configurations and quantities that solely manifest themselves as momenta.  
In this way, the scheme extends to cases where phase spaces are not just simple tangent bundles over configuration spaces, via operational distinguishibility now being tripartite \cite{Arelsusy}.
See this reference also for how, were supersymmetry to exist in nature, it could impose limitations on the applicability of Relationalism 3) 
and its development using less-than-phase-space constructs such as RigPhase.

As regards primality, firstly I only wish to consider {\sl direct} measurements rather than inferences (which could e.g. involve inferring distances from momentum-type measurements connected to photon momenta as in sight or spectral lines). 
One can dissociate from this problem by thinking in terms of simple `blind feeling out' measurements on a scale small enough that one can reach out to feel (for all that there will often be strong practical limitations on this).   
The virtue of this approach is that one's configuration measurements for particle mechanics are all {\sl geometrical}: local angles, adjoining rulers to the separations between objects; the blind Geometer's work is free from the impact connotations that enter his sighted colleague's work. 
[The latter's methodology will however often have greater {\sl practicality}, e.g. if both worked as surveyors...]

Next, in GR in geometrodynamical form, Riemannian 3-geometry measurements of space are again cleanly separated out from gravitational momentum/extrinsic curvature.  
Note how in the mechanics--geometrodynamics sequence of physical theories, one can pass from a simple notion of relational geometry measurements being operationally primary to a more complicated one (argued for e.g. in \cite{Christodoulou1, Christodoulou3}. 
A theme that may then support part of the argument for the primality of configurational measurements is that geometrical measurements 
are primary.\footnote{Moreover, this is {\sl not} to be because of geometry constituting a layer of structure prior to the account of the 
%%%%%%%%%%%%%%%%%%%%%%%%%%%%%%%%%%%%%%%%%%%%%%%%%%%%%%%%%%%%%%%%%%%%%%%%%%%%%%%%%%%%%%%%%%%%%%%%%%%%%%%%%%%%%%%%%%%%%%%%%%%%%%%%%%%%%%%%%%%%%%%%%%%%%
physical objects themselves, since the context in which configurations are being considered as primary is a relational one and one which is 
specifically to encompass GR, for which there is indeed no prior fixed notion of Riemannian geometry.}
%%%%%%%%%%%%%%%%%%%%%%%%%%%%%%%%%%%%%%%%%%%%%%%%%%%%%%%%%%%%%%%%%%%%%%%%%%%%%%%%%%%%%%%%%%%%%%%%%%%%%%%%%%%%%%%%%%%%%%%%%%%%%%%%%%%%%%%%%%%%%%%%%%%%%

From this perspective, it is somewhat incongruous for triad information to be considered as momentum-like within the Ashtekar variables type approaches given that this is another form of space-geometric information.    
It would likewise be somewhat incongruous if the Ashtekar $\underline{A}$ were to be only an indirectly inferrable quantity.\footnote{Here one 
%%%%%%%%%%%%%%%%%%%%%%%%%%%%%%%%%%%%%%%%%%%%%%%%%%%%%%%%%%%%%%%%%%%%%%%%%%%%%%%%%%%%%%%%%%%%%%%%%%%%%%%%%%%%%%%%%%%%%%%%%%%%%%%%%%%%%%%%%%%%%%%%%
usually argues (see e.g. \cite{PullinGambini} instead for the progression from Electromagnetism to Yang--Mills Theory to GR in Ashtekar variables 
form as theories based on notions of connection.  
As regards the current Appendix's theme of operational distinction and most primary types of measurements, firstly I note that in classical 
electromagnetism it is $\underline{B}$ that is measured (rather than $\underline{A}$), and that this is operationally distinct from  measuring 
$\underline{E}$.  
Secondly, mentioning Yang--Mills Theory brings one face to face with problems with centring one's thinking in terms of classical measurements; 
of course, QM offers much wider challenges than this the idea of operational primality of measurements of configurations.   
[This idea may only reflect a simple top-down thinking, c.f. Sec 19.]
As regards measuring Ashtekar's $\underline{A}$, one usually argues that one can make do with measuring curvature quantities and eventually inferring $\underline{A}$ from them.

Moreover, the solidity of the above sequence as a first-principles scheme partly rests on the notion of `connection' receiving a conceptually 
clear treatment (to be more precise, the corresponding notion of holonomy is particularly important for LQG).  
For Ashtekar variables, this is itself a geometrical perspective, for a yet more complicated notion of geometry than GR's usual (semi)Riemannian geometry. 
[Both Ashtekar's $\underline{E}$ and $\underline{A}$ can be construed of as geometrical objects.  
However, it is the $\underline{E}$ whose sense of geometry (triad variable) ties more directly to the more straightforwardly realized spatial Riemannian geometry.  
The $\underline{B}$-field does have more purely geometrical connotations by its being the curvature associated with the connection, 
but the physical realization of this geometry is more subtle than that of mechanics, 3-metric geometries or their triad counterpart.]

Moreover, the Ashtekar variables case does run into some contentions (e.g. \cite{Samuel}) as regards which features are appropriate for a connection serving this purpose, though Thiemann \cite{Thiemann} terms this `aesthetic').  
On the other hand, the solidity of this scheme as obtained from rearrangement of the usual geometrodynamical conceptualization of canonical GR rests on whether it is appropriate to extend the classical phase space to include degenerate configurations, as well as on canonical transformations.}  
%%%%%%%%%%%%%%%%%%%%%%%%%%%%%%%%%%%%%%%%%%%%%%%%%%%%%%%%%%%%%%%%%%%%%%%%%%%%%%%%%%%%%%%%%%%%%%%%%%%%%%%%%%%%%%%%%%%%%%%%%%%%%%%%%%%%%%%%%%%%%%%%%
%
On the other hand, the foundations of geometrodynamics are {\sl not} incongruous in this way, though the above-linked footnote does place limitations on, and give alternatives to, this particular sense of incongruence fostered by this Sec's tentative expansion of the point of view that $\fQ$ is primary.  
\noindent The debate about whether lengths, impacts, times, frequencies... are the most primary things to measure is indeed an old and 
unsettled one.  
Rods are arguably a less deep notion/apparatus than clocks e.g. via Bondi's argument \cite{Bondi59} that they are made out of quantum 
matter which is ultimately underlaid by frequencies.  
Also, by their nature and function, rods are necessarily macroscopic \cite{SW} and so interact with their 
environment in uncontrollable ways, whilst microscopic clocks are indeed possible.  
%
%A good question then, however, is: what is the smallest possible clock?  How does accuracy scale with size for a clock?  
%
I wish to make a somewhat different point too: that, regardless of primality, one can tell apart whether 
what one is measuring is instantaneous-configuration information or impact information.  

\mbox{ }

\noindent RigPhase Motivation 2) One might doubt canonical transformations due to their clash with the common notion of `adding in a potential'.
If this is to be considered as a structurally minor change, as is often done, then one has a problem due to the asymmetry between the complication caused by adding a $Q^4$ potential, say, as compared to the major structural upheaval of adding a $P^4$ kinetic term to one's Hamiltonian.  
Of course, this can be dealt with by accepting that canonical transformations imply that `adding a potential' is as great a structural upheaval (e.g. the above two additions to an HO are identical to each other under canonical transformation). 
This translates to changing the habitual simplicity requirements on Hamiltonians and Lagrangians due to their canonical disparity, and that doubtlessly cuts down on widely applicable theorems that depended on such canonically-disparate simplicity requirements. 
One might then hold all such results to have been empty anyway, but one might attempt to keep them by allotting particular significance to the physics in its presentation centred about $\fQ$.  
By the above, it is also clear that canonical transformations are at odds with statements of simplicity 
such as `at most quadratic in the momenta'.  
Yet further motivations for weakening Phase space at the QM level are given in Sec 18, along with some theoretical consequences.

%\mbox{ } 

\noindent As continuation of MRI and MPI formulations, APhase, DAPhase, ARigPhase and DARigPhase options are relevant alternatives at this point .  
The A stands for `almost', referring to it almost being Phase, except that the unphysical auxiliary variables' velocities (rather than their 
momenta) remain present; thus its physical content is equivalent to that of Phase.  
The DA stands for `differential almost' i.e. the parametrization-irrelevant form in which the unphysical auxiliary variables' differentials
d$\ttg$ are present rather than their velocities $\dot{\ttg}$.
These then have defined on them of almost-Hamiltonians and differential almost-Hamiltonians; this amounts to a fullest distinction at the level 
of `total' \cite{Dirac} objects: 
$\fH_{\sT\so\st\sa\sll} = \ttN\fH + \ttg_{\sfA}{\cal L} \si\sn^{\sfA}$, versus 
$\fA_{\sT\so\st\sa\sll} = \dot{\ttI}\fH + \dot{\ttg}_{\sfA}{\cal L} \si\sn^{\sfA}$ and
$\d\fA_{\sT\so\st\sa\sll} = \d\ttI\fH + \d \ttg_{\sfA}{\cal L} \si\sn^{\sfA}$.   
\noindent Upon performing classical reduction -- taking out Lin to pass to a tilded $\widetilde{\fQ}$ with just a $\widetilde{H}$ on it, as per Fig 2 -- there is no longer any DA-Hamiltonian/A-Hamiltonian/Hamiltonian distinction here, though there does remain a distinction in smeared objects: 
$\fH_{\sss\sm\se\sa\sr} = \ttN \fH$ versus $A_{\sss\sm\se\sa\sr} = \dot{\ttI} \fH$ and $\d \fA_{\sss\sm\se\sa\sr} = \d\ttI \fH$.

\mbox{ }  

\noindent I also mention a set-back  as regards the distinctiveness of the LMB relational approach (Hamiltonian Collapse problem). 
In passing to the Hamiltonian formalism prior to quantization, a lot of LMB-relational versus nonrelational formalism differences are ironed out \cite{FileR}.  
Now both have sum-form total ((D)A)Hamiltonians whereas the actions were product-form versus difference-form; moreover the objects of interest 
are Hamiltonians rather than the smeared objects that exhibit a `DA' trichotomy.    
However, as we shall see in Secs 16-21, the LMB-relational {\sl ideas}, if not precisely the same implementations of these that occur at the classical level, can be recycled at the quantum level. 

\mbox{ } 

\noindent{\bf Alternative: Histories Theory} holds that, instead of configurations being primary entities, it is {\bf histories that are the primary entities}.  
This amounts to supplanting Relationalism 3) with the consequences given in Sec 16.4.

%=================================================================================================================================================
%=================================================================================================================================================
\section{Categorizing extension in anticipation of quantizing?}
%=================================================================================================================================================
%=================================================================================================================================================

One possible motivation for this extension is that quantization can be formally understood as a (bad) functor, 
linking the classical to the quantum-mechanical.
I argue against this in a number of ways in Sec 19.  

\mbox{ } 

\noindent{\bf Categories} $(\MFO, \MFM)$ consist of objects $\MFO$ and {\it morphisms} $\MFM$ (the maps between the objects, 
$\MFM: \MFO \longrightarrow \MFO$), obeying a number of axioms (domain and codomain assignment, identity relations, associativity relations and book-keeping relations, see e.g. \cite{Lawvere, LawRos, MacLane} for details.)

\mbox{ } 

\noindent {\bf Functors} are then maps $\MFF: (\MFO_1, \MFM_1) \longrightarrow (\MFO_2, \MFM_2)$ that obey various further axioms concerning domain, codomain, identity and action on composite morphisms.  

\mbox{ }

\noindent By the above, while so far we have only been studying objects [Relationalism 3), Sec 12], but we should have also been studying morphisms.
In particular the morphisms corresponding to $\fQ$ are the so-called {\it point transformations} (p. 15 of \cite{Lanczos}) 
{\bf Point}: $\fQ \longrightarrow \fQ$.  
The morphisms of phase space are  the canonical transformations Can: Phase $\longrightarrow$ Phase.

I agree that momenta and Poisson brackets are part of the conception of this world. 
Thus  I am not contesting the passage to \noindent ($T^*(\fQ)$,\{ , \}). 
However, I do at least {\sl consider} resisting the suggestion that what mathematically preserves the Poisson bracket should be the associated morphisms of this, due to the previous SSec's argument of configuration variables and momenta being operationally distinguishable. 
Thus it is questionable to take bracket preserving morphisms that do not respect this physical insight.  
An alternative would be a rigged version RigPhase of the phase space that preserves this distinction.
Here, the $\fQ$-first interpretation of this is that the $\ttQ$'s are fundamental and each $\ttP$ then follows by conjugation 
and only transforms on $\fQ$ are primarily meaningful 
Here, if the $\ttQ$'s change, then the $\ttP$'s follow suit by being the new conjugates, without any extra freedom in doing so.   
Thus the morphisms of this are just Point again, the conjugate momenta being held to follow whatever the coordinates are rather than having their own morphisms (this approach's intent {\sl is} to put configurations first, so one should not be surprised at these and not the momenta having the primary transformation properties.   
See Fig \ref{RigPhase} for the ensuing options. 

%FFFFFFFFFFFFFFFFFFFFFFFFFFFFFFFFFFFFFFFFFFFFFFFFFFFFFFFFFFFFFFFFFFFFFFFFFFFFFFFFFFFFFFFFFFFFFFFFFFFFFFFFFFFFFFFFFFFFFFFFFFFFFFFFFFFFFFFFFFFFFFFFFF
{            \begin{figure}[ht]
\centering
\includegraphics[width=0.95\textwidth]{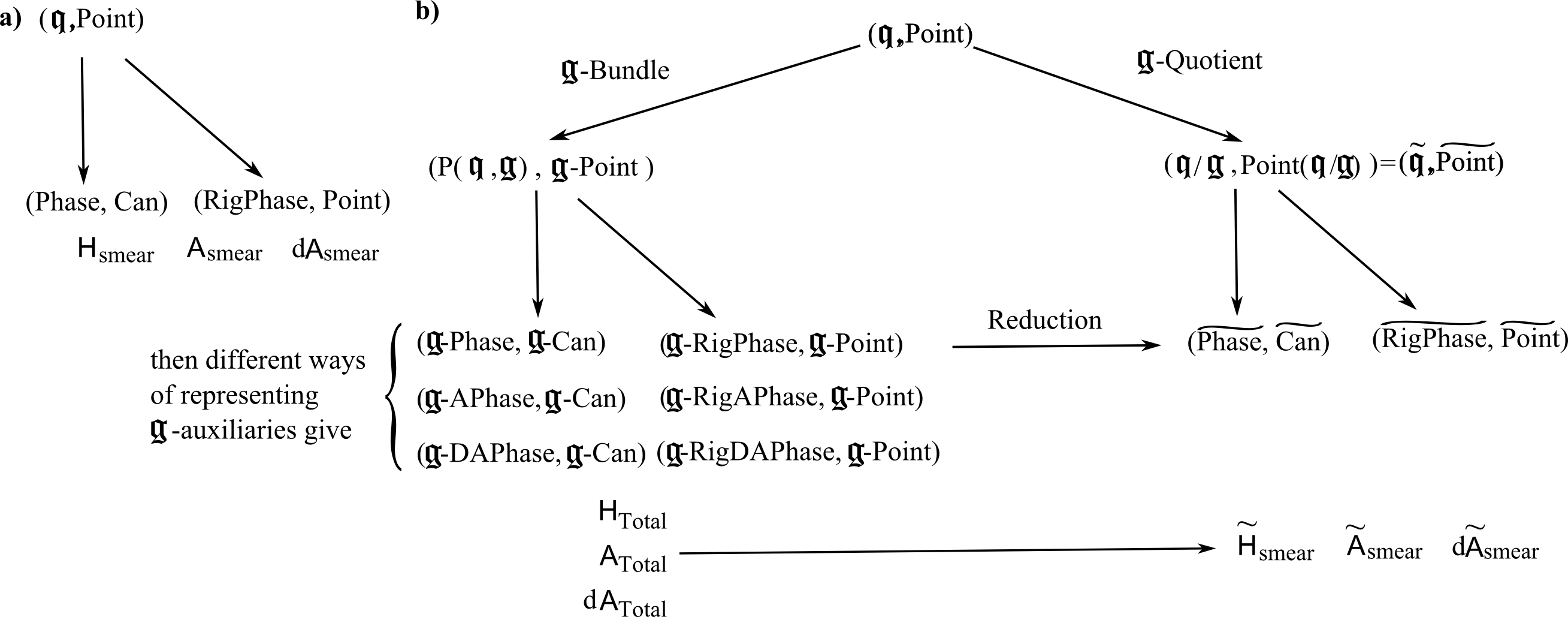}
\caption[Text der im Bilderverzeichnis auftaucht]{        \footnotesize{The options without and with the involvement of a $\fG$. 
[Once a $\fG$ is involved, manifestly temporally relational APhase and DAPhase alternatives arise. } }
\label{RigPhase} \end{figure}          }
%FFFFFFFFFFFFFFFFFFFFFFFFFFFFFFFFFFFFFFFFFFFFFFFFFFFFFFFFFFFFFFFFFFFFFFFFFFFFFFFFFFFFFFFFFFFFFFFFFFFFFFFFFFFFFFFFFFFFFFFFFFFFFFFFFFFFFFFFFFFFFFFFFF

\noindent As regards further involvement of categories, we shall see below that Perspecting and Propositioning, which are themselves at least partly inter-related, lead toward Topos Theory, a Topos being a category armed with some extra structures that make it behave much like the category of sets, $\underline{\mbox{Sets}}$ (see Sec 14 for more details).

%========================================================================================================
%========================================================================================================
\section{Perspectival extension at the classical level}\label{cl-pers}
%========================================================================================================
%========================================================================================================

Rovelli uses relationalism in the sense of objects not being located in spacetime (or space) but being located with respect to each other (though he does not develop this to the extent of the above LMB-A relational postulates).
Rovelli and Crane have, between them, put forward 4 perspectival postulates.  
Not all have substantial classical forms since some involve observers in (something like) Quantum Theory. 
Thus there is a gap in my classical line-up of these.

\mbox{ } 
	
\noindent I use Persp($\fQ$) to mean \{all Sub$\fQ \mbox{ } \underline{\subset} \mbox{ } \fQ$\}.
`Parts'  \cite{LawRos} is a similar categorical concept; I do not use this name since it sounds too much like partial (as in partial observable).
For $\underline{\mbox{Sets}}$, Parts is embodied by the notion of power set.  

\mbox{ }  

\noindent Moreover, in my own presentation, `all' has the connotation `all physical'.  
E.g. in 2-$d$ the subset comprising of the x-component of one particle and the y-component of another lacks in physical significance as a combination; there are some ways in which $\mathbb{R}^{nd}$ can be taken to remember its blockwise origin as $\{\mathbb{R}^d\}^n$  .

\mbox{ } 

\noindent One role for subconfigurations is in modelling records as localized subconfigurations of a single instant. 
Additionally, Histories Theory has an obvious localized subhistories counterpart of this.    

\mbox{ } 

\noindent Perspectivalism 3) {\bf Rovelli's partial observables} \cite{Rov90sConcat, Rovellibook} implement the conceptually-desirable idea of  physics being about correlations as well as being regarded as freeing oneself from complete/Dirac observables (which are much harder to find).    
The partial observables approach is contextual through requiring pairs of partial observables in order to extract entirely physical information.
This is a possible alternative to the Dirac or \K treatments of observables that involves considerably less structure-finding due to 
aiming to bypass rather than resolve the Problem of Observables.  
It is not however clear exactly what objects to compute in this approach in order to investigate whether subsystems exhibit correlation, 
or how a number of other aspects of the Problem of Time can be addressed via these \cite{Kuchar92, I93, Kuchar93}.

\mbox{ }

\noindent Perspectivalism 4) {\bf Use anything as a time for anything else's motion}.  
This can be seen as a `Relationalism 7W)' alternative to Relationalism 7) as regards how to abstract time from change; the W is for `weakened'.  
I point out its similarity to `{\it tot tempora quot motus}' (so many times as there are motions, which is the Scholastics' view of Aristotle, 
see e.g. p 54 of \cite{Jammerteneity}.  
Thus I also term this view of time {\sl AMR time} after Aristotle, Mach and Rovelli.
\mbox{ } 

\noindent Note 1) Perspectivalism 3-4) both have underlying senses of `democracy' and of `not needing to detailedly construct', 
though I argue against these in Secs 21 and 22 and in future works \cite{ARel2}.    

\mbox{ } 

\noindent One should also take stock that a lot of uses of the word `relational' in Ashtekar variables/LQG are what this article terms `perspectival', which additionally explains a number of other senses in which the word `relational' has been used over the years and to some extent fits various of the uses together.  
Thus e.g. GR in general and the Ashtekar variables formulation do have manifest many other  senses of `relational' outside of the more 
usually-ascribed perspectival ones.  
I note that the present Ashtekar variables/LQG literature does not systematically apply LMB(-A) relationalism (though there is some evidence for individual pieces of such thinking in various formalisms and techniques used).  
Doing so could give the subject further conceptual awareness.  
More generally, the extent and manner of the combineability of the Rovelli, Crane and LMB(-A) types of relationalism is an interesting one and 
LQG is a reasonable arena in which to study this question. 
The nature of the Problem of Time facets and strategies covered in this article are generally relevant here too.  
I continue with this in \cite{ARel2}.

%========================================================================================================
%========================================================================================================
\section{Propositions in the classical context}
%========================================================================================================
%========================================================================================================

Physics involves asking and answering questions, i.e. propositions and therefore some kind of {\sl propositional logic}.
As far as I am aware, the idea that physics {\sl is} questions which are equivalent to logic was first put 
forward by Mackey \cite{Mackey} (see also \cite{IL2} for suggestions of this for specific Problem of Time settings).  
\noindent [The relationalist will be interested here in those propositions concerning tangible physics.]  
I view this as an interesting observer-centred idea that is worth considering whether to elevate to a general physical principle. 
I term this `{\bf Mackey's Principle}'
Isham and Linden \cite{IL2} previously used this line of thought as a physical principle for Histories Theory; my new consideration is to speculate whether this is a {\sl general} principle.)

While this plays a bigger, more ambiguous, more subtle role at the quantum level (Sec 21), one can already consider 
it at the classical level.
\noindent From the $\fQ$-centric perspective of Relationalism 3) or some $\fS$-centric extension, this is an extension to 
Prop(Persp($\fS$)), where Prop denotes the set of all possible physical propositions about $\fS$.  
I am not claiming my treatment of this is exhaustive.   
(E.g. I will take `property' to be an unproblematic word, for all that is not, see e.g. p. 67 of \cite{Ibook} for more.)
My point is that the general theory of questions involves, a priori, both nontrivial temporal content and nontrivial atemporal content.
Then in strategies in which one can in fact purge questions of temporal content, one's remaining theory should parallel the nontrivial atemporal 
content of questions, so in that sense the structure of the atemporal content of questions is also important for this article.  

\mbox{ } 

\noindent Implementing Propositioning in at least quite general terms involves {\bf Topos Theory}.
(A {\it topos} is a category with three extra structures: 1) finite limits and colimits 2) power objects and 3) a subobject classifier \cite{LawRos, ToposRev}.  
This renders Topos Theory a  more advanced topic than Category Theory itself, possessing also has ties to geometry and to logic \cite{ToposRev, LawRos, MM68}.)
These allow a topos to bear further similarities to $\underline{\mbox{Sets}}$ (the familiar category of sets) and Boolean logic pinned on that.  
Then {\it ``constructing a theory of physics is equivalent to finding a representation in a topos of a certain formal language that is attached to the system"} \cite{ToposI} (see also \cite{JLBell}).  
This approach also renders regions of configuration space [or more generally of $\fS$, Sub$(\fS)$] primary, rather than the points out of which one usually considers the regions to be constituted of. 

\noindent At the classical level, Propositioning is implemented simply as $\underline{\mbox{Sets}}$, with Boolean logic itself entering as subobject classifier.  
One can consider this to be in terms of regions of configuration space or some extension $\fS$ or Sub($\fS$), with union, intersection 
and complement playing the roles of AND, OR and NOT.  

\noindent Classically, the situation is straightforward \cite{ToposI, ToposRev}: one is to use $\underline{\mbox{Sets}}$ as one's topos, which 
comes with the obvious Boolean logic entering as subobject classifier.   
Propositions about configuration space regions fit well within this classical scheme.  
This is based on quantities having values in (Borel) sets within some state space such as $\fQ$ or Phase.   
Open (or clopen: simultaneously closed and open) subsets are used. 
One is considering maps from state objects in $\Sigma$ to quantity-value objects in $R$.  
Here, $\Sigma$ is the linguistic precursor of $\fS$.  
There is a language which is independent of theory type (classical/quantum) though $R$ clearly does depend on this: the propositional language 
PL($\fS$).

%========================================================================================================
%========================================================================================================
\section{More general configurational relationalism}
%========================================================================================================
%========================================================================================================

Finally, for further use at both the classical and quantum levels, I make this extension of the 
arbitrary $\fG$-frame method/best matching procedure.

\mbox{ } 

\noindent One {\bf builds up} compound objects as bona fide combinations of one's primary objects ({\bf tensor/bundle} structure).
This applies to the relational action (`Relationalism 4'), but also more generally as a `building up' move, producing all of actions, notions of distance, notions of information and correlation, and quantum operators.  
The actions in question are Jacobi or Jacobi--Synge type actions.  

\mbox{ } 

\noindent {\bf $\fG$-act $\fG$-all move}.  
Whilst $\fG$'s physical irrelevance  can occasionally be directly implemented by $\fG$-independent objects, far more usually one has to implement it indirectly via incorporating $\fG$-{\bf act}, $\fG$-{\bf all} pair of moves into the sequence of maps in one's build-up.  
This builds $\fG$-bundle objects/bibundle objects with one bundling involving $\fG$.
Schematically, for continuous or discrete isometries (or a mixture), this is a metric background invariantizing (MBI) map
\beq
\mbox{MBI} = \mbox{\Large S}_{\sFG} \circ \mbox{Maps } \circ \stackrel{\rightarrow}{\fG} O
\eeq
for whichever object $O$ upon which $\fG$ is able to act and whichever Maps render this into a physically-interesting combination 
such as an action, a QM operator or one of the notions of distance/information/correlation of Sec 14 of \cite{FileR}.  
See \cite{FileR} for a topological-level counterpart.
\noindent This is a generalization of how one might go about implementing configurational relationalism.  
That was done in Sec 6 and 7's best matching via the `all' being a variation and the act an infinitesimal continuous isometry. 
However, the generalization also covers group-averaging (sum and divide by order is `all', with `act' being a suitable group action), 
or the similar group-summing/group-integrating, or inf-taking (e.g. in Gromov--Hausdorff type constructions \cite{Gromov, FileR}).   
And it is useable not only on actions, but also in the study of both classical and quantum notions of information, notions of distance, in 
Halliwell-type class functions \cite{AHall} (alongside combinations of these that form decoherence functionals \cite{H03, H09}), and for the quantum operators in Sec 14.6.   

\mbox{ }  

\noindent $\fG$-act, $\fG$-all also plays a role as regards classical notions of distance and of information, for quantum operators \cite{FileR}
and for class functions and decoherence functionals in histories/Halliwell-type approaches \cite{AHall} (Halliwell-type approaches are combined 
Histories-Semiclassical-Records approaches to the Problem of Time).  

\mbox{ } 

\noindent The above is all at the level of the configuration space metric; for brief discussion of topological relationalism, see \cite{FileR}.

%========================================================================================================
%========================================================================================================
\section{QM-Level relationalism}\label{QRel}
%========================================================================================================
%========================================================================================================

%========================================================================================================
\subsection{Outline of a quantization scheme for $\fG$-trivial theories}
%========================================================================================================

I consider the quantization scheme along the lines of Isham's \cite{I84}, which, whilst quite general, is by no means all-embracing.  
I take this to consist of the following steps as composed in Fig 3.   

\mbox{ }  

\noindent {\bf Select} is the selection of a set of classical objects that are to be promoted to quantum operators, alongside passing from the classical Poisson brackets to some commutation relations, algebra ComAl.  
It is very obvious in the simplest example: $\mq^{\sfA} \longrightarrow \hat{q}^{\sfA}$, $p_{\sfA} \longrightarrow \hat{p}_{\sfA}$ alongside the 
usual (`correspondence principle')
\beq
\mbox{from } \{\mq^{\sfA}, \mp_{\sfB}\} = \delta^{\sfA}_{\sfB} \mbox{ to } \mbox{ } 
[\hat{\mq}^{\sfA} \mbox{ } , \mbox{ } \hat{\mp}_{\sfB}] = i\hbar{\delta_{\sfA}}^{\sfB} \mbox{ } .  
\label{common}
\eeq
However, in general it exhibits a number of subtleties \cite{I84}.   
\noindent More generally, ComAl := (a chosen set of f($\hat{q}, \hat{p})$ alongside a bracket structure).   
Here, one has to in general make a {\it choice} of a preferred subalgebra of Phase objects  \cite{I84} to promote to QM operators.
 
\noindent {\bf Assoc} then associates a pre-Hilbert space PreHilb of pre-wavefunctions for these operators to act on. 

\noindent {\bf KinQuant}:= Select $\circ$ Assoc is {\it kinematical quantization}.   

\mbox{ } 

\noindent{\bf H-rep} is how to promote the Hamiltonian $\fH$ to a function $\hat{\fH}$ of KinQuant's operators, yielding a wave equation.  

\noindent {\bf HSolve} is then solving this: passing from PreHilb to Hilb: the Hilbert space within that is annihilated by $\hat{\fH}$.  
\noindent Note that Hilb's inner product restricts what are valid operators by self-adjointness. 

\noindent {\bf DynQuant} := HSolve $\circ$ HRep, the dynamical quantization.  

\mbox{ } 

\noindent {\bf Quant} := DynQuant $\circ$ KinQuant is, finally, this approach's basic notion of {\it quantization}. 
N.B. as formally defined here, this acts on a {\sl triple} such as (Phase, Can, $\fH$) or (RigPhase, Point, $\fH$).  

%FFFFFFFFFFFFFFFFFFFFFFFFFFFFFFFFFFFFFFFFFFFFFFFFFFFFFFFFFFFFFFFFFFFFFFFFFFFFFFFFFFFFFFFFFFFFFFFFFFFFFFFF
{            \begin{figure}[ht]
\centering
\includegraphics[width=0.62\textwidth]{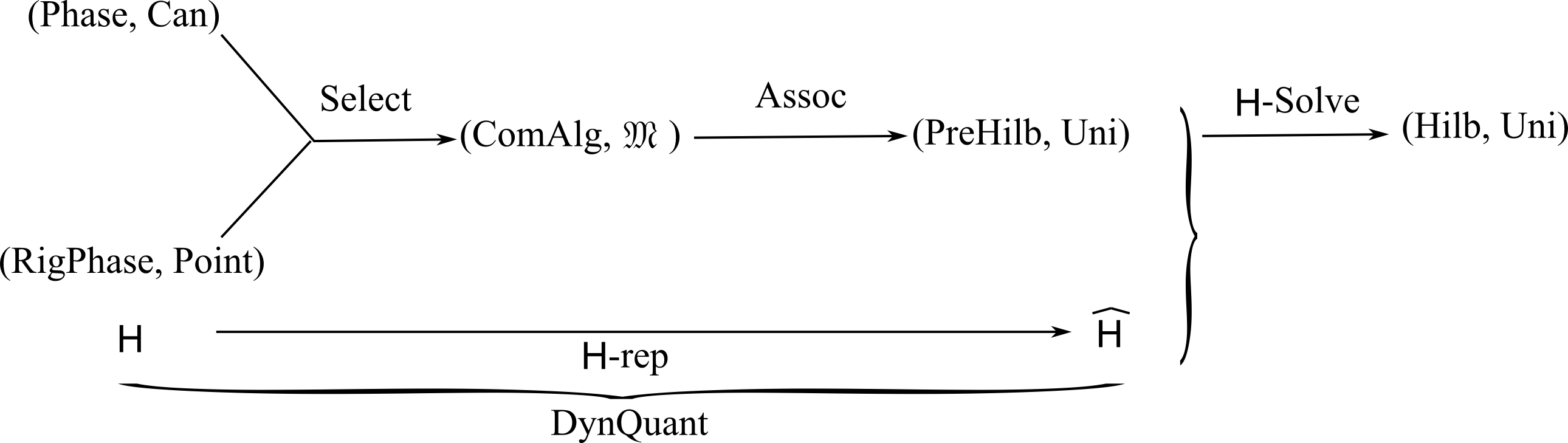}
\caption[Text der im Bilderverzeichnis auftaucht]{        \footnotesize{Breakdown of a simple quantization scheme in the 
absense of redundancies.} }
\label{Mewl}\end{figure}            }
%FFFFFFFFFFFFFFFFFFFFFFFFFFFFFFFFFFFFFFFFFFFFFFFFFFFFFFFFFFFFFFFFFFFFFFFFFFFFFFFFFFFFFFFFFFFFFFFFFFFFFFFF

%========================================================================================================
\subsection{Relational properties at the quantum level}
%========================================================================================================

\noindent Tangibility, temporal relationalism and configurational relationalism ideas remain relevant in the quantum arena. 
However, {\sl implementations} are different now, for at the classical level these were implemented at the level of actions, whereas in Quantum Theory there usually are no actions.\footnote{There are, however, still actions within which one can implement temporal relationalism as MPI
%%%%%%%%%%%%%%%%%%%%%%%%%%%%%%%%%%%%%%%%%%%%%%%%%%%%%%%%%%%%%%%%%%%%%%%%%%%%%%%%%%%%%%%%%%%%%%%%%%%%%%%%%%%%%%%%%%%%%%%%%%%%%%%%%%%%%%%%%%%%%%%%
in path integral approaches to Quantum Theory and, furthermore, in Histories Theory approaches to the Problem of Time.}
%%%%%%%%%%%%%%%%%%%%%%%%%%%%%%%%%%%%%%%%%%%%%%%%%%%%%%%%%%%%%%%%%%%%%%%%%%%%%%%%%%%%%%%%%%%%%%%%%%%%%%%%%%%%%%%%%%%%%%%%%%%%%%%%%%%%%%%%%%%%%%%%

One widespread and simple implementation of tangibility at the quantum level is to ultimately consider not wavefunctions or operators $\widehat{O}$ but rather combinations $\langle \psi_1 | \widehat{O} | \psi_2 \rangle$ (which include as subcases {\it expectations}: $\psi_1 = \psi_2$, 
and {\sl overlap integrals} $\widehat{O}$ = id).

%========================================================================================================
\subsection{Temporal relationalism at the quantum level: from Leibniz to the Frozen Formalism}
%========================================================================================================

\noindent I note that ordinary QM itself is not Leibniz-timeless.  
It is whole-universe QM that is, at least at first glance at conventional forms for its equations.  
In more detail, temporal relationalism $\Rightarrow$ MPI actions $\Rightarrow$ homogeneous quadratic constraint ${\cal H}$  $\Rightarrow$ at the 
quantum level the 

\mbox{ }

\noindent {\bf Frozen Formalism Facet} of the Problem of Time, i.e. that one has a timeless wave equation $\hat{\cal H}\Psi = 0$, 
in contradistinction from such as a 
time-dependent Schr\"{o}dinger equation $i\pa\psi/\pa t = \hat{H}\Psi$, 
Klein--Gordon equation $\Box\psi = \triangle\psi - \pa_{tt}\psi = m^2\psi$ or 
Dirac equation $i\{\gamma^t_{\sfA\sfB}\pa_t\psi^{\sfB} - \gamma^{\mu}_{\sfA\sfB}\}\pa_{\mu}\psi^{\sfB} = m\psi^{\sfA}$.\footnote{Here, I use 
%%%%%%%%%%%%%%%%%%%%%%%%%%%%%%%%%%%%%%%%%%%%%%%%%%%%%%%%%%%%%%%%%%%%%%%%%%%%%%%%%%%%%%%%%%%%%%%%%%%%%%%%%%%%%%%%%%%%%%%%%%%%%%%%%%%%%%%%%%%%%%%%%%
$\psi$ as wavefunction versus $\Psi$ as wavefunction {\sl of the universe}.  
$\psi^{\tfB}$ is a Dirac spinor and $\gamma^{\Gamma}_{\tfA\tfB}$ are Dirac matrices.}      
%%%%%%%%%%%%%%%%%%%%%%%%%%%%%%%%%%%%%%%%%%%%%%%%%%%%%%%%%%%%%%%%%%%%%%%%%%%%%%%%%%%%%%%%%%%%%%%%%%%%%%%%%%%%%%%%%%%%%%%%%%%%%%%%%%%%%%%%%%%%%%%%%%

\mbox{ } 

\noindent Each of the latter set of equations are common in ordinary QM but entail absolute time notions 
(whether Newtonian in the first example or Minkowskian in the other two).

\noindent Note also how in the relational approach the Frozen Formalism facet readily follows from the Leibniz temporal relationalism principle, 
rather than coming as a surprise or being considered to arise from MRI, which is, in the LMB-relational approach, not an explanatory principle 
but, rather a mathematical implementation of the actual Leibniz time principle.  

\mbox{ }  

\noindent The Problem of Time \cite{Kuchar92, I93, APOT, FileR, APOT2} occurs because `time' takes a different meaning in each of GR and 
ordinary Quantum Theory.  
This incompatibility underscores a number of problems with trying to replace these two branches with a single framework in situations in which 
the premises of both apply, such as in black holes or the very early universe.  
Other Problem of Time facets are introduced below.

%========================================================================================================
\subsection{Frozen Formalism Facet resolutions and Mach's time principle}
%========================================================================================================

There are approaches to QG in which time is not primary [thus implementing Leibniz's Relationalism 5)] but which then have an emergent time 
or a semblance of time  arising at the quantum level.  
Note how Mach provides a suggestion for how to resolve timelessness.  
Variants of the Mach time principle are discussed in Secs \ref{Mach-1}, \ref{SSSec: RWR}, the Conclusion, and \cite{ARel2, APOT2}.  
The last three of these references permit finer Machian assessment of Problem of Time strategies than are given below.  
I next analyze the usual set of Problem of Time strategies from a simple Machian time perspective.  

\mbox{ }

\noindent A) Perhaps one is to find a hidden time at the classical level \cite{Kuchar92} by performing a canonical transformation under which the quadratic constraint is sent to 
\beq
p_{t^{\th\ti\td\td\te\tn}} + \fH_{\st\sr\su\se} = 0
\label{Casta}
\eeq 
(for $p_{t^{\th\ti\td\td\te\tn}}$ the momentum conjugate to some new coordinate $t^{\sh\si\sd\sd\se\sn}$ that is a candidate timefunction).  
This is then promoted to a $t^{\sh\si\sd\sd\se\sn}$-dependent Schr\"{o}dinger equation,  
\beq
i\hbar\pa\Psi/\pa{t_{\sh\si\sd\sd\se\sn}} = \hat{\fH}_{\st\sr\su\se}\Psi \mbox{ } .
\eeq
Particular cases of this are as follows. 

\noindent A.1) Scale as internal time.

\noindent A.2) Superspace time \cite{Kuchar92, I93} is a further variant on this, following from the scale contribution giving the indefiniteness in GR kinetic term, a feature not found elsewhere in physics and which causes a number of difficulties. 
(E.g. invalidation of the usual Schr\"{o}dinger interpretation of the inner product, and a whole new theory for the dynamical meaning of zeros 
in the potential factor.)  

\noindent However, scale fails to be globally monotonic in general and the superspace time approach fails due to the GR potential term being incompatible with the GR kinetic term's conformal Killing vector.

\noindent A.3) York time (the conjugate of a particular scale variable). 
However, for this the practical scheme breaks down, and the split of $h_{\mu\nu}$ into embedding and true degrees of freedom has a global impasse.

\noindent A.4) Matter appended to the geometrodynamics might provide a timefunction (matter time).

\mbox{ } 

\noindent As criticisms of approach A) in general, this is not in line with Leibniz timelessness (since the time in use is taken to a priori exist in general at the classical level: Tempus Ante Quantum \cite{I93}), nor with the LMB-type take on Mach's time principle, 
since it involves using one particular change as the time for all the other changes.  
Moreover, matter time schemes themselves are typically additionally suspect along the lines of intangibility, via often involving such as  `reference fluids', and often endowed with unphysical properties to boot. 
[The intangibility might be taken as cover for the unphysicality, but is itself a conceptually suspect way of handling the Problem of Time 
along the lines exposited in the present article.]

JBB time itself is Tempus ante Quantum, but is now classically emergent and thus in line with Leibniz-timelessness and the LMB stance 
on Mach's time principle.  
One particular scheme for this has it arise to zeroth approximation from slow heavy h degrees of freedom, to which light fast l degrees 
of freedom then contribute to first order; see below for physical discussion of the nature of h and l degrees of freedom.  
On the other hand, this approach does {\sl not} gives a Frozen Formalism resolution.  
However, we shall see it is approximately recovered by the next entry, which does provide a Frozen Formalism resolution.  

\mbox{ }

\noindent B) Perhaps instead there is no time in general but a notion of time does emerge under certain circumstances. 
The most notable example of this is the Semiclassical Approach. 
Here, one has slow heavy `$h$'  variables that provide an approximate timestandard with 
respect to which the other fast light `$l$' degrees of freedom evolve \cite{Banks, HallHaw, Kuchar92, Kieferbook}.  
In the Halliwell--Hawking \cite{HallHaw} scheme for GR Quantum Cosmology, $h$ is scale (and homogeneous 
matter modes) and $l$ are small inhomogeneities.
\noindent h--l alignment with scale--shape removes ambiguity of how to allot h and l roles.

The semiclassical approach involves making the Born--Oppenheimer ansatz $\Psi(h, l) = \psi(\mh)
|\chi(h, l) \rangle$ and the WKB ansatz $\psi(h) = \mbox{exp}(iW(h)/\hbar)$.  
Next, one forms the $h$-equation ($\langle\chi| \hat{H} \Psi = 0$ for RPM's), which, under a number 
of simplifications, yields a Hamilton--Jacobi\footnote{For simplicity, I 
%%%%%%%%%%%%%%%%%%%%%%%%%%%%%%%%%%%%%%%%%%%%%%%%%%%%%%%%%%%%%%%%%%%%%%%%%%%%%%%%%%%%%%%%%%%%%%%%%%%%%%%%% 
present this in the case of 1 $h$ degree of freedom and with no linear constraints.} 
%%%%%%%%%%%%%%%%%%%%%%%%%%%%%%%%%%%%%%%%%%%%%%%%%%%%%%%%%%%%%%%%%%%%%%%%%%%%%%%%%%%%%%%%%%%%%%%%%%%%%%%%%
equation
\beq
\{\pa W/\pa h\}^2 = 2\{\fE - \fV(h)\} \mbox{ } 
\label{HamJac} 
\eeq
for $\fV(h)$ the $h$-part of the potential. 
Next, one way of solving this is for an approximate emergent semiclassical time $t^{\se\sm} = t^{\se\sm}(h)$. 
Finally, the $l$-equation $\{1 - |\chi\rangle\langle\chi|\}\hat{\H}\Psi = 0$ can be recast (modulo further 
approximations) into an emergent-

\noindent time-dependent Schr\"{o}dinger equation for the $l$ degrees of freedom,  
\beq
i\hbar\pa|\chi\rangle/\pa t^{\te\tm}  = \widehat{H}_{l}|\chi\rangle \mbox{ } .  
\label{TDSE2}
\eeq
(Here the left-hand side arises from the cross-term $\pa_{h}|\chi\rangle\pa_{h}\psi$ and 
$\widehat{H}_{l}$ is the piece of $\widehat{H}$ that survives).  
Note that the working leading to such a time-dependent wave equation ceases to work in the absense of making the WKB ansatz and approximation, which, additionally, in the quantum-cosmological context, is not known to be a particularly strongly supported ansatz and approximation to make.    

Then the scale part contributes an approximate timestandard with respect to which the shape part runs according to usual 
positive-definite kinetic term physics (`scale-indefiniteness alignment' exploited by aligning h with both).  
This can then be fed into more promising Histories-Records-semiclassical combination strategies.  

\noindent Note that the identification of the $t$ is unambiguous if one uses scale = h and shape = l split.  
This puts that `cursed' heterogeneity to a use -- it is the cure to approximate time selection in cosmological models.
Need to see the crudest estimate of t here is not very Machian, but the corrected version does allow for l-changes to contribute too, 
thus rendering it very Machian.  

\noindent Overall, emergent semiclassical time is Machian once back-reaction of the l-system on the chroniferous h-system is incorporated.  
This is qualitatively similar to the emergent JBB time scheme at the classical level.  

\mbox{ }  

\noindent C) A number of approaches take timelessness at face value. 
The most natural follow-on for Relationalism 3) is the `Tempus nihil est' postulate.
One considers only questions about the universe `being', rather than `becoming', a certain way.  
Considering physical questions about a system amounts to an algebra of propositions; this is implemented by regions or, better, by projectors at QM level.
Particular versions of timeless approaches that enter this article's discussions are as follows.

\noindent C.I) the {\it \NSI}  \cite{HP86,UW89} concerns the `being' probabilities for universe properties, such as `what is the probability 
that the universe is flat?' 
One obtains these via consideration of the probability that the universe belongs to region R of the 
configuration space that corresponds to a quantification of a particular such property, 
\beq 
\mbox{Prob(R)} \propto \int_{\sR}|\Psi|^2\d\Omega \mbox{ } , 
\label{NSI}
\eeq 
for $\d\Omega$ the configuration space volume element.

\noindent C.II) The {\it \CPI} \cite{PW83} goes further by addressing conditioned questions of `being' such as `what is the probability that the universe is flat given that it is isotropic'?  

\noindent C.III) {\it Records Theory} \cite{PW83, GMH, B94II, EOT, H99, Records} involves localized subconfigurations of a single instant.  
More concretely, it concerns whether these contain useable information, are correlated to each other, and a semblance of dynamics or history arises from this.  
[I do not use emergence and semblance as synonyms.]  
This requires notions of localization in space and in configuration space as well as notions of information; in fact, this article adds to existing criteria as regards specifying a mathematically-precise Records Theory.  
There is trickery to reduce temporal logic to atemporal logic \cite{Records}, and ID itself is then a fine study of atemporal logic 
(though this Records-Theoretic application of ID is, as far as I know, new to the present article).  

\noindent Analysis of attitudes to time permits classification of the times that arise from timeless approaches. 
This involves \NSII, \CPII, Page, Barbour, Gell-Mann--Halliwell and Hartle Records, my records, the combined approaches of Halliwell \cite{H03, H09} and of Gambini--Porto--Pullin \cite{GPP01, GPP04a, GPP4b, PGP1}, alongside extensions of these that include notions of observables/beables \cite{AHall, GPPT}, and also the Rovelli approach, as well as suggesting a number of further hybrid programs, and is to be covered in more depth in a subsequent paper \cite{ARel2}.  

\mbox{ }  

\noindent D) Perhaps instead it is the histories that are primary ({\it Histories Theory} \cite{GMH, Hartle}); this is an alternative to `$\fQ$ is primary' or `phase space is primary'.  
More generally, I note that temporal relationalism is implemented via MRI into suitable formulations in terms of path integrals.  
Additionally, one can use emergent JBB/semiclassical time to formulate Histories Theory.

%========================================================================================================
\subsection{Outline of a quantization scheme for $\fG$-nontrivial, classically-unreduced theories}
%========================================================================================================

Barbour-relationalism's indirect implementation at the classical level gives constraints ${\cal Q}$uad and ${\cal L}$in.
The reduced/

\noindent relationalspace r-scheme gives a constraint $\widetilde{{\cal Q}\mbox{uad}}$. 
Both these schemes are relational, thus Dirac and reduced schemes are both relational.  
In the preceding literature, Barbour, Rovelli and Smolin \cite{RS-BB} had largely not considered the LMB brand of relationalism along the above lines beyond the general idea that Dirac Quantization is an indirect scheme that implements configurational relationalism. 
Barbour had on the other hand considered a timeless records approach \cite{B94I, EOT} which has some relational features.

\cite{FileR} used the $\fG$-act, $\fG$-all method for Timeless Records Theory, and \cite{AHall} used it along the lines of Halliwell's combined approach.

Methods such as the group integration approach (essentially a variant/particular implementation of the Dirac quantization approach) 
are fairly widely used in LQG (e.g. in the Master Constraint program \cite{Phoenix, DT04}).  
%
%p 20, p 23 of arXiv version of the former, and all over the place in the latter.  
%
In this sense, LQG is configurationally relational at the quantum level.  

%FFFFFFFFFFFFFFFFFFFFFFFFFFFFFFFFFFFFFFFFFFFFFFFFFFFFFFFFFFFFFFFFFFFFFFFFFFFFFFFFFFFFFFFFFFFFFFFFFFFFFFFFFFFFFFFFFFFFFFFFFFFFFFFFFFFFFFFFFFFFFFFF
{            \begin{figure}[ht]
\centering
\includegraphics[width=0.82\textwidth]{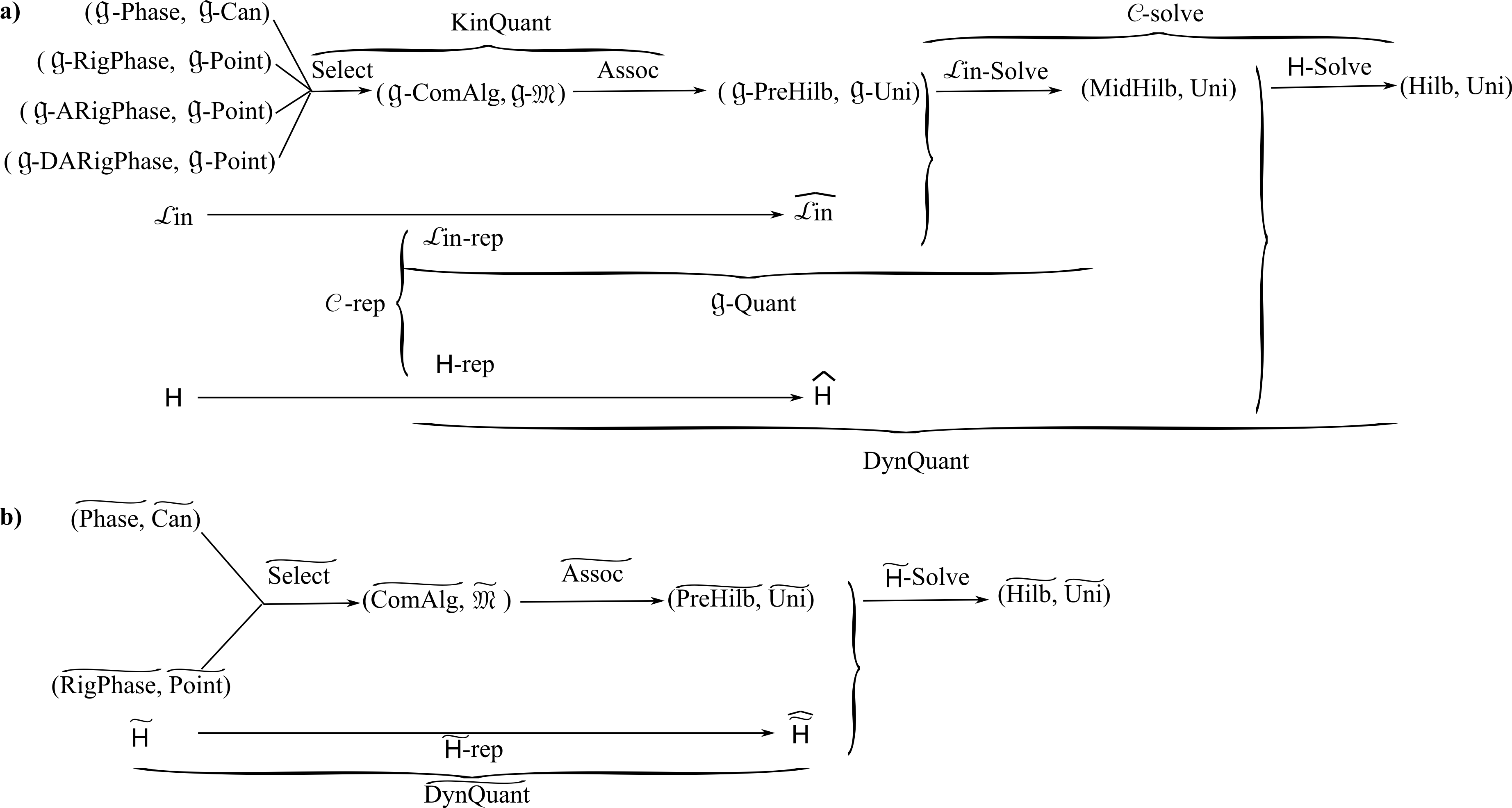}
\caption[Text der im Bilderverzeichnis auftaucht]{        \footnotesize{Breakdown of a simple quantization scheme in the presence 
of a group $\fG$ of physically-redundant transformations. a) The Dirac-type scheme. b) The reduced-type scheme.} }
\label{Rawr}\end{figure}            }
%FFFFFFFFFFFFFFFFFFFFFFFFFFFFFFFFFFFFFFFFFFFFFFFFFFFFFFFFFFFFFFFFFFFFFFFFFFFFFFFFFFFFFFFFFFFFFFFFFFFFFFFFFFFFFFFFFFFFFFFFFFFFFFFFFFFFFFFFFFFFFFFF

\mbox{ }

\noindent In this more general setting [Fig \ref{Rawr}.a)], {\bf Quant} := DynQuant $\circ$ $\fG$-Quant $\circ$ KinQuant.

\noindent $\fG$-{\bf Quant} := $\fG$-Solve $\circ$ $\fG$-Rep, each of which individual steps is defined in obvious parallel to $H$-Rep and $H$-Solve.  
$\fG$-{\bf Solve} sends PreHilb to MidHilb, standing for  `middlingly physical Hilbert space'.  

\noindent I also define ${\cal C}$-{\bf Rep} = H-Rep $\circ$ $\fG$-Rep (these maps commute), and ${\cal C}$-{\bf Solve} = H-Solve $\circ$ $\fG$-Solve.  
These are particularly useful in approaches that do not discriminate between linear and quadratic constraints.

%==================================================================================================================================================
\subsection{Outline of a quantization scheme for reduced theories}
%==================================================================================================================================================

\noindent Reduced-type formulations also embody configurational relationalism.  
These are essentially tilded versions of Fig 3, available whenever one has the good fortune to, at the classical level, 
beable to directly formulate or reduce away all the redundancies from the indirect formulation.
This is covered in Fig \ref{Rawr}b) and amounts to a tilde-ing of Fig 3 (itself following on from the classical reduction diagram of Fig 1).

%=====================================================================================================================================
\subsection{A further element of configurational relationalism at the quantum level}
%=====================================================================================================================================

\noindent QM makes use of further operators.  
Group averaging/refined algebraic quantization looks to be a good {\sl quantum} implementation of configurational relationalism. 
Given a non-configurationally-relational operator $\widehat{O}$, 
\beq
\int_{\sFG} \mbox{exp}
\left(
i\sum\mbox{}_{\mbox{}_{\sfZ}} \stackrel{\rightarrow}{\fG_{\sttg^{\tfZ}}}
\right) 
\, \widehat{O} \, 
\mbox{exp}
\left(
- i\sum\mbox{}_{\mbox{}_{\sfZ}} \stackrel{\rightarrow}{\fG_{\sttg^{\tfZ}}}
\right) 
\mathbb{D}\ttg
\eeq
(for $\fG_{\sttg^{\tfZ}}$ the action of infinitesimal generators) is a relational counterpart, which is another example of $\fG$-act, $\fG$-all 
procedure based on the exponentiated adjoint group action followed by integration over the group.  
One can do similarly to pass from non-configurationally relational states to configurationally-relational ones.
Admittedly, there are well-definedness issues in general as regards the measure $\mathbb{D}\ttg$ over the group $\fG$. 
In particular, treating the diffeomorphisms in this way is problematic.
See Sec 14 of \cite{FileR} for further discussion.

%================================================================================================================================================
\subsection{Composing configurational and temporal relationalism at the quantum level}
%================================================================================================================================================

Configurational relationalism is linked to a number of further Problem of Time facets.
Although specific properties of the diffeomorphisms are very important, quite a few features of the Problem of Time follow from relationalism alone 
rather than additional insistence of $\fG$ containing (more or less) Diff(3).

\mbox{ } 

\noindent {\bf Best Matching Problem}.  I discussed in Sec 9.5 how this is the generalization and configurational-relational conceptualization of 
the usually so-called thin sandwich problem.  
It is usually taken to be a classical facet of the Problem of Time, though additionally Sec 9.5's set-up of the thin sandwich set-up is in analogy with the QM set-up of transition amplitudes between states at two different times; this was indeed Wheeler's original motivation for it \cite{WheelerGRT}.    

\mbox{ }

\noindent{\bf Functional Evolution Problem}.  This concerns anomalies in the quantum commutator algebra, i.e. the further quantum 

\noindent counterpart of Relationalism 10).  
More specifically, a subset of anomalies are time-related e.g. through the anomalous terms being foliation-dependent.  

\mbox{ }

\noindent {\bf Problem of Observables}. This, and its resolutions, are as stated at the classical level in Sec 9.4 but now involving quantum objects 
and commutator algebra in place of Poisson brackets algebra.  

\mbox{ } 

\noindent{\bf Global Problem of Time}.  
Quantum mechanically, this is rather more involved than the classical meshing condition, now being some kind of largely unexplored statement about meshing together not of coordinate systems but of {\sl unitary quantum evolutions}.

\mbox{ }

\noindent {\bf Multiple Choice Problem} This is partly due to coordinate choices (which the Global Problem of Time's need 
for multiple coordinate patches compounds), or due to choices of functions to promote to an operator algebra (see Sec \ref{QMQPrim}).  
It then manifests itself at the quantum level in the form of different choices being capable of leading to unitarily-inequivalent 
quantum theories.  

\mbox{ }

\noindent These last two facets are not particularly LMB relational; we shall see, however that they are at least partially RC relational.

\mbox{ }

\noindent Once one's $\fG$ includes diffeomorphisms, the {\bf Foliation Dependence Problem} arises as a further possible source of background dependence (which would be in discord with LMB relationalism) and there is also rather more of a 
{\bf Spacetime Reconstruction/Replacement Problem} than in Sec 9.7's classical treatment.

%================================================================================================================================================
%================================================================================================================================================
\section{Examples of relational theories at the quantum level}
%================================================================================================================================================
%================================================================================================================================================

%================================================================================================================================================
\subsection{Indirect presentation of RPM triangleland}
%================================================================================================================================================

Here, the frozen wave equation is
\beq
\widehat{\cal E}\Psi = -\frac{\hbar^2}{2}\delta^{\alpha\beta}\delta^{ij}
\frac{\pa}{\pa \rho^{i\alpha}}\frac{\pa}{\pa \rho^{j\delta}}\Psi + V(\rho^{k\gamma})\Psi = E\Psi
\label{zut}
\eeq
summed from 1 to 3 in i,j and 1 to 2 in $\mu$, alongside the quantum zero total angular momentum equation 
\beq
\underline{\widehat{{\cal L}}}\Psi = \frac{\hbar}{i}\sum\mbox{}_{\mbox{}_{\mbox{\scriptsize $i = 1$}}}^{3} 
\underline{\rho}^i \cr \frac{\pa}{\pa \underline{\rho}^i}\Psi = 0   \mbox{ } ,
\label{QZAM}
\eeq
This working, unlike that of the next SSec, readily generalizes to all dimensions and particle numbers.  
Here, $\Psi$ is the wavefunction of the (model) universe.  
RPM Dirac-QM study was developed in \cite{RS-BB, 06I, FileR}.  
  
\mbox{ }  

\noindent Note 1) relational triangleland cannot distinguish between dimensions 2 and 3; in \cite{FileR} I used this to 
argue that dimension is not an a priori given in RPM's.

\noindent Note 2) $\fG$-Quant restricts the AM quantum numbers of the base and apex subsystems to counterbalance each other \cite{06I}, 
thus causing a substantial collapse in the physically admissible spectrum of states for this theory. 

\noindent 
As regards the indirect definition of rotation-averaged operators,  triangleland's $SO(2) = U(1)$ is a particularly simple case, for which  
\beq
\int_{g \in \sFG}\mathbb{D} g = \int_{\omega \in \mathbb{S}^1} \mathbb{D} \mathbb{\omega} = \int_{\omega = 0}^{2\pi}\d\omega
\eeq
and $\stackrel{\rightarrow}{\fG_g}$ is the infinitesimal 2-d rotation action via the matrix $\underline{\underline{R}}\mbox{}_{\omega}$ acting on the vectors of the model.
$\omega$ is here the absolute rotation.  

\mbox{ } 

\noindent Note 1) clearly (\ref{zut}) exhibits the Frozen Formalism Facet of the Problem of Time.  

\noindent Note 2) RPM's exhibit no Functional Evolution Problem nor have any place for discussions of foliations or of GR-like spacetime reconstruction, thus evading two further Problem of Time Facets.  
They clearly do exhibit the other Problem of Time facets.

%================================================================================================================================================
\subsection{Direct r-presentation of scaled RPM triangleland}
%================================================================================================================================================

Direct r-QM study for RPM's was developed in \cite{08II, MGM, ScaleQM, 08III, Banal, SemiclIII, FileR, QuadIII}.
Looking at specific simple solvable examples such as in these last 2 SSecs, I find (contrary to suggestions in e.g. 
\cite{EOT}) a {\sl lack} of non-standard QM.  
By this standardness of the actual mathematics involved (just new relational meanings for well-known equations and special functions), 
plenty of machinery is available for solving, and performing further calculations with, simple examples of quantum RPM's.  
Note however that these models are appropriate as whole-universe models and for detailed analysis of the Problem of Time.  
Thus this coincidence is a triumph as regards finding tractable toy models for the investigation of closed-universe questions 
(Problem of Time, foundations of Quantum Cosmology...)

For scaled triangleland,  the frozen wave equation is

\noindent 
\beq
\widehat{\widetilde{H}} \Psi = 
-\frac{\hbar^2}{2} 
\left\{ 
\frac{1}{I^2}
\left\{
\frac{\pa}{\pa I}
\left\{
\mI^2\frac{\pa\check{\Psi}}{\pa I}
\right\} + 
\frac{1}{\mbox{sin}\Theta}\frac{\pa\check{\Psi}}{\pa\Theta}
\left\{
\mbox{sin}\Theta\frac{\pa\check{\Psi}}{\pa\Theta}
\right\} 
+ \frac{1}{\mbox{sin}^2\Theta}\frac{\pa^2\check{\Psi}}{\pa\Phi^2}  
\right\} 
\right\}\Psi + \breve{\fV} = \breve{\fE}{\Psi} \mbox{ } .  
\label{spheTISE}
\eeq
\noindent Note 1) Some differences between absolutist and relational QM are as follows (these are well understandable from within 
the usual kind of a mathematical framework for Quantum Theory).  
There are differences of the subsystem versus the closed whole-system sort between absolutist QM and relational QM
Subsystem physics in a relational universe would be expected to be linked by global restrictions due to its  constraint equations.
Such differences were already found to be present in the GR case by DeWitt \cite{DeWitt67}.  
There is no a priori time, but many parts of QM do not depend on time, and \cite{FileR} explains how the remaining parts can be accessed using emergent time dependent wave equations.  

\noindent Note 2)  As regards the remaining Problem of Time facets for RPM's, in 1- and 2-$d$, the Best Matching problem is resolved 
(by the reduction process itself!)  
Also, the \K observables are known \cite{QuadII} and geometrically natural, and Dirac observables can then be constructed along the lines of \cite{AHall}, so the Problem of Observables is here well under control.

%===============================================================================================================================================
\subsection{The Dirac scheme for geometrodynamics}
%===============================================================================================================================================

\beq
\hat{\cal H}\Psi := -\hbar^2 \mbox{`}\left\{\frac{1}{\sqrt{{\fM}}}\frac{\delta}{\delta h^{{\mu\nu}}}
\left\{
\sqrt{{\fM}}{\fN}^{\mu\nu\rho\sigma}\frac{\delta\Psi}{\delta h^{{\rho\sigma}}}
\right\} 
- \xi \,\mbox{Ric}(h;{\bfM}]\right\}\mbox{'}\,\Psi -  \sqrt{h}\mbox{Ric}(x;h]\Psi + {\sqrt{h}2\Lambda   }\Psi  = 0  \label{WDE3} \mbox{ }  , \mbox{ and }       
\eeq
\beq
{\cal H}_{\mu} := -2D_{\nu}\pi^{\nu}\mbox{}_{\mu} = 0 \mbox{ } \mbox{ which becomes, at the quantum level, } \mbox{ } 
\widehat{\cal H}_{\mu}\Psi := -2D_{\rho}h^{\rho\nu}\delta\Psi/\delta h_{\mu\nu} = 0 \mbox{ }  .
\label{5}
\eeq 
The latter implements configurational relationalism, signifying that the wavefunction of the universe is to be a function of the 3-geometry alone.

\mbox{ } 

\noindent Whilst classical GR had its constraint algebra close and the refoliation invariance property by Notes 1) and 2) of Sec 9.4, there are no known corresponding results at the quantum level, constituting the Functional Evolution Problem and Foliation Dependence Problem facets of the Problem of Time.

%===============================================================================================================================================
\subsection{Comments on LQG and its Dirac-type quantization scheme}
%===============================================================================================================================================

One now has a new reformulation of the quadratic constraint and of the momentum constraint, alongside a new linear 
$SU(2)$-Yang--Mills constraint that reflects the extra degeneracy of the presentation's $\fG = $ Diff(3) $\times SU(2)$.   
The loop representation takes back this last degeneracy, whilst considering and knots formally uses up the Diff(3) constraint too. 
This amounts to performing the $\fG$-Quant steps of a Dirac-type approach.

I also comment that one source of good names for physical programs that purport to be relational and thus concern tangible entities, is to name the theories after the tangible entities rather than after convenient but meaningless parts of the formalism that happen to be in use in the mathematical study of that theory.  
By this criterion, `geometrodynamics' and `shape theory' \cite{Kendall}/`dynamics of pure shape' \cite{B03, Piombino}/`shape dynamics' \cite{B11} are very good relational names, whereas the name `Loop Quantum Gravity' (LQG) leaves quite a lot to be desired from a relational perspective.  
For, in this program the $SU(2)$ gauge group is put in by hand with the particular choice of spinor/bein variables that happen to simplify GR's canonical formalism.  
It is only the subsequent removal of its significance that brings loops into the program's formalism, but the lion's share of the work in 
{\sl any} canonical quantum GR program resides, rather, in removing Diff($\Sigma$) and in interpreting the Hamiltonian constraint. 
Even without the `put in by hand' issue, the parallel naming for RPM's would be `sphere theory' or `sphere dynamics' from this being the form the preshape space takes before the lion's share of the work is done in removing Rot($d$) (and, in the dynamics case, interpreting the energy-type constraint).  
But this is rather clearly {\sl not} enlightening naming because the preshape space spheres do {\sl not} play a deep conceptual role in shape theory.
As such, I would suggest names along the lines of `Knot Quantum Gestalt' or `quantum nododynamics'.
This is also particularly appropriate via the Gestalt conceptualization of GR being, at least at the level of metric structure, very much emphasized in the `LQG' program itself, just not in the LQG program's {\sl name}.

Thiemann's alternative name `Quantum Spin Dynamics (QSD)' is somewhat confuseable with such as the Ising model and its ilk, and also the spin connotations come from the $SU(2)$ gauge groups.  
On the other hand, letting the S stand for `spin-net' disambiguates the nature of the theory, though a spin-net still does have but a similar ontological status to a loop.  
Taking relational primality further, then, the final suggestion is that the S should stand for S-knot (spin-knot).

A useful further project would be to provide a LMB(-A) relational account of the quantum part of Thiemann's book \cite{Thiemann} (and of other schemes for modern LQG, there being a variety of such schemes).

%===============================================================================================================================================
%===============================================================================================================================================
\section{Extending primality of $\BigfQ$ to QM?} \label{QMQPrim}
%===============================================================================================================================================
%===============================================================================================================================================

\noindent QM $\fQ$-Primality Motivation 1) On the one hand, one often hears that QM unfolds on configuration space, on the other hand most hold that it unfolds equally well on whichever polarization within phase space, of which configuration space is but one.   

\noindent QM $\fQ$-Primality Motivation 2) I note that Isham's procedure \cite{I84} for kinematic quantization itself favours $\fQ$: the canonical group comes from just this and then the $\mathfrakV$ is controlled by $\fQ$ and its associated $\fG$ and $\fG_{\sc\sa\sn}$ so everything is controlled by configuration space too.    

\noindent QM $\fQ$-Primality Motivation 3) I expect then also for Point to be relevant to Pre-Hilbert space, and for the operator-orderings of the 
constraints likewise. 
This enables a connection with a conjecture of DeWitt concerning operator-ordering in Chapter 6 of \cite{FileR}.  

\mbox{ }  

\noindent Note 1) This Sec's somewhat outlandish suggestion is not the only possible reason behind Motivations 2 and 3; indeed if Can plays a role, Can is an enlargement of Point, and Point is then the `Can that act on $\fM_{\sfA\sfB}(\ttQ^{\sfC})$' restriction, which suffices to cover DeWitt's issue.

\noindent Note 2) (RigPhase, Point), as classically motivated in Sec 11 and argued to be a sensible implementation of $\fQ$-primality, has the same objects and brackets as (Phase, Can) and hence the same KinQuant.  
But in restricting Can to Point as morphisms preserving Poisson brackets and the rigging, then ComAl's morphisms would likewise be restricted.  
A possible problem with this view is that QM's $\ttP$'s and $\ttQ$'s may be more alike than their classical counterparts.  

\mbox{ } 

\noindent QM $\fQ$-Primality Motivation 4) A further reason to question the licitness of all canonical transformations is that classical equivalence under canonical transformations is in general broken in the passage to QM as per e.g. the {\bf Groenewold--Van Hove phenomenon} \cite{GVH1, GVH2}.  
Moreover, the Groenewold--Van Hove theorem continues to apply in the RigPhase approach, since one can envisage choosing 
$Q^3$ and $P^3$ there instead of $Q$ and $P$.  
Thus, whilst RigPhase is an illustration of weakening the canonical transformations, it is not the {\sl correct} weakening to take into account Gronewold--Van Hove; I leave as an open question what the set of canonical transformations that do become unitary transformations at the quantum level is characterized by.  

\mbox{ } 

\noindent Note 2) Using just Point does {\sl not} resolve out all aspects of unitary inequivalence, though it may help/may be a step in the right direction.
A deeper question is which weakening of the Can's or Point's at the classical level form up into classes that are preserved as unitary equivalence at the quantum level. 
(RigPhase, Point) itself is at the bottom of the hierarchy, whereas the solution of the currently posed problem is likely to involve a rather larger proportion of the canonical transformations.   

\mbox{ } 

\noindent Consequences of using RigPhase in place of Phase are as follows. 
Holding canonical transformations in doubt affects Internal Time and Histories Theory approaches to the Problem of Time, as well as Ashtekar variables and the recent Linking Theory approach \cite{Kos2, Kos1, Merc, Kos3, Gomes, Kos4, GrybT, Gom2}. 
The second and fourth of these make {\sl more} than the usual amount of use of canonical transformations;
this Sec suggests, moreover, the possibility of developing physics in the {\sl opposite} direction.    

\mbox{ } 

\noindent {\bf Histories primality alternative} Further reasons for Histories Theory appear at the QM level \cite{Hartle, GMH11}.  
This follows on from the path integral perspective, and additionally represents various modified interpretations of QM that are more suitable in the whole-universe context.

%====================================================================================================================================================
%====================================================================================================================================================
\section{Arguments against quantization in term of functoriality}
%====================================================================================================================================================
%====================================================================================================================================================

I analyze these by `taking this functor apart' into various simpler constituents as per Figs 3 and 4.
A number of these steps already have individual difficulties.  

\noindent Two reasons for the KinQuant functor to be `bad' (or, at least, ambiguous) are  
1) preferred subalgebra choice (and the nontriviality thereof by e.g. Groenewold--Van Hove  and 2) global considerations.
 
\noindent 3) A further reason for badness is anomalies arising in the passage from the Poisson brackets algebra to the commutator algebra. 

\noindent 4) In the case of H-rep for geometrodynamics, there are also well-definedness issues (functional derivatives, combinations with no functional-analytic reasons to be well-behaved) and regularization issues.  

\noindent 5) On top of these individual difficulties, the order in which some of these procedures are performed affects the outcome (e.g. reduce and then conformal-order does not match up with conformal-order and then reduce); this is covered in detail in Chapter 6 of \cite{FileR}.  

\noindent This bad functor conclusion, if not all the details of why, matches with Baez's \cite{Baez12} point of view, which was probably  
arrived at considering other/more quantization approaches than Isham 1984.  

\mbox{ } 

\noindent A different line of objection to quantization itself is that it is only desirable in the incipient parts of conceiving a quantum 
theory (so it is tied to an understood classical theory, but QM is more fundamental so one needs to pass to first principles Quantum Theory 
which one then has the sometimes fatal hope of recovery of a recognized  classical limit). 
[For some particular examples, the functor does work out, but the above demonstrates that for suitably general theories there is not at all a standard prescription for quantizing, and formatting this as a functor does not in any way help with that.]    

\mbox{ } 

\noindent Moreover, further and distinct use of categories in quantization are as follows.   

\mbox{ } 

\noindent Option 1) quantizing other categories - treating their objects as one usually treats $\fQ$.
An simple example of this is in some Histories Theory work, in which one considers the `history brackets' of histories and their conjugate momenta \cite{IL2}.

\mbox{ }

\noindent More general Example 1).  This is how Isham \cite{Isham03} considered quantization on quite general categories, with the 

\noindent morphisms/arrows now playing the role of momenta.

%====================================================================================================================================================
%====================================================================================================================================================
\section{Perspectival postulates at the QM level}
%====================================================================================================================================================
%====================================================================================================================================================

Localized subconfigurations furthermore play an underlying role in Crane's thinking \cite{Crane}.  
This involves the following postulate (my nomenclature).  

\mbox{ } 

\noindent Perspectivalism 1) {\bf Quantum Theory ONLY makes sense for subsystems} (my caps).

\noindent In the quantum GR arena, Crane deals with this by considering splits of the universe with the observer residing on the surface of that split.\footnote{In fact, Crane \cite{Crane93} defines the observer to be such a boundary of a localized region, though I caution that not all 
%%%%%%%%%%%%%%%%%%%%%%%%%%%%%%%%%%%%%%%%%%%%%%%%%%%%%%%%%%%%%%%%%%%%%%%%%%%%%%%%%%%%%%%%%%%%%%%%%%%%%%%%%%%%%%%%%%%%%%%%%%%%%%%%%%%%%%%%%%%%%%%%%%%%%
boundaries will have actual observers upon them and that sizeable boundaries would need to be populated by many observers, forming `shells of observers'/an array of detectors.
I also note that, at least as far as I can tell, Crane's view (and Rovelli's less specific one \cite{Rov96}) of observers both strip them of 
any connotations of animateness or manifestly capacitated for processing information.
I term this the {\it introspective implementation}, as it involves a shell of inward-looking observers.}
%%%%%%%%%%%%%%%%%%%%%%%%%%%%%%%%%%%%%%%%%%%%%%%%%%%%%%%%%%%%%%%%%%%%%%%%%%%%%%%%%%%%%%%%%%%%%%%%%%%%%%%%%%%%%%%%%%%%%%%%%%%%%%%%%%%%%%%%%%%%%%%%%%%%%
%
Each of these splits then has its own Hilbert space.
Crane then maintains that the whole universe has no Hilbert space (so his version of what I term Persp($\fQ$) has $\subset$ in place 
of my $\underline{\subset}$),  though a Hilbert space is to be recovered in a semiclassical limit.  

\mbox{ } 

\noindent I dissect this as follows due to not agreeing will all parts of it.  

\noindent Crane 1) QM makes sense for subsystems (true, and obvious); each has its own Hilbert space, within which the standard interpretation of QM applies.  

\noindent Crane 2) Almost all QM concerns subsystems (true at an elementary level, but sometimes forgotten in Quantum Cosmology).  

\noindent Crane 3) What would be QM for the whole universe does not possess a Hilbert space or the standard interpretation of QM. 

\noindent A--Crane Difference 1) Here I disagree about the lack of Hilbert space (shrunken as it may be due to closed-universe effects as per Secs 7 and 10), but agree about the subsequent non-applicability of the standard interpretation (though this is well known and plenty of different groups of authors have been willing to work under such conditions). 
That is why I use $\underline{\subset}$ and not $\subset$; this difference also accounts for my replacing `all' by `almost'.  
I envisage the scalefactor of the universe as a possible whole-universe property that can plausibly enter one's physical propositions; at the very least, scale is not a {\sl localized} subsystem.  

\noindent A--Crane Difference 2) Finally, Crane did not stipulate his  Sub$\fQ$ to carry the same local physical connotations as mine (as per Sec 14.1.1).  

\noindent Crane 4) QM makes sense for the universe as a whole in some kind of semiclassical limit. 
Here I agree with Crane, though in detail we may well not both mean exactly the same thing by a `semiclassical limit'.
Moreover Crane 4) is less necessary from my perspective (it applies just to interpretation, rather than to the possession of a Hilbert space as well).

\mbox{ } 

\noindent Perspectivalism 1W) (W for `weakened') is the name by which I shall refer to my above version [Crane 1, 2) and my side of 
A-Crane Differences 1, 2)].  

\mbox{ } 

\noindent Note 1) Perspectivalism 1W) is what I posit as the perspectival expansion of Relationalism 3) [though one {\sl could} instead elect to use Perspectivalism 1) in such a role]. 

\noindent Note 2) Whilst I have no evidence for such a connection, I comment that on the one hand one has multiplicity by inequivalent quantum theories and on the other hand multiplicity by multiple observer perspectives.  Are there ever any connections between these two multiplicities?  

\mbox{ }

\noindent Perspectivalism 2) allows for the QM of {\bf observers observing other observers observing subsystems}. 
This provides a further level of structure between the plethora of Hilbert spaces associated with all the perspectives of Perspectivalism 1).
This approach contends that  \cite{Rov96} {\bf QM propositions only makes sense in the context of particular observers} (my rewording).  
(This has no nontrivial classical counterpart.)

\mbox{ } 

\noindent Note 3) In RPM's, the perspectival approach can be toy-modelled using the notion of a localized fictitious test-observers who can measure nearby inter-particle distances and perhaps relative angles about their particular position, archetypes being considering the base subsystem of a triangleland or the base alongside the relative angle (the `fictitious test-observer more or less at the centre of mass of a localized binary'). 
This can accommodate Perspectivalism 1) to some extent, but cannot meaningfully incorporate Perspectivalism 2) which explicitly requires 
non-negligibility of observer material properties.

\noindent Note 4) The Crane set-up then involves `the entire set of SubQuant's', with Perspectivalism 2 necessitating yet further structure 
beyond that, as well as nontrivial modelling of observers as quantum systems [an idea nobody can really calculate with, which also features 
in the thinking of Page \cite{Page1, Page2} and of Hartle \cite{IGUS}; the state of the art here is Hartle's information gathering and utilizing system (IGUS) model for observers \cite{IGUS, GMH11}.

\mbox{ }

\noindent My preceding SSec makes the possible advance that `the entire set' requires physical Sub$\fQ$'s rather than the greater mathematical generality 
of Sub$\fQ$'s, and that it is additionally contingent on a definition of localization in space to be obeyed as a further practical consideration 
in the selection of a set.  
I add that these desirable features would not appear to be conducive to the implementing sets being mathematically simple; the notion of localization in space in particular would introduce a source of ambiguity into the model: {\sl how} local and by what criterion.

\mbox{ } 

\noindent The concepts embodied by Perspectivalism 3) and 4) carry over straightforwardly to the quantum level.  
A form of Perspectivalism 4) purely at the QM level had previously been proposed by Page and Wootters \cite{PW83}, and this position (if not accompanying technical details) was also taken up by Gambini--Porto--Pullin \cite{GPP03, GPP04a}.

%===================================================================================================================================================
%===================================================================================================================================================
\section{Quantum Propositioning}
%===================================================================================================================================================
%===================================================================================================================================================

\noindent Represent propositions at the quantum level by projectors is a key move.
In ordinary Quantum Theory, for state $\Rho$ and proposition $P$ implemented by projector $\widehat{\mP}$, Prob($P$; $\Rho$) = tr($\widehat{\Rho}\widehat{\mP}$) with Gleason's theorem providing strong uniqueness criteria for this choice of object from the perspective of satisfying the basic axioms of probabilities (see e.g. \cite{Ibook}).
Furthermore, in some approaches to the Problem of Time, one goes beyond \cite{IL2} 
the usual context and interpretation that are ascribed to projectors in ordinary Quantum Theory. 
Propositions were used from the inception in the \CPI \cite{PW83}.
Desiring a projector implementation of propositions led to Isham and Linden's reformulation of Histories Theory. 
The Records Theory within Histories Theory is, more trivially, satisfactory from the perspective of having its propositions represented by 
projectors.  
On the other hand, the \NSI and Halliwell approaches are still using configuration space regions rather than proposition-projector association
\cite{FileR, AHall}. 
Prima facie, this is suspect.  
Moreover, one might be able to pin a topos-type interpretation on such practises (though this has indeed not been worked out yet).  

\mbox{ }  

\noindent One possible more general mathematical implementation of propositioning is in {\bf Isham--Doering (ID)-type approaches} \cite{ToposI, ToposRev} to using Topos Theory  in physics provides a large amount of candidate structure as regards realizing Prop($\fS$) at the quantum level. 
\noindent Quantum-mechanically, there are distributivity issues with the logic, insofar as quantum logic \cite{Qlog} is not distributive.  
However, alternatively, internal logic, which ID term `intuitionistic logic' \cite{bigcite2,ToposI,ToposTalk,ToposRev} may be more applicable than quantum logic, and this is now distributive again.  
Given a state, truth values are assigned to all propositions about the quantum system, in which sense this is a neo-realist approach.  
N.B. this is independent of measurements/observers; this is to be compared with the Crane-type characterization of subsystems in terms of boundary observers (Sec 15) in Sec 16.6.   
 
\noindent In fact, the ID study has 2 languages at the quantum level: as well as the propositional language PL($\fS$), it has a higher-order typed language, L($\fS$).  
The propositional language is simpler and more directly related to standard quantum logic. 
However, standard quantum logic comes with various problems, so that being an intuitionistic logic distinct from standard quantum logic is an 
advantageous feature of $\mL(\fS$) [as opposed to $\mP\mL(\fS)$].  

Here, there is a map 
that maps projectors to subobjects of the spectral presheaf, which is the QM analogue of a classical state space.\footnote{A presheaf 
%%%%%%%%%%%%%%%%%%%%%%%%%%%%%%%%%%%%%%%%%%%%%%%%%%%%%%%%%%%%%%%%%%%%%%%%%%%%%%%%%%%%%%%%%%%%%%%%%%%%%%%%%%%%%%%%%%%%%%%%%%%%%%%%%%%%%%%%%%%%%%%%%%%
is a mathematical implementation of the idea of attaching local data to a structure; as such it ought to be of considerable interest 
in Records Theory and in Histories Theory, for all that hitherto these have been modelled with more (metric) structure.} 
%%%%%%%%%%%%%%%%%%%%%%%%%%%%%%%%%%%%%%%%%%%%%%%%%%%%%%%%%%%%%%%%%%%%%%%%%%%%%%%%%%%%%%%%%%%%%%%%%%%%%%%%%%%%%%%%%%%%%%%%%%%%%%%%%%%%%%%%%%%%%%%%%%%
%
Moreover, the topos perspective itself favours the higher-order language L($\fS$) as the natural kind of language that can be represented in a topos; it is `internal' to it. 

\mbox{ } 

\noindent Note 1) This approach does require radical revision of QM itself \cite{ToposI, ToposRev}, amounting to avoiding the obstruction in standard QM due to the Kochen--Specker theorem.

\noindent Note 2) The logic role played in classical physics by the Boolean algebra is played more generally in QM-relevant topoi by 
a Heyting algebra \cite{ToposRev} (this and the Boolean algebra are both distributive lattices, the difference being that the Heyting 
Algebra has no law of the excluded middle).  

\noindent Note 3) Implementing propositions in this fashion is more sophisticated than the \NSI or Halliwell's combined approach, for which one 
does so by thinking about classical regions that have limitations as regards implementing the entirety of QM propositions and relations between them. 

\noindent Note 4) QM ends up involving the topos $\underline{\mbox{Sets}}^{V(H)^{\to\tp}}$, i.e. the topos of contravariant (`opposite') valued functors on the poset category $V(H)$ of commutative subalgebras of the algebra of bounded operators on the Hilbert space of the system, $H$.

%==================================================================================================
%==================================================================================================
\section{Conclusion}
%==================================================================================================
%==================================================================================================

This article concerned relationalism.  
I contrasted and composed the very different LMB (Leibniz--Mach--Barbour) and RC (Rovelli--Crane) uses of this word.  
Moreover, I argued strongly that relationalism is relevant to background independence and that background independence is an important 
part of many of the leading `Quantum Gravity' programs, to the extent that I took QG to stand, rather, for {\sl Quantum Gestalt}. 
This refers to the Relativistic Gravity--Background Independence Gestalt as exemplified by GR (taken to be not only a set of relativistic field equations for gravitation, but also a {\it philosophical perspective} about freeing physics of background structures held by Einstein among others).  
Finally, I argued that Background Independence is wont to lead to many of the facets of the Problem of Time, so that valuing Background 
Independence means having to, among other concerns, face up to this conceptually and technically difficult issue.

In the LMB approach, Jacobi--Synge actions are more relationally primary than Euler--Lagrange ones or equations of motion, since the first are Leibniz-timeless whilst the last two are, from the relational perspective, in terms of a Machian emergent time and thus not possibly relationally primary.
Leibniz timelessness implies no primary notion of velocity and thus that the usual canonical definition of momentum needs reformulating; 
I provide how to do this in terms of the Jacobi--Synge line element. 
I gave further $\fQ$ and $\fG$ compatibility requirements.
I considered QM counterparts of the classical relational implementation.

I then provided four expansions on Barbour's `$\fQ$ is primary': to phase space (or possibly some slightly weaker structure), 
categorization, propositioning and perspecting.
Whilst categorization might be argued for as a prerequisite to quantization, I argued against this point of view in general; however, a more complicated version of categorization ends up underlying propositioning and perspecting.
I argued that the ideas of Isham, Doering and Butterfield concerning applying Topos Theory to Quantum Theory are one possible implementation of 
perspecting.  
Finally, I showed how perspecting is how the LMB relational approach makes contact with the Rovelli--Crane (RC) type of relational approach.

Moreover, there is tension between some parts of LMB and RC relationalism, since LMB takes Mach's `time is to be abstracted from change' to involve {\sl all} change, whilst Rovelli's approach takes it to involve {\sl any} change (which I term AMR time: Aristotle--Mach--Rovelli).  
I argue then that it is better to hold to an intermediate interpretation: a {\sl sufficient totality of locally relevant change}, 
% 
%(STLRC)
%
as judged at the level of which relative forces significantly affect the equations of motion, which I term LMB-A time, is new to this article and is further elaborated upon in \cite{ARel2} (see Sec 22.1 for an outline).  
This covers further detail of this approach, and how each of these three approaches leads to distinct attitudes to the theory of time-keeping and 
to distinct timeless approaches to the Problem of Time in Quantum Gestalt.

%========================================================================================================
\subsection{Application 1) 3 interpretations of Mach's `time is to be abstracted from change' principle}
%========================================================================================================

Each of these three has a supporting argument.
Rovelli's interpretation applies generically (regardless of physical regime or material composition of the system in question).  
On the other hand, LMB's interpretation, by using the totality of the change, produces an {\sl incontestable} time, in that there is no more 
change elsewhere that can run in suspiciously irregular ways that cause of one to doubt one's timestandard.  
However, there is plenty of middle ground between these two positions.  

\noindent Leibniz--Barbour's statement of classical bestness implies the following more general notion.  

\mbox{ }

\noindent [LBA] {\bf Some times are better than others}.

\mbox{ }

\noindent Note 1) this holds without having to go to the whole-universe extreme, which is impractical, and operationally undesirable -- we know little of the motions and constitutions of very distant bodies.  
Also, since Leibniz's day, the notion of `universe' has gotten a whole lot more complicated with GR -- were this taken seriously, does it refer to the Hubble radius or to the actual entirely of the universe including unobservable portions?  
N.B. that this Leibniz--GR discrepancy in the nature of the universe becomes a red herring in the below discussion.  

\noindent Note 2) This also raises a partial objection to the Rovelli scheme, in that whilst Rovelli does not state he is looking for `good clocks', those are a fair notion to ask about, and yet clash conceptually with Rovelli-type notions that any time, or any clock, will do.  
On the other hand, Barbour's approach is very conducive toward asking and conceptually answering this question.    

\noindent Note 3) in counterpoint, it might be argued that this is a question which in general has no answer, due to Rovelli's genericity.  
However, I propose a distinct LMB-A' resolution to this issue below.  

\mbox{ }

\noindent Relationalism 7M) {\bf Time is best in practise abstracted from a sufficient totality of locally relevant change.}  

\mbox{ } 

\noindent This is `middling' (M) between LMB Relationalism 7) and AMR Relationalism 7W); it conforms to what consists sufficiently accurate timekeeping in practise.      

\noindent In comparison to AMR, it has the extra element of acknowledging some times are locally more useful to consider than others.  
Even in a generic situation, one can have a local ranking procedure for candidate times, alongside a refining procedure 
until acceptable accuracy of predictability is obtained.  

\noindent In comparison with LMB time, the practically-used ephemeris-type method for the solar system is entirely adequate without having to consider the net physical effect (tidal effect) of distant massive bodies such as Andromeda (thus also rendering irrelevant our increasing lack of precise knowledge of the masses and positions as one considers more and more distant objects). 

\noindent I refer to 6) and 7M) together as the LMB-A view of time.  

\noindent 7M) is built on the following wider view of Barbour and I.   

\mbox{ } 

\noindent {\bf MORE change is MORE reliable as a timestandard provider than less change}.  

\mbox{ } 

\noindent  Moreover, each of AMR, LMB and LMB-A entails a different attitude to timekeeping.  

\mbox{ } 

\noindent {\bf AMR timekeeping}: using any system as a clock, including very localized subsystems being taken to be able to work very well as clocks in isolation from wider observations of change.  

\mbox{ } 

\noindent {\bf LMB timekeeping}: the universe is only perfect clock. 
\mbox{ }  

\noindent {\bf LMB-A timekeeping}: is like astronomers' ephemeris actually was, or atomic clock timekeeping from the perspective including  recalibration by comparison with solar system dynamics.  
This more faithful representation of the astronomers' actual procedure (such as ephemeris time or atomic time recalibrated in accord with solar-system physics) is why I take inquiring in detail into what is meant by `universe' in the LMB approach to be a red herring.  

\noindent Relationalism 7M) is to be implemented by a {\bf Generalized Local Ephemeris Procedure}. 
I.e., do not just use a change to abstract a time, but also check whether using this time in the equations of motion for other 
changes suffices to predict these to one's desired accuracy.  
Further detail of this approach, and of how each of the above three approaches leads to distinct attitudes to the theory of time-keeping and 
to distinct timeless approaches to the Problem of Time in QG with distinct notions of emergent time is the subject of \cite{ARel2}.

\mbox{ } 

\noindent N.B. that the Relationalism 7 versus Relationalism 7W clash means that one cannot just add the RC postulates to the LMB ones; 
Relationalism 7M then represents a compromise between these [and is also the most accurate encapsulation of astronomical (based) timestandards].

%================================================================================================================================================
\subsection{Other applications} 
%================================================================================================================================================

This article's main applications are as follows.

\noindent Application 2) is the wider relational assessment of the Problem of Time's facets and strategies.
\noindent As regards Frozen Formalism Problem resolving strategies, Leibniz-time, Mach-time and tangibility arguments hold against most
Tempus ante Quantum strategies (internal time, matter time...).
On the other hand, emergent JBB time and the approximately equivalent emergent semiclassical time comply with these criteria.  
Timeless and Histories Theory approaches generally do so too.  
Many other Problem of Time facets have also been exposited as relational in this article. 
In particular, I have argued that the Thin Sandwich Problem facet is generalized by, and relationally interpreted as, the Best Matching Problem, 
with other LMB-relational ties being established to the Foliation Dependence Problem, Functional Evolution Problem, Problem of Beables and Spacetime Reconstruction Problem.
On the other hand, the Global and Multiple Choice Problems of time have been found to have at least some ties to Rovelli's approach to 
relationalism (which is also an alternative stance on the Problem of Beables and the most sympathetic of the relational positions considered 
in this article as regards internal time and matter time approaches).

Application 1)'s finer distinctions between the LMB, LMB-A, and AMR attitudes to Mach's time principle serve as further distinction between various 
timeless approaches and semiclassical-histories-timeless approach combinations, though this is largely left to a subsequent article \cite{ARel2}.  
Application 6) below also concerns a way of enhancing Timeless Records Theory/composites of at least Histories Theory and Records Theory.

Finally, this article's more minimalist option of not considering (all) canonical transformations at least partly selects against Internal Time, Histories Brackets, Ashtekar Variables and Linking Theory based approaches.

\mbox{ } 

\noindent Application 3) concerns the status of scale in physical theories.  
Barbour argues against it, in good part due to its heterogeneity as regards all the other degrees of freedom (the shapes). 
However, I argue that this heterogeneity can be used to obtain an approximate timestandard based on the scale or some closely-related variable.  
Whilst this in itself does not comply with my or Barbour's interpretation of the Mach time principle, allowing the shape degrees of freedom 
to contribute perturbatively to the more accurate form of resulting timestandard does comply thus.  
Moreover, this looks to be in particular an accurate rendition of classical and quantum cosmology (due to the good fortune of anisotropies and inhomogeneities being much smaller than scale effects in that setting).

\mbox{ } 

\noindent Application 4) concerns further foundations for Ashtekar variables/LQG.
Ashtekar variables/LQG are often stated to be relational; more presicsely, that means Rovelli-relational. 
In the present article, however, I laid out how Ashtekar Variables are also LMB(-A) relational.  
This is of foundational importance in widening the use of `relational' to describe, motivate, structure and further develop the 
Ashtekar variables/LQG program.  
I pointed out how RPM's and geometrodynamics form a natural sequence of theories of geometrical configurations, whilst regarding the 
Astekar bein variables as a momentum is then somewhat incongruous.
On the other hand, electromagnetism, Yang--Mills Theory and Ashtekar variables form a {\sl rival} natural sequence.  
I also pointed out that whilst LQG is a relational approach in all of this article's senses, it does not presently bear a relational name, 
(unlike geometrodynamics or shape dynamics, which are named for their configurationally relational contents).
I suggested Knot Quantum Gestalt, Quantum Nododynamics or Quantum Spin-Knot Dynamics (knots being configurationally relational 
whilst loops include a Diff(3)'s worth of physical redundancy).  

\mbox{ } 

\noindent Application 5) is Relationalism versus Supersymmetry, which shall be continued in \cite{Arelsusy}.  

I emphasize that by the comments I provide in theis article on Ashtekar variables/LQG, supersymmetry, String Theory and M-Theory, I demonstrate  contact between LMB relationalism and much of modern Theoretical Physics.  

\mbox{ } 

\noindent Application 6) concerns bringing together ID and timeless approaches (some of which also involve elements of RC relationalism).  
All of these areas of study principally concern subsystems.  
In the case of ID, this is via at least the great majority of QM propositions that can be asked are about subsystems.
ID has the broadest mathematization (Topos Theory), whilst Crane uses categories. 
Records Theory does not conventionally make use of these structures, and yet ID's development supports Histories Theory (which contains, and 
shared many similarities with, Records Theory) and sheaves \cite{MM68, JLBell} as a means of locally attaching information may well provide an 
insightful new formulation for Records Theory.

There is however a chasm to bridge.  
In Records Theory, records are local in a {\sl metric space} sense, whilst topoi and sheaves have a structurally weaker {\sl topological space} 
notion of locality.  
Thus one would need to extend/specialize (provide an extra layer of structure) to ID's work or consider the extent to which one can conceptualize with records using solely a topological notion of locality.  

\mbox{ }

\noindent {\bf Acknowledgements}: I thank: my wife Claire, Amelia, Sophie, Sophie, Anya, Bryony, Amy, Hettie, Hannah, Duke, Becky and 
one other for being supportive of me whilst this work was done.   
Professors Julian Barbour, Enrique Alvarez, Louis Crane, Jonathan Halliwell, Chris Isham, Karel Kucha\v{r}, Marc Lachi\`{e}ze-Rey and Jeremy 
Butterfield, Dr's Cecilia Flori, Sean Gryb, Tim Koslowsk and Ioannis Alevizos, and Mr Eduardo Serna for discussions between 2008 and the present.  
Professors Belen Gavela, Marc Lachi\`{e}ze-Rey, Malcolm MacCallum, Don Page, Reza Tavakol and Jeremy Butterfield for support with my career.  
I was funded by a grant from the Foundational Questions Institute (FQXi) Fund, 
a donor-advised fund of the Silicon Valley Community Foundation on the basis of proposal FQXi-RFP3-1101 to the FQXi.  
I thank also Theiss Research and the CNRS for administering this grant.  
This work was started at DAMTP, Centre for Mathematical Sciences, Cambridge,   
and continued at F\'{\i}sica Te\'{o}rica, Universidad Autonoma de Madrid.

%=====================================================BIBLIOGRAPHY==========================================================================

\end{document}